\documentclass[12pt]{article}
\usepackage{color}
\usepackage{amssymb,amsmath,comment}
\usepackage{graphicx}
\usepackage{braket}
\usepackage{epsfig}
\usepackage{here}
\graphicspath{{./Figures/}}
\newcommand{\ba}{\begin{align}}
\newcommand{\ea}{\end{align}}

\newcommand{\ov}{\overline}
\def\nn{\nonumber}

\def\bea{\begin{eqnarray}}
\def\eea{\end{eqnarray}}
 \makeatletter
\def\alt{\mathrel{\mathpalette\gl@align<}}
\def\agt{\mathrel{\mathpalette\gl@align>}}
\def\gl@align#1#2{\lower.6ex\vbox{\baselineskip\z@skip\lineskip\z@
\ialign{$\m@th#1\hfil##\hfil$\crcr#2\crcr\sim\crcr}}} \makeatother
\renewcommand{\thefootnote}{\fnsymbol{footnote}}
\begin{document}
\begin{flushright}
\end{flushright}
\vspace*{1.0cm}

\begin{center}
\baselineskip 20pt 
{\Large\bf 
Renormalizable $SO(10)$ GUT \\
with Suppressed Dimension-5 Proton Decays
}
\vspace{1cm}

{\large 
Naoyuki Haba$^a$,
\ Yukihiro Mimura$^{a,b}$
\ and \ Toshifumi Yamada$^a$
} \vspace{.5cm}

{\baselineskip 20pt \it
$^a$ Institute of Science and Engineering, Shimane University, Matsue 690-8504, Japan\\
$^b$ Department of Physical Sciences, College of Science and Engineering, \\
Ritsumeikan University, Shiga 525-8577, Japan
}

\vspace{.5cm}

\vspace{1.5cm} {\bf Abstract} \end{center}

We study a renormalizable SUSY $SO(10)$ GUT model where the Yukawa couplings of 
 single ${\bf 10}$, single ${\bf \overline{126}}$ and single ${\bf 120}$ fields, $Y_{10},Y_{126},Y_{120}$,
 account for the quark and lepton Yukawa couplings and the neutrino mass.
We pursue the possibility that $Y_{10},Y_{126},Y_{120}$ reproduce the correct quark and lepton masses, CKM and PMNS matrices
 and neutrino mass differences, and at the same time suppress dimension-5 proton decays (proton decays via colored Higgsino exchange)
 through their texture,
 so that the soft SUSY breaking scale can be reduced as much as possible without 
 conflicting the current experimental bound on proton decays.
We perform a numerical search for such a texture, and investigate implications of that texture on unknown neutrino parameters,
 the Dirac CP phase of PMNS matrix, the lightest neutrino mass and the $(1,1)$-component of the neutrino mass matrix in the charged lepton basis.
Here we concentrate on the case when the active neutrino mass is generated mostly by the Type-2 seesaw mechanism,
 in which case we can obtain predictions for the neutrino parameters from the condition that 
 dimension-5 proton decays be suppressed as much as possible.

\thispagestyle{empty}

\newpage
\renewcommand{\thefootnote}{\arabic{footnote}}
\setcounter{footnote}{0}
\baselineskip 18pt
\section{Introduction}

The $SO(10)$ grand unified theory (GUT)~\cite{Georgi:1974my,Fritzsch:1974nn} is a well-motivated scenario beyond the Standard Model (SM), since
 it unifies the SM gauge groups into an anomaly-free group, 
 it unifies the SM matter fields and the right-handed neutrino of each generation
 into one {\bf 16} representation, and it accommodates the seesaw mechanism for the tiny neutrino mass~\cite{seesaw1,seesaw2,GellMann:1980vs,seesaw3,seesaw4}.
Renormalizable $SO(10)$ GUT models~\cite{Matsuda:2000zp}-\cite{Deppisch:2018flu}, 
 where the electroweak-symmetry-breaking Higgs field originates from ${\bf 10},\,{\bf \ov{126}},\,{\bf 120}$
 fields (or some of them) and the SM Yukawa couplings stem from renormalizable terms 
 $\tilde{Y}_{10}\,{\bf 16}\, {\bf 10}\, {\bf 16}+\tilde{Y}_{126}\,{\bf 16}\, {\bf \ov{126}}\, {\bf 16}+\tilde{Y}_{120}\,{\bf 16}\, {\bf 120}\, {\bf 16}$
 (or part of them),
 are particularly interesting, because the SM Yukawa couplings and the active neutrino mass are described in a unified manner
 with fundamental Yukawa couplings $\tilde{Y}_{10},\tilde{Y}_{126},\tilde{Y}_{120}$.
Specifically, the up-type quark, down-type quark, charged lepton and neutrino Dirac Yukawa matrices are derived as
$Y_u=Y_{10}+r_2Y_{126}+r_3Y_{120}$, $Y_d=r_1(Y_{10}+Y_{126}+Y_{120})$, $Y_e=r_1(Y_{10}-3Y_{126}+r_eY_{120})$, $Y_D=Y_{10}-3r_2Y_{126}+r_\nu Y_{120}$,
 with $Y_{10}\propto \tilde{Y}_{10}$, $Y_{126}\propto \tilde{Y}_{126}$, $Y_{120}\propto \tilde{Y}_{120}$,
 and $r_1,r_2,r_3,r_e,r_\nu$ being numbers.
The Majorana mass for right-handed neutrinos and the Type-2 seesaw~\cite{Schechter:1980gr,Lazarides:1980nt,Mohapatra:1980yp} contribution to the active neutrino mass
 are both proportional to $Y_{126}$.

Supersymmetric (SUSY) GUT models are currently severely constrained by the non-observation of proton decay through dimension-5 operators from colored Higgsino exchange~\cite{Weinberg:1981wj,Sakai:1981pk}, the most stringent bound being on the $p\to K^+ \nu$ mode~\cite{Abe:2014mwa}.
This constraint is imminent in SUSY renormalizable $SO(10)$ GUT models,
 because natural unification of the top and bottom quark Yukawa couplings requires $\tan\beta\sim50$.
For such large $\tan\beta$, 
 right-handed dimension-5 operators $E^cU^cU^cD^c$ give a significant contribution to the $p\to K^+\bar{\nu}_\tau$ decay~\cite{Goto:1998qg}, 
 and it is hard to realize a cancellation in the $E^cU^cU^cD^c$ operators' contribution and that of 
 left-handed dimension-5 operators $QQQL$ to the $p\to K^+\bar{\nu}_\tau$ decay
 {\it and} a cancellation in the $QQQL$ operators' contributions to the $p\to K^+\bar{\nu}_\mu$ decay.
Besides, it is impossible to enhance the colored Higgsino mass well above $2\cdot10^{16}$~GeV 
 (by some adjustment of the mass spectrum of GUT-scale particles that modifies the unification conditions)
 because the $SO(10)$ gauge coupling would become non-perturbative immediately above
 the thresholds of the components of rank-5 ${\bf 126}+{\bf \ov{126}}$ fields
\footnote{
We consider that field-theoretical description breaks down immediately above the GUT-scale and hence
 such non-perturbative behavior of the $SO(10)$ gauge theory is not realized in Nature.
Additionally, we assume that there are no higher-dimensional operators suppressed by the scale of the breakdown of field-theoretical description.
Therefore, we can neglect the contribution of higher-dimensional operators to the Yukawa coupling unification.
}. Although one can increase the soft SUSY breaking scale to suppress dimension-5 proton decays,
 the higher the SUSY particle masses, the more the naturalness of the electroweak scale is lost.
In this situation, it is worth recalling that it is the fundamental Yukawa couplings $\tilde{Y}_{10},\tilde{Y}_{126},\tilde{Y}_{120}$
 that determine the coefficients of the dimension-5 operators.
There may be a texture of the fundamental Yukawa couplings that suppresses dimension-5 proton decays
 and at the same time reproduces the correct quark and lepton Yukawa couplings and neutrino mass matrix.
Specifically, as the up quark Yukawa coupling is a specially small Yukawa coupling in the minimal SUSY Standard Model (MSSM)
 with $\tan\beta\sim50$, if those components of the Yukawa matrices $\tilde{Y}_{10},\tilde{Y}_{126},\tilde{Y}_{120}$
 responsible for dimension-5 proton decays
 are related to the up quark Yukawa coupling, then dimension-5 proton decays are maximally suppressed.
The above idea has been sought for in Refs.~\cite{Dutta:2004zh,Dutta:2005ni} based on the model that includes
 single ${\bf 10}$, single ${\bf \ov{126}}$ and single ${\bf 120}$ fields~\cite{Dutta:2004hp,Bertolini:2004eq,Yang:2004xt}.

In this paper, we perform a numerical search for such a texture in the model that includes
 single ${\bf 10}$, single ${\bf \ov{126}}$ and single ${\bf 120}$ fields, by the following steps.
First, we spot those components of the Yukawa matrices $Y_{10},Y_{126},Y_{120}$ (proportional to $\tilde{Y}_{10},\tilde{Y}_{126},\tilde{Y}_{120}$)
 which can be reduced to suppress dimension-5 proton decays without conflicting the requirement that they reproduce
 the correct quark and lepton Yukawa couplings and neutrino mass matrix.
Next, we numerically fit the experimental data on the quark and lepton masses,
 CKM and PMNS mixing matrices and neutrino mass differences in terms of $Y_{10},Y_{126},Y_{120}$,
 and meanwhile we minimize the components of $Y_{10},Y_{126},Y_{120}$ spotted above.
In this way, we numerically discover a texture of the fundamental Yukawa couplings that suppresses dimension-5 proton decays and
 reproduces the correct fermion data.
We further discuss implications of the texture on unknown neutrino parameters, in particular
  the Dirac CP phase of PMNS matrix, $\delta_{\rm pmns}$,
  the lightest neutrino mass, $m_1$, and the $(1,1)$-component of the neutrino mass matrix in the charged lepton basis, $m_{ee}$, that regulates
  the neutrinoless double beta decay.

The present paper focuses on the case when the active neutrino mass is dominated by the Type-2 seesaw contribution
 coming from the tiny vacuum expectation value (VEV) of ${\bf \ov{126}}$ field,
 whereas the Type-1 seesaw contribution resulting from integrating out right-handed neutrinos is assumed subdominant.
In this case, the neutrino mass matrix is directly proportional to $Y_{126}$
 and we can derive predictions for the neutrino parameters 
 from the condition that dimension-5 proton decays be suppressed as much as possible.


This paper is organized as follows:
In Section~\ref{section-model}, we review the renormalizable SUSY $SO(10)$ GUT model where 
 the electroweak-symmetry-breaking Higgs field originates from single ${\bf 10}$, single ${\bf \ov{126}}$ and single ${\bf 120}$ fields.
We also re-derive the dimension-5 proton decay partial widths, and clarify the relation between the dimension-5 proton decays
 and the Yukawa couplings $Y_{10},Y_{126},Y_{120}$.
In Section~\ref{section-inferring}, we spot those components of the Yukawa matrices $Y_{10},Y_{126},Y_{120}$
 which can be reduced to suppress dimension-5 proton decays without conflicting the requirement that they reproduce
 the correct quark and lepton Yukawa couplings and neutrino mass matrix.
In Section~\ref{section-fitting}, we perform a numerical search for a texture of $Y_{10},Y_{126},Y_{120}$
 that suppresses dimension-5 proton decays and at the same time reproduces the correct fermion data,
 and discuss a connection between the suppression of dimension-5 proton decays and the neutrino parameters.
Section~\ref{summary} summarizes the paper.
\\

\section{Renormalizable SUSY $SO(10)$ GUT}
\label{section-model}

We consider a SUSY $SO(10)$ GUT model that contains fields
 in ${\bf 10}$, ${\bf 126}$, ${\bf \overline{126}}$, ${\bf 120}$ representations, denoted by
 $H$, $\Delta$, $\overline{\Delta}$, $\Sigma$,
 and three matter fields in {\bf 16} representation, denoted by $\Psi_i$ ($i$ is the flavor index).
The model also contains fields in ${\bf 210}$, ${\bf 45}$ representations, denoted by $\Phi$, $A$, which are
  responsible for breaking $SU(5)$ subgroup of $SO(10)$.
The most general renormalizable Yukawa couplings are given by
\bea
W_{\rm Yukawa}\ =\ (\tilde{Y}_{10})_{ij}\,\Psi_i H\Psi_j+(\tilde{Y}_{126})_{ij}\,\Psi_i\overline{\Delta}\Psi_j+(\tilde{Y}_{120})_{ij}\,\Psi_i\Sigma\Psi_j
\eea
 where $\tilde{Y}_{10}$ and $\tilde{Y}_{126}$ are $3\times3$ complex symmetric matrices and $\tilde{Y}_{120}$ is a $3\times3$ complex antisymmetric matrix, 
 and $i,j$ are the flavor indices that correspond to that of $\Psi_i$.
The electroweak-breaking-Higgs fields of the Minimal SUSY Standard Model (MSSM), $H_u,H_d$, are linear combinations of 
 (${\bf 1}$, ${\bf2}$, $\pm\frac{1}{2}$) components of $H$, $\Delta$, $\overline{\Delta}$, $\Sigma$, $\Phi$.
Accordingly, the Yukawa coupling for up-type quarks, $Y_u$, that for down-type quarks, $Y_d$, and that for charged leptons, $Y_e$, 
 and the Dirac Yukawa coupling for neutrinos, $Y_D$, are derived as
\bea
W_{\rm Yukawa}\ \supset\ (Y_u)_{ij}\,Q_i H_u U_i^c+(Y_d)_{ij}\,Q_i H_d D_i^c+(Y_e)_{ij}\,L_i H_d E_i^c+(Y_D)_{ij}\,L_i H_u N_i^c
\label{susyyukawa}
\eea
 where $Y_u,\ Y_d,\ Y_e,\ Y_D$ are given by
\bea
&&Y_u\ =\ Y_{10}+r_2\,Y_{126}+r_3\,Y_{120},
\label{yu}\\
&&Y_d\ =\ r_1\left(Y_{10}+Y_{126}+Y_{120}\right),
\label{yd}\\
&&Y_e\ =\ r_1\left(Y_{10}-3Y_{126}+r_e\,Y_{120}\right),
\label{ye}\\
&&Y_D\ =\ Y_{10}-3r_2\,Y_{126}+r_D\,Y_{120}
\label{ydirac}
\eea
 at a $SO(10)$ breaking scale.
Here $Y_{10}\propto\tilde{Y}_{10},  \ Y_{126}\propto\tilde{Y}_{126},  \ Y_{120}\propto\tilde{Y}_{120}$ and 
 $r_1,r_2,r_3,r_e,r_D$ are numbers.
By a phase redefinition, we take $r_1$ to be real positive.

Majorana mass for the right-handed neutrinos is obtained as $(Y_{126})_{ij}\,\ov{v}_R\,N_i^c N_j^c$
 where $\ov{v}_R$ denotes $\overline{\Delta}$'s VEV.
Integrating out $N_i^c$ yields an effective operator $L_iH_uL_jH_u$, which we call the Type-1 seesaw contribution.
Additionally, the ({\bf 1}, {\bf3}, 1) component of $\overline{\Delta}$ mixes with that of $E$ after $SO(10)$ breaking.
Integrating out the ({\bf 1}, {\bf3}, 1) components yields an effective operator
 $L_iH_uL_jH_u$, which we call the Type-2 seesaw contribution.
This paper centers on the case where the Type-2 seesaw contribution dominates over the Type-1 one,
 in which case the Wilson coefficient of the Weinberg operator $(C_\nu)_{ij} L_iH_uL_jH_u$ satisfies
\bea
(C_\nu)_{ij} \ \propto \ (Y_{126})_{ij}
\label{cnu}
\eea
 at a $SO(10)$ breaking scale.
In Appendix~B, we present an example of VEV configurations that realize the dominance of the Type-2 seesaw contribution.
\\

$H$, $\Delta$, $\bar{\Delta}$, $\Sigma$, $\Phi$ contain pairs of
 ({\bf 3}, {\bf1}, $-\frac{1}{3}$) and (${\bf \overline{3}}$, {\bf1}, $\frac{1}{3}$) components,
 which we call `colored Higgs fields' and denote by $H_C^A$, $\ov{H}_C^B$ ($A,B$ are labels), respectively.
Exchange of $H_C^A,\ov{H}_C^B$ gives rise to dimension-5 operators inducing a proton decay.
Those couplings of $H_C^A,\overline{H}^B_C$ which contribute to such operators are
\bea
W_{\rm Yukawa}\ \supset\ \sum_A\left[ \ \frac{1}{2}(Y_L^A)_{ij}\,Q_i H_C^A Q_j +  (\overline{Y}^A_L)_{ij}\,Q_i \overline{H}^A_C L_j + (Y_R^A)_{ij}\,E^c_i H_C^A U^c_j +  (\overline{Y}_R^A)_{ij}\,U^c_i \overline{H}_C^A D^c_j  \ \right]
\label{coloredHiggsYukawa}
\eea
where $\ov{Y}_L^A,\,Y_R^A,\,\ov{Y}^A_R$ are proportional to $Y_{10},\,Y_{126}$ or $Y_{120}$,
 and $Y_L^A$ are proportional to $Y_{10}$ or $Y_{126}$.
After integrating out $H^A_C,\overline{H}^B_C$, we get effective dimension-5 operators contributing to proton decay,
\bea
-W_5 \ = \ \frac{1}{2}C_{5L}^{ijkl}\,(Q_k Q_{l})(Q_i L_{j}) + C_{5R}^{ijkl}\,E^c_k U^c_{l} U^c_i D^c_{j}
\eea
(in the first term, isospin indices are summed in each bracket) where
\begin{align}
C_{5L}^{ijkl}(\mu=\mu_{H_C})&= \left. \sum_{A,B}({\cal M}_{H_C}^{-1})_{AB}
\left\{(Y_L^A)_{kl}(\ov{Y}^B_L)_{ij}
-\frac{1}{2}(Y_L^A)_{li}(\ov{Y}^B_L)_{kj}-\frac{1}{2}(Y_L^A)_{ik}(\ov{Y}^B_L)_{lj}\right\}\right|_{\mu=\mu_{H_C}},
\label{c5lgeneral}\\
C_{5R}^{ijkl}(\mu=\mu_{H_C})&= \left. \sum_{A,B}({\cal M}_{H_C}^{-1})_{AB}
\left\{(Y_R^A)_{kl}(\ov{Y}_R^B)_{ij}-(Y_R^A)_{ki}(\ov{Y}_R^B)_{lj}\right\}\right|_{\mu=\mu_{H_C}},
\label{c5rgeneral}
\end{align}
 and ${\cal M}_{H_C}$ denotes the mass matrix for $H^A_C,\overline{H}^B_C$
 and $\mu_{H_C}$ is taken around the eigenvalues of ${\cal M}_{H_C}$.
 \\

We concentrate on the $(Q_k Q_{l})(Q_i L_{j})$ operators' contributions to 
 the $p\to K^+\bar{\nu}_\alpha$ $(\alpha=e,\mu,\tau)$ and $p\to K^0e_\beta^+$ $(e_\beta=e,\mu)$ decays and
 the $E^c_k U^c_{l} U^c_i D^c_{j}$ operators' contribution to the $p\to K^+\bar{\nu}_\tau$ decay.
For other decay modes, the $(Q_k Q_{l})(Q_i L_{j})$ operators' contributions to the $N\to \pi e_\beta^+$ and $p\to \eta e_\beta^+$ decays
 are suppressed in the same texture that suppresses the above contributions as we comment in Section~\ref{section-inferring}.
The rest of the decay modes are bounded only weakly~\cite{Tanabashi:2018oca} and so we do not discuss them in this paper.

The contribution of the $C_{5L}^{ijkl}(Q_k Q_{l})(Q_i L_{j})$ term to the $p\to K^+\bar{\nu}_\alpha$ ($\alpha=e,\mu,\tau$) decays is given by
\begin{align}
&\Gamma(p\to K^+\bar{\nu}_\alpha)\vert_{{\rm from} \,C_{5L}}
\ = \ {\cal C}\,
\left\vert \beta_H(\mu_{\rm had})\frac{1}{f_\pi}\left\{
\left(1+\frac{D}{3}+F\right)C_{LL}^{s\alpha \,ud}(\mu_{\rm had})+\frac{2D}{3}C_{LL}^{d\alpha \,us}(\mu_{\rm had})\right\}
\right\vert^2.
\label{ptoknuLformula}
\end{align}
Here ${\cal C}=\frac{m_N}{64\pi}\left(1-\frac{m_K^2}{m_N^2}\right)^2$ 
with $m_N$ denoting the nucleon mass and $m_K$ the kaon mass.
$\alpha_H,\beta_H$ denote hadronic matrix elements, $D,F$ are parameters of the baryon chiral Lagrangian, and
$C_{LL}$ are Wilson coefficients of the effective Lagrangian, $-{\cal L}_6\supset C_{LL}^{ijkl}(\psi_{u_Lk}\psi_{d_Ll})(\psi_{d_Li}\psi_{\nu_Lj})$ where $\psi$ denotes a SM Weyl spinor and spinor index is summed in each bracket.
The Wilson coefficients $C_{LL}$ satisfy
  \footnote{
  When writing $C_{5L}^{s\alpha\, du}$, we mean that $Q_i$ is in the flavor basis where the down-type quark Yukawa coupling is
  diagonal and that the down-type quark component of $Q_i$ is exactly $s$ quark (the up-type quark component of $Q_i$ is a mixture of $u,c,t$).
  Likewise, $Q_k$ is in the flavor basis where the down-type quark Yukawa coupling is
  diagonal and its down-type component is exactly $d$ quark, 
  and $Q_l$ is in the flavor basis where the up-type quark Yukawa coupling is
  diagonal and its up-type quark component is exactly $u$ quark.
  The same rule applies to $C_{5L}^{u\alpha \,ds}$ and others.
  }
\begin{align}
C_{LL}^{s\alpha\,ud}(\mu_{\rm had})= A_{LL}(\mu_{\rm had},\mu_{\rm SUSY})
\frac{M_{\widetilde{W}}}{m_{\tilde{q}}^2}
\,{\cal F}\,
g_2^2\left(C_{5L}^{s\alpha\,ud}-C_{5L}^{u\alpha\, ds}\right)\vert_{\mu=\mu_{\rm SUSY}},
\\
C_{LL}^{d\alpha\,su}(\mu_{\rm had})= A_{LL}(\mu_{\rm had},\mu_{\rm SUSY})
\frac{M_{\widetilde{W}}}{m_{\tilde{q}}^2}
\,{\cal F}\,
g_2^2\left(C_{5L}^{d\alpha\, us}-C_{5L}^{u\alpha\, ds}\right)\vert_{\mu=\mu_{\rm SUSY}}.
\end{align}
Here ${\cal F}$ is a loop function factor given by ${\cal F} =\frac{1}{x-y}(\frac{x}{1-x}\log x - \frac{y}{1-y}\log y)/16\pi^2 + \frac{1}{x-1}(\frac{x}{1-x}\log x+1)/16\pi^2$
  with $x=|M_{\widetilde{W}}|^2/m_{\tilde{q}}^2$ and $y=m_{\tilde{\ell}^\alpha}^2/m_{\tilde{q}}^2$.
Also, $M_{\widetilde{W}}$ denotes the Wino mass, $m_{\tilde{q}}$ the 1st and 2nd generation left-handed squark masses (which are usually degenerate),
 and $m_{\tilde{\ell}^\alpha}$ the mass of the left-handed slepton of flavor $\alpha$.
$A_{LL}(\mu_{\rm had},\mu_{\rm SUSY})$ accounts for SM renormalization group (RG) corrections in the evolution
  from soft SUSY breaking scale $\mu_{\rm SUSY}$ to a hadronic scale where the values of $\alpha_H,\beta_H$ are reported.
Here we neglect SM RG corrections involving the $u,d,s$-quark and charged lepton Yukawa couplings,
 and accordingly, quark flavor mixings along the RG evolution are neglected.
The Wilson coefficients $C_{5L}$ are related to the colored Higgs Yukawa couplings as
\begin{align}
&C_{5L}^{s\alpha\, ud}(\mu_{\rm SUSY})-C_{5L}^{u\alpha\, ds}(\mu_{\rm SUSY})
\nn\\
&=A_L^\alpha(\mu_{\rm SUSY},\mu_{H_C})\sum_{A,B}({\cal M}_{H_C}^{-1})_{AB}\,\left.
\frac{3}{2}\left\{(Y_L^A)_{ud}(\ov{Y}_L^B)_{s\alpha}-(Y_L^A)_{ds}(\ov{Y}_L^B)_{u\alpha}\right\}\right|_{\mu=\mu_{H_C}},
\label{c5l1}\\
&C_{5L}^{d\alpha\, us}(\mu_{\rm SUSY})-C_{5L}^{u\alpha\, ds}(\mu_{\rm SUSY})
\nn\\
&=A_L^\alpha(\mu_{\rm SUSY},\mu_{H_C})\sum_{A,B}({\cal M}_{H_C}^{-1})_{AB}\,\left.
\frac{3}{2}\left\{(Y_L^A)_{us}(\ov{Y}_L^B)_{d\alpha}-(Y_L^A)_{ds}(\ov{Y}_L^B)_{u\alpha}\right\}\right|_{\mu=\mu_{H_C}},
\label{c5l2}
\end{align}
 where $A_L^\alpha(\mu_{\rm SUSY},\mu_{H_C})$ accounts for MSSM RG corrections in the evolution from $\mu_{H_C}$ to $\mu_{\rm SUSY}$.

The contribution of the $C_{5L}^{ijkl}(Q_k Q_{l})(Q_i L_{j})$ term to the $p\to K^0e_\beta^+$ $(e_\beta=e,\mu)$ decays is given by
\begin{align}
&\Gamma(p\to K^0e_\beta^+)
\ = \ {\cal C}\,
\left\vert \beta_H(\mu_{\rm had})\frac{1}{f_\pi}
\left(1-D+F\right)\ov{C}_{LL}^{u\beta \,us}(\mu_{\rm had})
\right\vert^2.
\label{ptokmuformula}
\end{align}
Here $\ov{C}_{LL}$ are Wilson coefficients of the effective Lagrangian, $-{\cal L}_6\supset \ov{C}_{LL}^{ijkl}(\psi_{u_Lk}\psi_{d_Ll})(\psi_{u_Li}\psi_{e_Lj})$,
  which satisfy
\begin{align}
\ov{C}_{LL}^{u\beta\,us}(\mu_{\rm had})= A_{LL}(\mu_{\rm had},\mu_{\rm SUSY})
\frac{M_{\widetilde{W}}}{m_{\tilde{q}}^2}
\,{\cal F}\,
g_2^2\left(-C_{5L}^{u\beta\, us}+C_{5L}^{s\beta\, uu}\right)\vert_{\mu=\mu_{\rm SUSY}},
\end{align}
 where $A_{LL}(\mu_{\rm had},\mu_{\rm SUSY})$ accounts for SM RG corrections.
The Wilson coefficients $C_{5L}$ are related to the colored Higgs Yukawa couplings as
\begin{align}
&C_{5L}^{u\beta \,us}(\mu_{\rm SUSY})-C_{5L}^{s\beta \,uu}(\mu_{\rm SUSY})
\nn\\ 
&=A_L^\beta(\mu_{\rm SUSY},\mu_{H_C})\sum_{A,B}({\cal M}_{H_C}^{-1})_{AB}\,\left.
\frac{3}{2}
\left\{(Y_L^A)_{us}(\ov{Y}_L^B)_{u\beta}-(Y_L^A)_{uu}(\ov{Y}_L^B)_{s\beta}\right\}\right|_{\mu=\mu_{H_C}},
\label{c5l3}
\end{align}
 where $A_L^\beta(\mu_{\rm SUSY},\mu_{H_C})$ accounts for MSSM RG corrections.

The contribution of the $C_{5R}^{ijkl}\,E^c_k U^c_{l} U^c_i D^c_{j}$ term to the $p\to K^+\bar{\nu}_\tau$ decay is given by
\begin{align}
&\Gamma(p\to K^+\bar{\nu}_\tau)\vert_{{\rm from} \,C_{5R}}
= \  {\cal C} \,
\left\vert \alpha_H(\mu_{\rm had})\frac{1}{f_\pi}\left\{
\left(1+\frac{D}{3}+F\right)C_{RL}^{ud \,\tau s}(\mu_{\rm had})+\frac{2D}{3}C_{RL}^{us \,\tau d}(\mu_{\rm had})
\right\}\right\vert^2.
\label{ptoknuRformula}
\end{align}
Here $C_{RL}$ are Wilson coefficients of the effective Lagrangian,
 $-{\cal L}_6\supset C_{RL}^{ijkl}(\psi_{\nu_Lk}\psi_{d_Ll})(\psi_{u_R^ci}\psi_{d_R^cj})$,
 which satisfy
\footnote{
$y_t,\,y_\tau$ in Eqs.~(\ref{crl1}),(\ref{crl2}) are Yukawa couplings of MSSM and so already include the factors of
 $1/\sin\beta$ and $1/\cos\beta$, respectively.
 }
\begin{align}
C_{RL}^{ud\,\tau s}(\mu_{\rm had}) \ = \ A_{RL}(\mu_{\rm had},\mu_{\rm SUSY})
\frac{\mu_H}{m_{\tilde{t}_R}^2}\,
{\cal F}'\ (V^{\rm ckm}_{ts})^*\, y_t y_\tau
\,C_{5R}^{ud\tau t}\vert_{\mu=\mu_{\rm SUSY}},
\label{crl1}\\
C_{RL}^{us\,\tau d}(\mu_{\rm had}) \ = \ A_{RL}(\mu_{\rm had},\mu_{\rm SUSY})
\frac{\mu_H}{m_{\tilde{t}_R}^2}\,
{\cal F}'\ (V^{\rm ckm}_{td})^*\, y_t y_\tau
\,C_{5R}^{us\tau t}\vert_{\mu=\mu_{\rm SUSY}},
\label{crl2}
\end{align}
 where $V^{\rm ckm}_{ij}$ denotes $(i,j)$-component of CKM matrix.
Here ${\cal F}'$ is another loop function factor given by ${\cal F}' =\frac{1}{x-y}(\frac{x}{1-x}\log x - \frac{y}{1-y}\log y)/16\pi^2$
  with $x=|\mu_H|^2/m_{\tilde{t}_R}^2$ and $y=m_{\tilde{\tau}_R}^2/m_{\tilde{t}_R}^2$.
Also, $\mu_H$ denotes the $\mu$-term, $m_{\tilde{t}_R}$ the mass of the right-handed top squark, and $m_{\tilde{\tau}_R}$ the mass of the right-handed tau slepton.
$A_{RL}(\mu_{\rm had},\mu_{\rm SUSY})$ accounts for SM RG corrections.
The Wilson coefficients $C_{5R}$ are related to the colored Higgs Yukawa couplings as
\begin{align}
C_{5R}^{ud\tau t}(\mu_{\rm SUSY}) \ = \ 
A_R^{\tau t}(\mu_{\rm SUSY},\mu_{H_C})\sum_{A,B}({\cal M}_{H_C}^{-1})_{AB}\left.
\left\{(Y_R^A)_{\tau t}(\ov{Y}^B_R)_{ud}-(Y_R^A)_{\tau u}(\ov{Y}^B_R)_{td}\right\}\right|_{\mu=\mu_{H_C}},
\label{c5r1}\\
C_{5R}^{us\tau t}(\mu_{\rm SUSY}) \ = \ 
A_R^{\tau t}(\mu_{\rm SUSY},\mu_{H_C})\sum_{A,B}({\cal M}_{H_C}^{-1})_{AB}\left.
\left\{(Y_R^A)_{\tau t}(\ov{Y}^B_R)_{us}-(Y_R^A)_{\tau u}(\ov{Y}_R^B)_{ts}\right\}\right|_{\mu=\mu_{H_C}},
\label{c5r2}
\end{align}
where $A_R^{\tau t}(\mu_{\rm SUSY},\mu_{H_C})$ accounts for MSSM RG corrections.

The flavor-dependent terms in Eqs.~(\ref{c5l1}),(\ref{c5l2}),(\ref{c5l3}),(\ref{c5r1}),(\ref{c5r2}) are related to the fundamental Yukawa couplings $Y_{10},Y_{126},Y_{120}$ as follows.
Since $Y_L^A$ is defined as $W_{\rm Yukawa} \supset (Y_L^A)_{ij}\,Q_i H_C^A Q_j$, its flavor indices are symmetric
 and thus $Y_L^A$ is not proportional to $Y_{120}$.
Therefore, we can write without loss of generality ($\alpha=e,\mu,\tau$ and $\beta=e,\mu$)
\begin{align}
&\sum_{A,B}({\cal M}_{H_C}^{-1})_{AB}\left\{(Y_L^A)_{ud}(\ov{Y}^B_L)_{s\alpha}-(Y_L^A)_{ds}(\ov{Y}^B_L)_{u\alpha}\right\}
\nn\\
&=\frac{1}{M_{H_C}}
\left[
a\left\{(Y_{10})_{u_L d_L}(Y_{10})_{s_L\alpha_L}-(Y_{10})_{d_L s_L}(Y_{10})_{u_L\alpha_L}\right\}
+b\left\{(Y_{10})_{u_L d_L}(Y_{126})_{s_L\alpha_L}-(Y_{10})_{d_L s_L}(Y_{126})_{u_L\alpha_L}\right\}\right.
\nn\\
&+c\left\{(Y_{10})_{u_L d_L}(Y_{120})_{s_L\alpha_L}-(Y_{10})_{d_L s_L}(Y_{120})_{u_L\alpha_L}\right\}
\nn\\
&+d\left\{(Y_{126})_{u_L d_L}(Y_{10})_{s_L\alpha_L}-(Y_{126})_{d_L s_L}(Y_{10})_{u_L\alpha_L}\right\}
+e\left\{(Y_{126})_{u_L d_L}(Y_{126})_{s_L\alpha_L}-(Y_{126})_{d_L s_L}(Y_{126})_{u_L\alpha_L}\right\}
\nn\\
&\left.+f\left\{(Y_{126})_{u_L d_L}(Y_{120})_{s_L\alpha_L}-(Y_{126})_{d_L s_L}(Y_{120})_{u_L\alpha_L}\right\}\right],
\nn\\
\label{yyl1}
\\
&\sum_{A,B}({\cal M}_{H_C}^{-1})_{AB}\left\{(Y_L^A)_{us}(\ov{Y}^B_L)_{d\alpha}-(Y_L^A)_{ds}(\ov{Y}^B_L)_{u\alpha}\right\}
={\rm (Above \ expression \ with \ exchange} \ d_L\leftrightarrow s_L),
\label{yyl2}
\\
&\sum_{A,B}({\cal M}_{H_C}^{-1})_{AB}\left\{(Y_L^A)_{us}(\ov{Y}^B_L)_{u\beta}-(Y_L^A)_{uu}(\ov{Y}^B_L)_{s\beta}\right\}
\nn\\
&=\frac{1}{M_{H_C}}
\left[
a\left\{(Y_{10})_{u_L s_L}(Y_{10})_{u_L\beta_L}-(Y_{10})_{u_L u_L}(Y_{10})_{s_L\beta_L}\right\}
+b\left\{(Y_{10})_{u_L s_L}(Y_{126})_{u_L\beta_L}-(Y_{10})_{u_L u_L}(Y_{126})_{s_L\beta_L}\right\}\right.
\nn\\
&+c\left\{(Y_{10})_{u_L s_L}(Y_{120})_{u_L\beta_L}-(Y_{10})_{u_L u_L}(Y_{120})_{s_L\beta_L}\right\}
\nn\\
&+d\left\{(Y_{126})_{u_L s_L}(Y_{10})_{u_L\beta_L}-(Y_{126})_{u_L u_L}(Y_{10})_{s_L\beta_L}\right\}
+e\left\{(Y_{126})_{u_L s_L}(Y_{126})_{u_L\beta_L}-(Y_{126})_{u_L u_L}(Y_{126})_{s_L\beta_L}\right\}
\nn\\
&\left.+f\left\{(Y_{126})_{u_L s_L}(Y_{120})_{u_L\beta_L}-(Y_{126})_{u_L u_L}(Y_{120})_{s_L\beta_L}\right\}\right],
\label{yyl3}
\end{align}
 where $M_{H_C}$ denotes a typical value of the eigenvalues of ${\cal M}_{H_C}$, and $a,b,c,d,e,f,g,h,j$ are numbers
 determined from the colored Higgs mass matrix~\cite{Fukuyama:2004xs}-\cite{Bajc:2005qe}.
Here $(Y_{10})_{u_Ld_L}$ denotes the $(1,1)$-component of $Y_{10}$ in the term $(Y_{10})_{ij}\,\Psi_i H \Psi_j$
 in the flavor basis where the left-handed up-type quark component of $\Psi_i$ has the diagonalized up-type quark Yukawa coupling,
 and the left-handed down-type quark component of $\Psi_j$ has the diagonalized down-type quark Yukawa coupling.
$(Y_{10})_{d_L s_L}$, $(Y_{126})_{u_L d_L}$ and others are defined analogously.
Since each of $Y_R^A,\ov{Y}_R^A$ is proportional to $Y_{10}$, $Y_{126}$ or $Y_{120}$, we can write
\begin{align}
&\sum_{A,B}({\cal M}_{H_C}^{-1})_{AB}\left\{(Y_R^A)_{\tau t}(\ov{Y}^B_R)_{ud}-(Y_R^A)_{\tau u}(\ov{Y}^B_R)_{td}\right\}
\nn\\
&=\frac{1}{M_{H_C}}
\left[
a\left\{(Y_{10})_{\tau_R t_R}(Y_{10})_{u_Rd_R}-(Y_{10})_{\tau_R u_R}(Y_{10})_{t_Rd_R}\right\}
+b\left\{(Y_{10})_{\tau_R t_R}(Y_{126})_{u_Rd_R}-(Y_{10})_{\tau_R u_R}(Y_{126})_{t_Rd_R}\right\}\right.
\nn\\
&+c\left\{(Y_{10})_{\tau_R t_R}(Y_{120})_{u_Rd_R}-(Y_{10})_{\tau_R u_R}(Y_{120})_{t_Rd_R}\right\}
\nn\\
&+d\left\{(Y_{126})_{\tau_R t_R}(Y_{10})_{u_Rd_R}-(Y_{126})_{\tau_R u_R}(Y_{10})_{t_Rd_R}\right\}
+e\left\{(Y_{126})_{\tau_R t_R}(Y_{126})_{u_Rd_R}-(Y_{126})_{\tau_R u_R}(Y_{126})_{t_Rd_R}\right\}
\nn\\
&+f\left\{(Y_{126})_{\tau_R t_R}(Y_{120})_{u_Rd_R}-(Y_{126})_{\tau_R u_R}(Y_{120})_{t_Rd_R}\right\}
\nn\\
&+g\left\{(Y_{120})_{\tau_R t_R}(Y_{10})_{u_Rd_R}-(Y_{120})_{\tau_R u_R}(Y_{10})_{t_Rd_R}\right\}
+h\left\{(Y_{120})_{\tau_R t_R}(Y_{126})_{u_Rd_R}-(Y_{120})_{\tau_R u_R}(Y_{126})_{t_Rd_R}\right\}
\nn\\
&\left.+j\left\{(Y_{120})_{\tau_R t_R}(Y_{120})_{u_Rd_R}-(Y_{120})_{\tau_R u_R}(Y_{120})_{t_Rd_R}\right\}\right],
\nn\\
\label{yyr1}
\\
&\sum_{A,B}({\cal M}_{H_C}^{-1})_{AB}\left\{(Y_R^A)_{\tau t}(\ov{Y}^B_R)_{us}-(Y_R^A)_{\tau u}(\ov{Y}^B_R)_{ts}\right\}
={\rm (Above \ expression \ with \ replacement} \ d_R\to s_R),
\label{yyr2}
\end{align}
 where $a,b,c,d,e,f$ are the same numbers as those in Eqs.~(\ref{yyl1})-(\ref{yyl3}).
\\

\section{Components of the Yukawa matrices that can be reduced}
\label{section-inferring}

We spot those components of the Yukawa matrices $Y_{10},Y_{126},Y_{120}$
 which can be reduced to suppress dimension-5 proton decays without conflicting the requirement that they reproduce
 the correct quark and lepton Yukawa couplings and neutrino mass matrix.
Specifically, we attempt to reduce the pair-products of the components of $Y_{10},Y_{126},Y_{120}$
 that appear in Eqs.~(\ref{yyl1})-(\ref{yyr2}) (e.g. $(Y_{10})_{\tau_R t_R}(Y_{10})_{u_Rd_R}$)
 to the order of the up quark Yukawa coupling times the top quark Yukawa coupling $O(y_u\,y_t)$.
As a matter of fact, some pair-products cannot simultaneously be reduced
 because of the requirement that $Y_{10},Y_{126},Y_{120}$ reproduce the correct quark and lepton Yukawa couplings.
In this circumstance, we tune the colored Higgs mass matrix
 such that coefficients $a,b,c,d,e,f,g,h,j$ in Eqs.~(\ref{yyl1})-(\ref{yyr2}) 
 realize cancellations among the problematic pair-products.
Finally, we present ``those components of the Yukawa matrices $Y_{10},Y_{126},Y_{120}$ that can be reduced"
 as well as an example of the colored Higgs mass matrix that gives coefficients $a,b,c,d,e,f,g,h,j$ that realize the above-mentioned cancellations.
\\

\begin{itemize}

\item
Focus on Eq.~(\ref{yyr1}).
We have $(Y_{10})_{\tau_R t_R}+r_2(Y_{126})_{\tau_R t_R}+r_3(Y_{120})_{\tau_R t_R}=y_t\times$
($t_L$-$\tau_R$ part of the mixing matrix), 
 and since $t_L$ and $\tau_R$ are both 3rd generation components, 
$t_L$-$\tau_R$ part of the mixing matrix is almost maximal.
The component $(Y_{120})_{\tau_R t_R}$ is suppressed compared to $(Y_{10})_{\tau_R t_R},(Y_{126})_{\tau_R t_R}$ 
 because $Y_{120}$ is an antisymmetric matrix.
Consequently, one or both of $(Y_{10})_{\tau_R t_R}$ and $(Y_{126})_{\tau_R t_R}$
 are always on the order of the top quark Yukawa coupling $y_t$.
Hence, in order to reduce the Yukawa coupling pair-products in Eq.~(\ref{yyr1}), it is necessary to reduce
\bea
(Y_{10})_{u_R d_R}, \ \ \ (Y_{126})_{u_R d_R}, \ \ \ {\rm and} \ \ \ (Y_{120})_{u_R d_R}.
\eea
Note that although $Y_{120}$ is an antisymmetric matrix, $(Y_{120})_{u_R d_R}$ is not necessarily suppressed to $O(y_u)$ or below.
This is because $(Y_{120})_{u_R d_R}$ can be on the order of $(Y_{120})_{u_R s_R}$ times the Cabibbo angle $\lambda\simeq0.22$,
 and $(Y_{120})_{u_R s_R}$ can be on the order of $0.22\times \frac{y_t}{y_b}y_s$ as we see later,
 so that $(Y_{120})_{u_R d_R}$ can be as large as $0.22^2\times \frac{y_t}{y_b}y_s$, which is much greater than $y_u$.

Eq.~(\ref{yyr1}) also contains terms of the form $(Y_A)_{\tau_R u_R}(Y_B)_{t_R d_R}$ $(A,B=10,126,120)$.
They can be estimated to be $\sin^2\theta^{\rm ckm}_{13}\, y_t^2$ 
 ($\theta^{\rm ckm}_{ij}$ denotes the $(i,j)$-mixing angle of CKM matrix),
 which is numerically close to $y_u\,y_t$.
Hence, we do not need to reduce $(Y_A)_{\tau_R u_R}$ or $(Y_A)_{t_R d_R}$ further.

\item
Focus on Eq.~(\ref{yyr2}).
For the same reason as above, we have to reduce
\bea
(Y_{10})_{u_R s_R}, \ \ \ (Y_{126})_{u_R s_R}, \ \ \ {\rm and} \ \ \ (Y_{120})_{u_R s_R}.
\eea

Eq.~(\ref{yyr2}) also contains terms of the form $(Y_A)_{\tau_R u_R}(Y_B)_{t_R s_R}$ $(A,B=10,126,120)$,
 which are estimated to be $\sin\theta^{\rm ckm}_{13}\sin\theta^{\rm ckm}_{23}\,y_t^2$.
They contribute to the $p\to K^+ \bar{\nu}_\tau$ decay amplitude
 by a similar amount to the terms $(Y_A)_{\tau_R u_R}(Y_B)_{t_R d_R}$ in Eq.~(\ref{yyr1}),
 because these terms enter the decay amplitude in the form
 $V_{ts}^{\rm ckm}(Y_A)_{\tau_R u_R}(Y_B)_{t_R d_R}+V_{td}^{\rm ckm}(Y_A)_{\tau_R u_R}(Y_B)_{t_R s_R}$
 and CKM matrix satisfies $|V_{ts}^{\rm ckm}|\simeq\sin\theta^{\rm ckm}_{23}$ and 
 $|V_{td}^{\rm ckm}|\sim\sin\theta^{\rm ckm}_{13}$.
Therefore, we tolerate the terms $(Y_A)_{\tau_R u_R}(Y_B)_{t_R s_R}$ and do not reduce $(Y_A)_{\tau_R u_R}$ or $(Y_A)_{t_R s_R}$ further.

\item
As a matter of fact, it is impossible to simultaneously reduce $(Y_{10})_{u_R s_R}$, $(Y_{126})_{u_R s_R}$ and $(Y_{120})_{u_R s_R}$ to $O(y_u)$.
This is because Eq.~(\ref{yd}) gives
\begin{align}
(Y_{10})_{u_R s_R}+(Y_{126})_{u_R s_R}+(Y_{120})_{u_R s_R} \ &= \ \frac{1}{r_1}(Y_d)_{u_R s_R}
\nn\\
&\simeq \ \frac{y_t}{y_b}y_s\times (s_L\mathchar`-u_R \ {\rm part \ of \ the \ mixing \ matrix}),
 \label{ursr}
\end{align}
 where $r_1$ is estimated to be $y_b/y_t$ so that the top and bottom quark Yukawa couplings are reproduced.
$s_L\mathchar`-u_R \ {\rm part \ of \ the \ mixing \ matrix}$
 is estimated to be the Cabibbo angle $\lambda\simeq0.22$ and thus
 we get $(Y_{10})_{u_R s_R}+(Y_{126})_{u_R s_R}+(Y_{120})_{u_R s_R}\simeq0.22\times\frac{y_t}{y_b}y_s$,
 which is much greater than the up quark Yukawa coupling $y_u$.

A way out is to adjust the colored Higgs mass matrix such that
 coefficients $c,f$ in Eqs.~(\ref{yyl1})-(\ref{yyr2}) are zero,
\bea
c=f=0.
\eea
Then, we are exempted from reducing $(Y_{120})_{u_R s_R}$, because
 $(Y_{120})_{u_R s_R}$ appears only in the term $(Y_{120})_{\tau_R t_R}(Y_{120})_{u_Rs_R}$
 and the component $(Y_{120})_{\tau_R t_R}$ is suppressed because $Y_{120}$ is an antisymmetric matrix.
As a bonus, it is no longer necessary to reduce $(Y_{120})_{u_R d_R}$.

\item
Focus on Eqs.~(\ref{yyl1}),(\ref{yyl2}).
Since $\alpha$ ranges in the whole three flavors, it is difficult to reduce $(Y_{A})_{s_L \alpha_L}$, $(Y_{A})_{d_L \alpha_L}$ and $(Y_{A})_{u_L \alpha_L}$ $(A=10,126,120)$
 for all $\alpha$.
Hence, we leave these Yukawa couplings untouched and instead reduce $(Y_{B})_{u_L d_L}$, $(Y_{B})_{u_L s_L}$ and $(Y_{B})_{d_L s_L}$ $(B=10,126)$ (one side of the Yukawa coupling pair-products).

\item
Unfortunately, at least one of $(Y_{10})_{u_L s_L}, \, (Y_{10})_{d_L s_L}, \, (Y_{126})_{u_L s_L}, \, (Y_{126})_{d_L s_L}$ is on the order of
 $V^{\rm ckm}_{cd}\,\frac{y_t}{y_b}y_s$, and consequently, some of the Yukawa coupling pair-products in Eqs.~(\ref{yyl1}),(\ref{yyl2})
 cannot be suppressed to $O(y_u\,y_t)$ for all $\alpha$.
This is seen from two equalities,
\begin{align}
&(Y_{10})_{s_L c_L}+(Y_{126})_{s_L c_L}+(Y_{120})_{s_L c_L} \ \simeq \ \frac{y_t}{y_b}y_s
\times (c_L\mathchar`-s_R \ {\rm part \ of \ the \ mixing \ matrix)},
\label{clsl}
\end{align}
and
\begin{align}
&(Y_{10})_{d_L s_L}+(Y_{126})_{d_L s_L}-V^{\rm ckm}_{ud} \left\{ (Y_{10})_{u_L s_L}+(Y_{126})_{u_L s_L} \right\}
\nn\\
&=\ V^{\rm ckm}_{cd} \left\{ (Y_{10})_{c_L s_L}+(Y_{126})_{c_L s_L} \right\}
+V^{\rm ckm}_{td} \left\{ (Y_{10})_{t_L s_L}+(Y_{126})_{t_L s_L} \right\}.
\label{dlsl}
\end{align}
Since $c_L$ and $s_R$ are both 2nd generation components,
 $c_L\mathchar`-s_R \ {\rm part \ of \ the \ mixing \ matrix}$
 in Eq.~(\ref{clsl}) is nearly maximal.
Also, $(Y_{120})_{s_L c_L}$ is suppressed compared to $(Y_{10})_{c_L s_L},(Y_{126})_{c_L s_L}$ because $Y_{120}$ is an antisymmetric matrix.
Hence, we have $(Y_{10})_{c_L s_L}+(Y_{126})_{c_L s_L}\simeq\frac{y_t}{y_b}y_s(\mu_{H_C})$,
 and from Eq.~(\ref{dlsl}) we conclude that at least one of $(Y_{10})_{u_L s_L}$, $(Y_{126})_{u_L s_L}$, $(Y_{10})_{d_L s_L}$ and $(Y_{126})_{d_L s_L}$
 is on the order of $V^{\rm ckm}_{cd}\,\frac{y_t}{y_b}y_s$.
 \footnote{
One might hope that the term $V^{\rm ckm}_{td} \left\{ (Y_{10})_{t_L s_L}+(Y_{126})_{t_L s_L} \right\}$
 cancels the term $V^{\rm ckm}_{cd} \left\{ (Y_{10})_{c_L s_L}+(Y_{126})_{c_L s_L} \right\}$,
 but this is not compatible with the correct quark Yukawa couplings.}

A natural way out is to reduce $(Y_{10})_{u_L s_L}$ and $(Y_{126})_{u_L s_L}$,
 while tuning coefficients $a,b,d,e$ such that 
 $a\, (Y_{10})_{d_L s_L}+d\, (Y_{126})_{d_L s_L}=0$ and $b\, (Y_{10})_{d_L s_L}+e\, (Y_{126})_{d_L s_L}=0$ hold.
This choice is because $(Y_{10})_{u_L s_L}$ and $(Y_{126})_{u_L s_L}$ can more easily be related to the tiny up quark Yukawa coupling.

\item
Finally, focus on Eq.~(\ref{yyl3}).
Since we leave $(Y_{A})_{s_L e_L}$ and $(Y_{A})_{s_L \mu_L}$ untouched, we have to reduce $(Y_{10})_{u_L u_L}$ and $(Y_{126})_{u_L u_L}$.

\end{itemize}

To sum up, in order to suppress dimension-5 proton decays,
 we have to reduce the following Yukawa couplings:
\bea
&&(Y_{10})_{u_R d_R}, \ (Y_{126})_{u_R d_R}, \ (Y_{10})_{u_R s_R}, \ (Y_{126})_{u_R s_R}
\nn\\
&&(Y_{10})_{u_L d_L}, \ (Y_{126})_{u_L d_L}, \ (Y_{10})_{u_L u_L}, \ (Y_{126})_{u_L u_L}, \ (Y_{10})_{u_L s_L}, \ (Y_{126})_{u_L s_L}
\label{tobereduced}
\eea
Meanwhile, we have to adjust the colored Higgs mass matrix such that
 $c=f=0$,
 $a\, (Y_{10})_{d_L s_L}+d\, (Y_{126})_{d_L s_L}=0$ and $b\, (Y_{10})_{d_L s_L}+e\, (Y_{126})_{d_L s_L}=0$ hold.

We comment on the $N\to \pi e_\beta^+$ and $p\to \eta e_\beta^+$ decays.
Their decay amplitudes contain terms obtained by replacing $s$ with $d$ in Eq.~(\ref{yyl3}).
Therefore, by reducing $(Y_{10})_{u_L d_L}$, $(Y_{126})_{u_L d_L}$,
 $(Y_{10})_{u_L u_L}$ and $(Y_{126})_{u_L u_L}$,
 these decay modes are also suppressed.
\\

We present an example of the colored Higgs mass matrix that realizes $c=f=0$ and $a/d=b/e$.
The latter is a necessity condition for 
 $a\, (Y_{10})_{d_L s_L}+d\, (Y_{126})_{d_L s_L}=0$ and $b\, (Y_{10})_{d_L s_L}+e\, (Y_{126})_{d_L s_L}=0$.

To study the colored Higgs mass matrix, we have to write the superpotential for $H$, $\Delta$, $\ov{\Delta}$, $\Sigma$, $\Phi$, $A$ fields,
 introduce $SO(10)$-breaking VEVs, and specify the colored Higgs components of the fields.
To this end, we use the result of Ref.~\cite{Fukuyama:2004ps}.
The notation for fields is common for our paper and Ref.~\cite{Fukuyama:2004ps} except that {\bf 120} field is 
 written as $D$ in Ref.~\cite{Fukuyama:2004ps}.
We define the couplings, coupling constants and masses for the fields according to Eqs.~(2),(3) of Ref.~\cite{Fukuyama:2004ps}
 (our definition of the coupling constants and masses is reviewed in Appendix).
We employ the same notation for 
 the VEVs of $\Delta$, $\ov{\Delta}$, $\Phi$, $A$ as Ref.~\cite{Fukuyama:2004ps},
 and write the (${\bf 3}$, ${\bf1}$, $-\frac{1}{3}$) and (${\bf \ov{3}}$, ${\bf1}$, $\frac{1}{3}$) components
 as Table~3 of Ref.~\cite{Fukuyama:2004ps}.
\footnote{
We have confirmed that the mass matrix of the (${\bf 1}$, ${\bf2}$, $\pm\frac{1}{2}$) fields given in Eq.~(68) of Ref.~\cite{Fukuyama:2004ps}
 is correct. 
However, we argue that in the colored Higgs mass matrix given in Eq.~(69) of Ref.~\cite{Fukuyama:2004ps}, 
 the sign of the term $\frac{2}{5}\sqrt{\frac{2}{3}}\lambda_7A_2$ in $m_{77}^{(3,1,-\frac{1}{3})}$ should be minus.
Otherwise, we have confirmed that Eq.~(69) of Ref.~\cite{Fukuyama:2004ps} is correct.
We argue that in the superpotential of the VEVs given in Eq.~(27) of Ref.~\cite{Fukuyama:2004ps},
 the sign of the term $\lambda_6[A_1(-\frac{1}{5})+A_2(-\frac{3}{5\sqrt{6}})]$ should be flipped.
Otherwise, we have confirmed that Eq.~(27) of Ref.~\cite{Fukuyama:2004ps} is correct.
 }

Now we present the example of the colored Higgs mass matrix. It satisfies
\begin{align}
\lambda_{18} \ &= \ 0, \ \ \ \ \ 
\lambda_{20} \ = \ 0, \ \ \ \ \
\frac{\lambda_{21}}{\lambda_{19}}\ = \ 3\frac{\lambda_{17}}{\lambda_{16}},
\nn\\
i\,A_1 \ &= \ -\frac{1}{6}\frac{\lambda_{21}}{\lambda_{19}}\Phi_3, \  \ \ \ \ 
i\,A_2 \ = \ -\frac{\sqrt{3}}{6}\frac{\lambda_{21}}{\lambda_{19}}\Phi_2.
\label{texture}
\end{align}
The ``texture" of Eq.~(\ref{texture}) cannot be derived from any symmetry, and so fine-tunings of superpotential parameters are needed for its realization.
These fine-tunings are natural at the quantum level due to the non-renormalization theorem.
The VEV configuration in the second line of Eq.~(\ref{texture}) can satisfy the $F$-flatness conditions
 (displayed in Eq.~(28) of Ref.~\cite{Fukuyama:2004ps}).
Given Eq.~(\ref{texture}), the colored Higgs mass matrix is given by
\begin{align}
   &W \ \supset \ \begin{pmatrix} 
      H^{(\ov{3},1,\frac{1}{3})}  &  \ov{\Delta}^{(\ov{3},1,\frac{1}{3})}_{(6,1,1)}  & \Delta^{(\ov{3},1,\frac{1}{3})}_{(6,1,1)}  & \Delta^{(\ov{3},1,\frac{1}{3})}_{(\ov{10},1,3)}
      & \Phi^{(\ov{3},1,\frac{1}{3})} & \Sigma^{(\ov{3},1,\frac{1}{3})}_{(6,1,3)} & \Sigma^{(\ov{3},1,\frac{1}{3})}_{(\ov{10},1,1)}
   \end{pmatrix}{\cal M}_{H_C}
   \begin{pmatrix} 
H^{(3,1,-\frac{1}{3})}  \\  \Delta^{(3,1,-\frac{1}{3})}_{(6,1,1)}  \\ \ov{\Delta}^{(3,1,-\frac{1}{3})}_{(6,1,1)} \\ \ov{\Delta}^{(3,1,-\frac{1}{3})}_{(10,1,3)}
      \\ \Phi^{(3,1,-\frac{1}{3})} \\ \Sigma^{(3,1,-\frac{1}{3})}_{(6,1,3)} \\ \Sigma^{(3,1,-\frac{1}{3})}_{(10,1,1)}
   \end{pmatrix}
\end{align}
where
\begin{align}
&{\cal M}_{H_C}
=\begin{pmatrix} 
      m_3 &\frac{ \lambda_3\Phi_2}{\sqrt{30}}-\frac{ \lambda_3\Phi_1}{\sqrt{10}} & -\frac{\lambda_4\Phi_1}{\sqrt{10}}-\frac{\lambda_4\Phi_2}{\sqrt{30}}
      & -\sqrt{\frac{2}{15}}\lambda_4\Phi_3 & \frac{\lambda_4\ov{v}_R}{\sqrt{5}} & 0 & 0 \\
      \frac{\lambda_4\Phi_2}{\sqrt{30}}-\frac{\lambda_4\Phi_1}{\sqrt{10}} & m_2+i\frac{\lambda_{21}}{\lambda_{19}}\frac{\lambda_6\Phi_2}{30\sqrt{2}} & 0 & 0 &0 &0 & 0 \\
      -\frac{\lambda_3\Phi_1}{\sqrt{10}}-\frac{\lambda_3\Phi_2}{\sqrt{30}} & 0 & m_2-i\frac{\lambda_{21}}{\lambda_{19}}\frac{\lambda_6\Phi_2}{30\sqrt{2}} &\frac{\lambda_2\Phi_3}{15\sqrt{2}} & -\frac{\lambda_2\ov{v}_R}{10\sqrt{3}} & 0 & 0 \\
      -\sqrt{\frac{2}{15}}\lambda_3\Phi_3 & 0 & \frac{\lambda_2\Phi_3}{15\sqrt{2}} & m_{66}  & -\frac{\lambda_2\ov{v}_R}{5\sqrt{6}} & 0 & 0 \\
      \frac{\lambda_3v_R}{\sqrt{5}} & 0 & -\frac{\lambda_2v_R}{10\sqrt{3}} & -\frac{\lambda_2v_R}{5\sqrt{6}} & m_{77} & 0 & 0 \\
      -\lambda_{17}\frac{\Phi_3}{\sqrt{3}} & 0 & \lambda_{21}\frac{\Phi_3}{6\sqrt{5}} & \lambda_{21}\frac{\Phi_2}{3\sqrt{5}} & \frac{\lambda_{21}\ov{v}_R}{2\sqrt{15}} & m_{22} & \frac{2\lambda_{15}\Phi_3}{9} \\
      -\sqrt{\frac{2}{3}}\lambda_{17}\Phi_2 & 0 & \lambda_{21}\frac{\Phi_2}{3\sqrt{10}} & \lambda_{21}\frac{\Phi_3}{3\sqrt{10}} &
 \frac{\lambda_{21}\ov{v}_R}{2\sqrt{15}} & \frac{2\lambda_{15}\Phi_3}{9} & m_{33}
   \end{pmatrix}
 \label{texture-coloredhiggs}
 \\
&m_{66}= m_2+\lambda_2(\frac{\Phi_1}{10\sqrt{6}}+\frac{\Phi_2}{30\sqrt{2}})-i\frac{\lambda_{21}}{\lambda_{19}}\frac{\lambda_6\Phi_2}{30\sqrt{2}}
\\
&m_{77} = m_1+\lambda_1(\frac{\Phi_1}{\sqrt{6}}+\frac{\Phi_2}{3\sqrt{2}}+\frac{2\Phi_3}{3})-i\frac{\lambda_{21}}{\lambda_{19}}\frac{\sqrt{2}\lambda_7\Phi_2}{15}
\\
&m_{22} = m_6+\frac{1}{3}\sqrt{\frac{2}{3}}\lambda_{15}\Phi_1
\\
&m_{33} = m_6+\frac{\sqrt{2}}{9}\lambda_{15}\Phi_2
\end{align}
The Wilson coefficients of the terms $C_{5L}^{ijkl}(Q_k Q_{l})(Q_i L_{j})$, $C_{5R}^{ijkl}E^c_k U^c_{l} U^c_i D^c_{j}$,
 which appear after integrating out the colored Higgs fields, are given by
\begin{align}
&C_{5L}^{ijkl}(\mu=\mu_{H_C}) = C_{5R}^{ijkl}(\mu=\mu_{H_C})
\nn\\
=&\begin{pmatrix} 
      (\tilde{Y}_{10})_{kl}  &  0  &  (\tilde{Y}_{126})_{kl} & (\tilde{Y}_{126})_{kl}
      & 0 & (\tilde{Y}_{120})_{kl} & (\tilde{Y}_{120})_{kl}
   \end{pmatrix}{\cal M}_{H_C}^{-1}
   \begin{pmatrix} 
(\tilde{Y}_{10})_{ij}  \\  (\tilde{Y}_{126})_{ij}  \\ 0 \\ 0
      \\ 0 \\ (\tilde{Y}_{120})_{ij} \\ (\tilde{Y}_{120})_{ij}
   \end{pmatrix}
\end{align}
First, since the upper-right $5\times 2$ part of ${\cal M}_{H_C}$ is zero,
 the upper-right $5\times 2$ part of the inverse matrix ${\cal M}_{H_C}^{-1}$ is also zero.
It follows that the terms
 $(\tilde{Y}_{10})_{kl}(\tilde{Y}_{120})_{ij}$ and $(\tilde{Y}_{126})_{kl}(\tilde{Y}_{120})_{ij}$ do not appear in the Wilson coefficients $C_{5L}^{ijkl},C_{5R}^{ijkl}$, and hence $c=f=0$ in Eqs.~(\ref{yyl1})-(\ref{yyr2}).
Second, the upper-left $5\times5$ part of ${\cal M}_{H_C}^{-1}$ is exactly the inverse matrix of 
 the same part of ${\cal M}_{H_C}$.
It is possible to mathematically prove that the components of ${\cal M}_{H_C}^{-1}$ satisfy a relation
$({\cal M}_{H_C}^{-1})_{11}:({\cal M}_{H_C}^{-1})_{31}:({\cal M}_{H_C}^{-1})_{41}=
({\cal M}_{H_C}^{-1})_{12}:({\cal M}_{H_C}^{-1})_{32}:({\cal M}_{H_C}^{-1})_{42}$
when the (3,2),~(4,2) and (5,2)-components of ${\cal M}_{H_C}$ are zero as in Eq.~(\ref{texture-coloredhiggs}).
Then, since the numbers $a,d$ in Eqs.~(\ref{yyr1})-(\ref{yyl3}) are determined by $({\cal M}_{H_C}^{-1})_{11},({\cal M}_{H_C}^{-1})_{31},({\cal M}_{H_C}^{-1})_{41}$ and the numbers $b,e$ are determined by $({\cal M}_{H_C}^{-1})_{12},({\cal M}_{H_C}^{-1})_{32},({\cal M}_{H_C}^{-1})_{42}$,
 we get $a/d=b/e$.
We comment that if the model contains a {\bf 54}-representation field, its VEV must be 0 to realize the relation $a/d=b/e$.

We are yet to prove that Eq.~(\ref{texture}) is compatible with 
 the situation that \textit{all the fields have GUT-scale masses}
 except for one pair of (${\bf 1}$, ${\bf2}$, $\pm\frac{1}{2}$) fields that give the MSSM Higgs fields
 and a (${\bf 1}$, ${\bf3}$, 1) field that has mass slightly below the GUT scale to realize the Type-2 seesaw mechanism.
Also, Eq.~(\ref{texture}) must be consistent with the right value of $a/d$ that realizes 
 $a\, (Y_{10})_{d_L s_L}+d\, (Y_{126})_{d_L s_L}=0$, and with the right values of $r_1,r_2,r_3,r_e$
 that reproduce the correct fermion data.
(Note that common coupling constants enter the colored Higgs mass matrix and the mass matrix of the (${\bf 1}$, ${\bf2}$, $\pm\frac{1}{2}$) fields.)
We have numerically checked that under the restriction of Eq.~(\ref{texture}) and the condition that the 
 mass matrix of the (${\bf 1}$, ${\bf2}$, $\pm\frac{1}{2}$) fields have one zero eigenvalue,
 the ratio of the masses of various fields (other than the pair of (${\bf 1}$, ${\bf2}$, $\pm\frac{1}{2}$) fields)
 and the values of
 $a/d,\,r_1,r_2,r_3,r_e$ vary in a wide range and there is no correlation among them.
It is thus quite likely that the gauge coupling unification is achieved with a help of GUT-scale threshold corrections and
 the right values of $a/d$ and $r_1,r_2,r_3,r_e$ are obtained even with Eq.~(\ref{texture}).
\\

\section{Numerical search for the texture of $Y_{10}$, $Y_{126}$, $Y_{120}$}
\label{section-fitting}

We search for the texture of the Yukawa couplings $Y_{10}$, $Y_{126}$, $Y_{120}$ discussed in Section~\ref{section-inferring},
 i.e., the texture which reproduces the correct quark and lepton Yukawa couplings and neutrino mass matrix according to Eqs.~(\ref{yu})-(\ref{ye}),(\ref{cnu})
 and in which the components of Yukawa couplings
 $(Y_{10})_{u_R d_R}$, $(Y_{126})_{u_R d_R}$, $(Y_{10})_{u_R s_R}$, $(Y_{126})_{u_R s_R}$,
 $(Y_{10})_{u_L d_L}$, $(Y_{126})_{u_L d_L}$, $(Y_{10})_{u_L u_L}$, $(Y_{126})_{u_L u_L}$, $(Y_{10})_{u_L s_L}$, $(Y_{126})_{u_L s_L}$
 are reduced.
\\

\subsection{Procedures}

First, we numerically calculate the MSSM Yukawa coupling matrices $Y_u,Y_d,Y_e$ at scale $\mu=2\cdot10^{16}$~GeV in $\ov{{\rm DR}}$ scheme.
We also calculate 
 the {\bf flavor-dependent} part of RG corrections to 
 the coefficient of the $L_i ({\bf 1},{\bf 3},1) L_j$ operator 
 ($({\bf 1},{\bf 3},1)$ is a component of $\ov{\Delta}$ and 
 this operator originates from the $\Psi_i\ov{\Delta}\Psi_j$ operator)
 in the evolution from $\mu=2\cdot10^{16}$~GeV to $\mu=10^{13}$~GeV
 and to the coefficient of the Weinberg operator 
 in the evolution from $\mu=10^{13}$~GeV to $\mu=M_Z$,
 written as $R_{ij}$ and defined as $(C_\nu)_{ij}|_{\mu=M_Z}=c\sum_{k,l}R_{ik}R_{jl}(Y_{126})_{kl}|_{\mu=2\cdot10^{16}~{\rm GeV}}$ where $c$ is a flavor-independent constant.
Here $10^{13}$~GeV is a typical scale of the mass of a $({\bf 1},{\bf 3},1)$ particle that is integrated out to obtain the Weinberg operator (see Appendix~B).
\footnote{
In our analysis, we neglect RG corrections involving the coupling of the $L_i ({\bf 1},{\bf 3},1) L_j$ operator,
 since it is much smaller than 1 in all the fitting and minimization results.
}
In the calculation of the RG equations, we assume the following spectrum of the pole masses of SUSY particles for concreteness:
\bea
&&m_{\widetilde{q}}=m_{\widetilde{u}^c}=m_{\widetilde{d}^c}=m_{\widetilde{\ell}}=m_{\widetilde{e}^c}=m_{H^0}=m_{H^\pm}=m_A=20~{\rm TeV},
\nn\\
&&M_{\widetilde{g}}=M_{\widetilde{W}}=\mu_H=2~{\rm TeV},
\ \ \ \ \ \ \tan\beta \ = \ 50.
\label{massspectrum}
\eea
However, we caution that the values of $Y_u,Y_d,Y_e$ at $\mu=2\cdot10^{16}$~GeV and $R_{ij}$ only logarithmically depend 
 on the SUSY particle mass spectrum and so the texture of $Y_{10}$, $Y_{126}$, $Y_{120}$ we search is not sensitive to the spectrum;
 for example, multiplying the spectrum with factor 10 does not change our results.
We adopt the following input values for quark masses and CKM matrix parameters:
The isospin-averaged quark mass and strange quark mass in $\ov{{\rm MS}}$ scheme are obtained from lattice calculations in
 Refs.~\cite{lattice1,lattice2,lattice3,lattice4,lattice5,lattice6} as
 $\frac{1}{2}(m_u+m_d)(2~{\rm GeV})=3.373(80)~{\rm MeV}$ and $m_s(2~{\rm GeV})=92.0(2.1)~{\rm MeV}$.
The up and down quark mass ratio is obtained from an estimate in Ref.~\cite{latticereview} as $m_u/m_d=0.46(3)$.
The $\ov{{\rm MS}}$ charm and bottom quark masses are obtained from QCD sum rule calculations in Ref.~\cite{cb} as
  $m_c(3~{\rm GeV})=0.986-9(\alpha_s^{(5)}(M_Z)-0.1189)/0.002\pm0.010~{\rm GeV}$
  and $m_b(m_b)=4.163+7(\alpha_s^{(5)}(M_Z)-0.1189)/0.002\pm0.014~{\rm GeV}$.
The top quark pole mass is obtained from $t\bar{t}$+jet events measured by ATLAS~\cite{Aad:2019mkw} as
 $M_t=171.1\pm1.2$~GeV.
The CKM mixing angles and CP phase are calculated from the Wolfenstein parameters in the latest CKM fitter result~\cite{ckmfitter}.
For the QCD and QED gauge couplings, we use $\alpha_s^{(5)}(M_Z)=0.1181$ and $\alpha^{(5)}(M_Z)=1/127.95$.
For the lepton and W, Z, Higgs pole masses, we use the values in Particle Data Group~\cite{Tanabashi:2018oca}.

The result is given in terms of the singular values of $Y_u,Y_d,Y_e$ and the CKM mixing angles and CP phase at $\mu=2\cdot10^{16}$~GeV,
 as well as $R_{ij}$ in the flavor basis where $Y_e$ is diagonal ($R_{ij}$ is also diagonal in this basis),
 tabulated in Table~\ref{values}.
For each singular value of $Y_u,Y_d$, we present 1$\sigma$ error that has propagated from experimental error of
 the corresponding input quark mass.
For the CKM mixing angles and CP phase, we present 1$\sigma$ errors that have propagated from experimental errors of the input Wolfenstein parameters.
\begin{table}[H]
\begin{center}
  \caption{The singular values of MSSM Yukawa couplings $Y_u$, $Y_d$, $Y_e$, and the mixing angles and CP phase of CKM matrix,
  at $\mu=2\cdot10^{16}$~GeV in $\ov{{\rm DR}}$ scheme.
  Also shown is the flavor-dependent RG correction $R_{ij}$ for the Weinberg operator $C_\nu$,
   defined as $(C_\nu)_{ij}|_{\mu=M_Z}=c\sum_{k,l}R_{ik}R_{jl}(Y_{126})_{kl}|_{\mu=2\cdot10^{16}~{\rm GeV}}$ 
   ($c$ is a flavor-independent constant),
   in the flavor basis where $Y_e$ is diagonal ($R_{ij}$ is also diagonal in this basis).
  For each singular value of the quark Yukawa matrices, we present 1$\sigma$ error that has propagated from experimental error of 
   the corresponding input quark mass,
  and for the CKM parameters, we present 1$\sigma$ errors that have propagated from experimental errors of the input 
   Wolfenstein parameters.}
  \begin{tabular}{|c||c|} \hline
                                      & Value with Eq.~(\ref{massspectrum})  \\ \hline
    $y_u$           &2.69(14)$\times10^{-6}$          \\
    $y_c$           &0.001384(14)                               \\ 
    $y_t$            &0.478(98)                                    \\ \hline
    $y_d$           &0.0002908(92)                             \\
    $y_s$           &0.00579(13)                                 \\ 
    $y_b$           &0.3552(23)                                   \\ \hline
    $y_e$           &0.00012202                                 \\
    $y_\mu$           &0.025766                                \\
    $y_\tau$           &0.50441                                  \\ \hline
    $\cos\theta_{13}^{\rm ckm}\sin\theta_{12}^{\rm ckm}$            & 0.22474(25)      \\
    $\cos\theta_{13}^{\rm ckm}\sin\theta_{23}^{\rm ckm}$           & 0.0398(10)         \\
    $\sin\theta_{13}^{\rm ckm}$                                                              & 0.00352(21)      \\
    $\delta_{\rm km}$~(rad)                                                                       &1.147(33)            \\ \hline
    $R_{ee}$ & 1.00  \\
    $R_{\mu\mu}$ & 1.00 \\
    $R_{\tau\tau}$ & 0.961 \\ \hline
  \end{tabular}
  \label{values}
  \end{center}
\end{table}

Next, we fit the MSSM Yukawa couplings $Y_u,Y_d,Y_e$ and the neutrino mixing angles and mass differences 
 with the fundamental Yukawa couplings $Y_{10},Y_{126},Y_{120}$ and the numbers $r_1,r_2,r_3,r_e$ according to Eqs.~(\ref{yu})-(\ref{ye}),(\ref{cnu}).
Meanwhile, we minimize the following quantity:
\bea
\sum_{A=10,126}\,\left\{ \ 
\left|(Y_A)_{u_R d_R}\right|^2+\left|(Y_A)_{u_R s_R}\right|^2+\left|(Y_A)_{u_L d_L}\right|^2+\left|(Y_A)_{u_L u_L}\right|^2+\left|(Y_A)_{u_L s_L}\right|^2 \ \right\}
\label{tobeminimized}
\eea
To facilitate the analysis, we concentrate on the parameter region where $r_3=0$ (which is compatible with any values of $r_1,r_2,r_e$
 and Eq.~(\ref{texture})).
Then, we obtain $(Y_{10})_{u_R\, j}+r_2(Y_{126})_{u_R\, j}\leq y_u$ and $(Y_{10})_{u_L\, j}+r_2(Y_{126})_{u_L\, j}\leq y_u$ for any flavor index $j$, and it becomes easier to reduce Eq.~(\ref{tobeminimized}) to the order of the tiny up quark Yukawa coupling $y_u$.
Given $r_3=0$, Eqs.~(\ref{yu})-(\ref{ye}) can be rearranged as follows:
We fix the flavor basis such that the left-handed down-type quark components in $\Psi_i$ have diagonal $Y_d$ Yukawa coupling
 with real positive diagonal components.
$Y_u$, which is still symmetric, is then written as
\footnote{
Note that $Y_u$ in Eq.~(\ref{susyyukawa}) is the complex conjugate of $Y_u$ in SM defined as $-{\cal L}=\bar{q}_L Y_u u_R H$.
}
\bea
Y_u=V_{\rm CKM}^T   
\begin{pmatrix} 
      y_u & 0 & 0 \\
      0 & y_c \,e^{2i \,d_2} & 0 \\
      0 & 0 & y_t \,e^{2i \,d_3}\\
   \end{pmatrix}
   V_{\rm CKM}
   \label{yu3}
\eea
 where $d_2,d_3$ are unknown phases.
In the same flavor basis, $Y_d$ becomes
\bea
Y_d=\begin{pmatrix} 
      y_d & 0 & 0 \\
      0 & y_s & 0 \\
      0 & 0 & y_b \\
   \end{pmatrix}
   V_{dR}
   \label{yd3}
 \eea
 where $V_{dR}$ is an unknown unitary matrix.
From Eqs.~(\ref{yd}),(\ref{ye}) and the fact that $Y_{10},Y_{126}$ are symmetric and $Y_{120}$ is antisymmetric, we get
\bea
&&Y_{126}=\frac{1}{1-r_2}\left\{\frac{1}{r_1}\frac{1}{2}\left(Y_d+Y_d^T\right)-Y_u\right\},
\\
&&\frac{1}{r_1}Y_e=Y_u-(3+r_2)Y_{126}+r_e\frac{1}{r_1}\frac{1}{2}\left(Y_d-Y_d^T\right).
\eea
We perform the singular value decomposition of $Y_e$ as
\bea
Y_e = U_{eL}\begin{pmatrix} 
      y_e & 0 & 0 \\
      0 & y_\mu & 0 \\
      0 & 0 & y_\tau \\
   \end{pmatrix}
   U_{eR}^\dagger,
 \eea
 and calculate the active neutrino mass matrix in the charged-lepton-diagonal basis as
\bea
(M_\nu)_{\ell \ell'} \ \propto \ R_{\ell \ell} \ (U_{eL}^TY_{126}U_{eL})_{\ell\ell'} \ 
R_{\ell' \ell'}, \ \ \ \ \ \ \ell,\ell'=e,\mu,\tau,
\label{mnumatching2}
\eea
 where $\ell,\ell'$ denote flavor indices for the left-handed charged leptons.
Utilizing Eqs.~(\ref{yu3})-(\ref{mnumatching2}), we perform the fitting as follows.
We fix $y_u,y_c,y_t$ and CKM matrix by the values in Table~\ref{values},
 while we vary $y_d/r_1,\,y_s/r_1,\,y_b/r_1$, unknown phases $d_2,d_3$, unknown unitary matrix $V_{dR}$
 and complex numbers $r_2,r_e$.
Here we eliminate $r_1$ by requiring that the central value of the electron Yukawa coupling $y_e$ be reproduced.
In this way, we try to 
 reproduce the correct values of $y_d,y_s,y_\mu,y_\tau$, $\theta_{12}^{\rm pmns},\theta_{13}^{\rm pmns},\theta_{23}^{\rm pmns}$
 and neutrino mass difference ratio $\Delta m_{21}^2/\Delta m_{32}^2$.
Specifically, we require $y_d,y_s$ to fit within their respective 3$\sigma$ ranges,
 while we do not constrain $y_b$ because $y_b$ may be subject to sizable GUT-scale threshold corrections.
We impose stringent restrictions on the values of neutrino mixing angles and mass differences,
 because we are primarily interested in
 the prediction for the neutrino Dirac CP phase from the condition that dimension-5 proton decays be suppressed
 as much as possible, and so it is essential to suppress variation of the other neutrino parameters.
In particular, we require $\sin^2\theta_{12}^{\rm pmns},\sin^2\theta_{13}^{\rm pmns}$, $\Delta m_{21}^2/\Delta m_{32}^2$ 
 to fit within their respective 1$\sigma$ ranges reported by NuFIT~4.1~\cite{Esteban:2018azc,nufit}.
We assume two narrow benchmark ranges of $\sin^2\theta_{23}^{\rm pmns}$, 
 since the current experimental error of $\sin^2\theta_{23}^{\rm pmns}$ is too large.
Only the normal hierarchy of the neutrino mass is considered because no good fitting is obtained with the inverted hierarchy.
Finally, since the experimental errors of $y_\mu,y_\tau$ are tiny, 
 we only require their reproduced values to fit within $\pm$0.1\% ranges of their central values.
The constraints are summarized in Table~\ref{fitting}.
\begin{table}[H]
\begin{center}
  \caption{Allowed ranges of quantities in the analysis.}
  \begin{tabular}{|c|c|} \hline
     & Allowed range \\ \hline
    $y_u$           &2.73$\times10^{-6}$ \ \ (fixed)\\
    $y_c$           &0.001406 \ \ (fixed)\\
    $y_t$            &0.4842 \ \ (fixed)\\ \hline
    $y_d$           &0.0002953$\pm0.0000093\cdot3$\\
    $y_s$           &0.00588$\pm0.00013\cdot3$\\
    $y_b$           &unconstrained\\ \hline
    $y_e$           &0.00012288 (used to fix $r_1$)\\
    $y_\mu$           &0.025948$\pm 0.1$\%\\
    $y_\tau$           &0.50625$\pm 0.1$\% \\ \hline
    $\cos\theta_{13}^{\rm ckm}\sin\theta_{12}^{\rm ckm}$            & 0.22474 (fixed)\\
    $\cos\theta_{13}^{\rm ckm}\sin\theta_{23}^{\rm ckm}$           & 0.0399 (fixed)\\
    $\sin\theta_{13}^{\rm ckm}$                                                              & 0.00352 (fixed)\\
    $\delta_{\rm km}$~(rad)           &1.147 (fixed)\\ \hline
    $\sin^2\theta_{12}^{\rm pmns}$ & $0.310\pm0.012$\\
    $\sin^2\theta_{13}^{\rm pmns}$ & $0.02237\pm0.00065$\\
    $\sin^2\theta_{23}^{\rm pmns}$ & $0.45\pm0.01$ \ \ {\rm or} \ \ $0.55\pm0.01$\\
    $\Delta m_{21}^2/\Delta m_{32}^2$ & $0.02923\pm0.00084$ \\
    $\delta_{\rm pmns},\ \alpha_2,\ \alpha_3,\ m_1$ & unconstrained \\ \hline
    $d_2,\ d_3,\ V_{dR}$ & unconstrained \\
    $r_1$ & eliminated in favor of $y_e$ \\
    $r_3 $ & 0 (fixed) \\
    $r_2,\ r_e$ & unconstrained \\ \hline
  \end{tabular}
  \label{fitting}
  \end{center}
\end{table}
Within the constraints of Table~\ref{fitting}, we minimize the quantity~Eq.~(\ref{tobeminimized}) repeatedly
 starting from different random values of $y_d/r_1,\,y_s/r_1,\,y_b/r_1$, $d_2,d_3$, $V_{dR}$, $r_2,r_e$.
Each fitting and minimization result is plotted on the planes of 
 the neutrino Dirac CP phase $\delta_{\rm pmns}$, the lightest neutrino mass $m_1$, and the absolute value of the (1,1)-component of the neutrino mass matrix in the charged-lepton-diagonal basis $|m_{ee}|$, versus the ``maximal proton decay amplitude" defined in the next subsection.
\label{procedures}
\\

\subsection{Results}

We present the plots of fitting and minimization results obtained by the procedures of Section~\ref{procedures},
 on the planes of $\delta_{\rm pmns}$, $m_1$, $|m_{ee}|$ versus ``maximal proton decay amplitude".
The ``maximal proton decay amplitude" of the $p\to K^+\nu$ mode, $\tilde{A}(p\to K^+ \nu)$, is defined as
\begin{align}
\tilde{A}(p\to K^+ \nu)^2 \ = \ 
&\left\{\tilde{A}(p\to K^+ \bar{\nu}_\tau)|_{{\rm from} \ C_{5R}}+\tilde{A}(p\to K^+ \bar{\nu}_\tau)|_{{\rm from} \ C_{5L}}\right\}^2
\nn\\
& \ \ \ \ \ \ \ \ \ \ \ \ +\tilde{A}(p\to K^+ \bar{\nu}_\mu)^2|_{{\rm from} \ C_{5L}}+\tilde{A}(p\to K^+ \bar{\nu}_e)^2|_{{\rm from} \ C_{5L}},
\nn\\
\label{maximal}
\end{align}
 where
\begin{align}
\tilde{A}(p\to K^+ \bar{\nu}_\tau)|_{{\rm from} \ C_{5R}} \ = \ &y_ty_\tau\sum_{A,B}
\left|(1+\frac{D}{3}+F)\,V_{ts}^{\rm ckm}\,\{(Y_A)_{\tau_R t_R}(Y_B)_{u_Rd_R}-(Y_A)_{\tau_R u_R}(Y_B)_{t_Rd_R}\}\right.
\nn\\
&\left. \ \ \ \ \ \ \ \ \ \ \ \ \ \ \ \ \ \ +\frac{2D}{3}\,V_{td}^{\rm ckm}\,\{(Y_A)_{\tau_R t_R}(Y_B)_{u_Rs_R}-(Y_A)_{\tau_R u_R}(Y_B)_{t_Rs_R}\}\right|
\nn\\
{\rm ( \ sum \ over} \ A,B=&(10,10),\,(10,126),\,(126,10),\,(126,126),\,(120,10),\,(120,126),\,(120,120) \ ),
\nn\\
\\
\tilde{A}(p\to K^+ \bar{\nu}_\alpha)|_{{\rm from} \ C_{5L}} \ = \ &g_2^2\sum_{A,B=10,126} \frac{3}{2}
\left|(1+\frac{D}{3}+F)(Y_A)_{u_L d_L}(Y_B)_{s_L \alpha_L}+\frac{2D}{3}(Y_A)_{u_L s_L}(Y_B)_{d_L \alpha_L}\right|
\nn\\
\end{align}
 with $\alpha=e,\mu,\tau$, and $y_t,y_\tau,g_2$ denoting the top and tau Yukawa couplings and the weak gauge coupling at soft SUSY breaking scale $\mu_{\rm SUSY}$.
The ``maximal proton decay amplitudes" of the $p\to K^0 e^+_\beta$ modes, $\tilde{A}(p\to K^0 e^+_\beta)$, are defined as
\begin{align}
\tilde{A}(p\to K^0 e^+_\beta) \ = \ &g_2^2\sum_{A,B=10,126}\frac{3}{2}
(1-D+F)\left|(Y_A)_{u_L s_L}(Y_B)_{u_L \beta_L}-(Y_A)_{u_L u_L}(Y_B)_{s_L \beta_L}\right|
\nn\\
\end{align}
 with $\beta=e,\mu$.
The above $\tilde{A}(p\to K^+ \nu)^2$ is related to the $p\to K^+\nu$ decay width when
 coefficients $a,b,d,e,g,h,j$ have similar absolute values\footnote{
 Remember that we are setting $c=f=0$.
 } and the terms with these coefficients interfere maximally constructively
 under the condition of $a\, (Y_{10})_{d_L s_L}+d\, (Y_{126})_{d_L s_L}=0$ and $b\, (Y_{10})_{d_L s_L}+e\, (Y_{126})_{d_L s_L}=0$
 and when the contributions from left-handed dimension-5 operators to the $p\to K^+\bar{\nu}_\tau$ mode and those from right-handed ones interfere maximally constructively.
Likewise, $\tilde{A}(p\to K^0 e^+_\beta)^2$ are related to the $p\to K^0 e^+_\beta$ decay widths
 when coefficients $a,b,d,e$ have similar absolute values and the terms with these coefficients interfere maximally constructively
 under the condition of $a\, (Y_{10})_{d_L s_L}+d\, (Y_{126})_{d_L s_L}=0$ and $b\, (Y_{10})_{d_L s_L}+e\, (Y_{126})_{d_L s_L}=0$.
Therefore, $\tilde{A}(p\to K^+ \nu)$ and $\tilde{A}(p\to K^0 e^+_\beta)$ allow us to estimate 
 how much the texture of the GUT-scale Yukawa couplings $Y_{10},Y_{126},Y_{120}$ contributes to the suppression of dimension-5 proton decays.

As a matter of fact, we have found that $\tilde{A}(p\to K^0 e^+_\beta)$ $(\beta=e,\mu)$ are smaller than $\tilde{A}(p\to K^+ \nu)$ in all the fitting and minimization results.
Considering that the current experimental bound is more severe for the $p\to K^+ \nu$ decay than for the $p\to K^0 e^+_\beta$ decays,
 it is phenomenologically more important to study the suppression of $\tilde{A}(p\to K^+ \nu)$ than that of $\tilde{A}(p\to K^0 e^+_\beta)$.
Therefore, we present the plots of $\delta_{\rm pmns},m_1,|m_{ee}|$ versus $\tilde{A}(p\to K^+ \nu)$ only,
 and solely discuss the suppression of $\tilde{A}(p\to K^+ \nu)$.

As a reference, we also present ``minimal proton partial lifetime" $1/\tilde{\Gamma}(p\to K^+ \nu)$ that corresponds to $\tilde{A}(p\to K^+ \nu)$
 and is computed for a sample SUSY particle mass spectrum. It is defined as
\begin{align}
\tilde{\Gamma}(p\to K^+ \nu) &= \frac{m_N}{64\pi}\left(1-\frac{m_K^2}{m_N^2}\right)^2 \times
\nn\\
&\left[ \ \left\{
\left|
\frac{\alpha_H(\mu_{\rm had})}{f_\pi}
A_{RL}(\mu_{\rm had},\mu_{\rm SUSY}) \frac{\mu_H}{m_{\tilde{t}_R}^2} {\cal F}' \, A_R^{\tau t}(\mu_{\rm SUSY},\mu_{H_C}) 
\frac{1}{M_{H_C}}\tilde{A}(p\to K^+ \bar{\nu}_\tau)|_{{\rm from} \ C_{5R}}\right|
\right.\right.
\nn\\
&\left. \ \ \ \ \ +\left|
\frac{\beta_H(\mu_{\rm had})}{f_\pi}
A_{LL}(\mu_{\rm had},\mu_{\rm SUSY}) \frac{M_{\widetilde{W}}}{m_{\tilde{q}}^2} {\cal F} \, A_L^\tau(\mu_{\rm SUSY},\mu_{H_C}) 
\frac{1}{M_{H_C}}\tilde{A}(p\to K^+ \bar{\nu}_\tau)|_{{\rm from} \ C_{5L}}\right|
\right\}^2
\nn\\
&+\left|\frac{\beta_H(\mu_{\rm had})}{f_\pi} A_{LL}(\mu_{\rm had},\mu_{\rm SUSY}) \frac{M_{\widetilde{W}}}{m_{\tilde{q}}^2} {\cal F}\,
A_L(\mu_{\rm SUSY},\mu_{H_C}) \frac{1}{M_{H_C}}\tilde{A}(p\to K^+ \bar{\nu}_\mu)|_{{\rm from} \ C_{5L}}\right|^2
\nn\\
&\left. +\left|\frac{\beta_H(\mu_{\rm had})}{f_\pi} A_{LL}(\mu_{\rm had},\mu_{\rm SUSY}) \frac{M_{\widetilde{W}}}{m_{\tilde{q}}^2} {\cal F}\,
A_L(\mu_{\rm SUSY},\mu_{H_C}) \frac{1}{M_{H_C}}\tilde{A}(p\to K^+ \bar{\nu}_e)|_{{\rm from} \ C_{5L}}\right|^2 \ \right]
\label{samplepartiallifetime}
\end{align}
 where the following sample spectrum of the pole masses of SUSY particles is assumed:
\bea
&&m_{H^0}=m_{H^\pm}=m_A=({\rm sfermion \ mass})=300~{\rm TeV},
\nn\\
&&M_{\widetilde{g}}=M_{\widetilde{W}}=\mu_H=10~{\rm TeV},
\ \ \ \ \ \ \tan\beta \ = \ 50.
\label{massspectrum-new}
\eea
We also assume $M_{H_C}=2\cdot10^{16}$~GeV.
Symbols in Eq.~(\ref{samplepartiallifetime}) have been defined in the paragraphs containing Eqs.~(\ref{ptoknuLformula})-(\ref{yyr2})
 \footnote{We have neglected the muon and electron Yukawa couplings in the calculation of $A_{L}^\mu,\,A_{L}^e$ and rewritten them as $A_{L}$.},
 and $m_K$ denotes the kaon mass, $m_N$ the nucleon mass and $f_\pi$ the pion decay constant in the chiral limit.
The ``minimal proton partial lifetime" $1/\tilde{\Gamma}(p\to K^+ \nu)$ is the partial lifetime when
 the absolute values of coefficients $a,b,d,e,g,h,j$ are all 1 and the terms with these coefficients interfere maximally constructively
 under the condition of $a\, (Y_{10})_{d_L s_L}+d\, (Y_{126})_{d_L s_L}=0$ and $b\, (Y_{10})_{d_L s_L}+e\, (Y_{126})_{d_L s_L}=0$
 and when the contributions from left-handed dimension-5 operators and those from right-handed ones interfere maximally constructively,
 hence the name.
We consider $1/\tilde{\Gamma}(p\to K^+ \nu)$ to be a good measure for 
 how much the texture of $Y_{10},Y_{126},Y_{120}$ contributes to the suppression of the $p\to K^+\nu$ decay.
We caution that the mass spectrum Eq.~(\ref{massspectrum-new}) is assumed for reference purposes and has no intrinsic meaning.
We also comment that 
 although the mass spectrum Eq.~(\ref{massspectrum-new}) is different from the one assumed for the RG calculation of the GUT-scale Yukawa couplings,
 the resulting difference in the GUT-scale Yukawa couplings is less significant than the one stemming from experimental uncertainties of the input parameters.

In the evaluation of $\tilde{A}(p\to K^+ \nu)$ and $1/\tilde{\Gamma}(p\to K^+ \nu)$, we employ the following numerical values and formulas:
The baryon chiral Lagrangian parameters are given by $D=0.804$, $F=0.463$,
 and the pion decay constant in the chiral limit is $f_\pi=0.0868$~GeV~\cite{Borsanyi:2012zv}.
The hadronic form factors for proton decay amplitudes are taken from Ref.~\cite{Aoki:2017puj} as
 $\alpha_H(\mu_{\rm had})=-\beta_H(\mu_{\rm had})=-0.0144$~GeV$^3$ for $\mu_{\rm had}=2$~GeV.
The RG corrections represented by $A_{RL}$, $A_{LL}$, $A_R^{\tau t}$, $A_L^\tau$, $A_L$ are calculated 
 by using 1-loop RG equations~\cite{Hisano:1992jj,Goto:1998qg}.
We take $\mu_{H_C}=2\cdot10^{16}$~GeV and $\mu_{\rm SUSY}=300$~TeV.

\begin{figure}[H]
\begin{center}
\includegraphics[width=80mm]{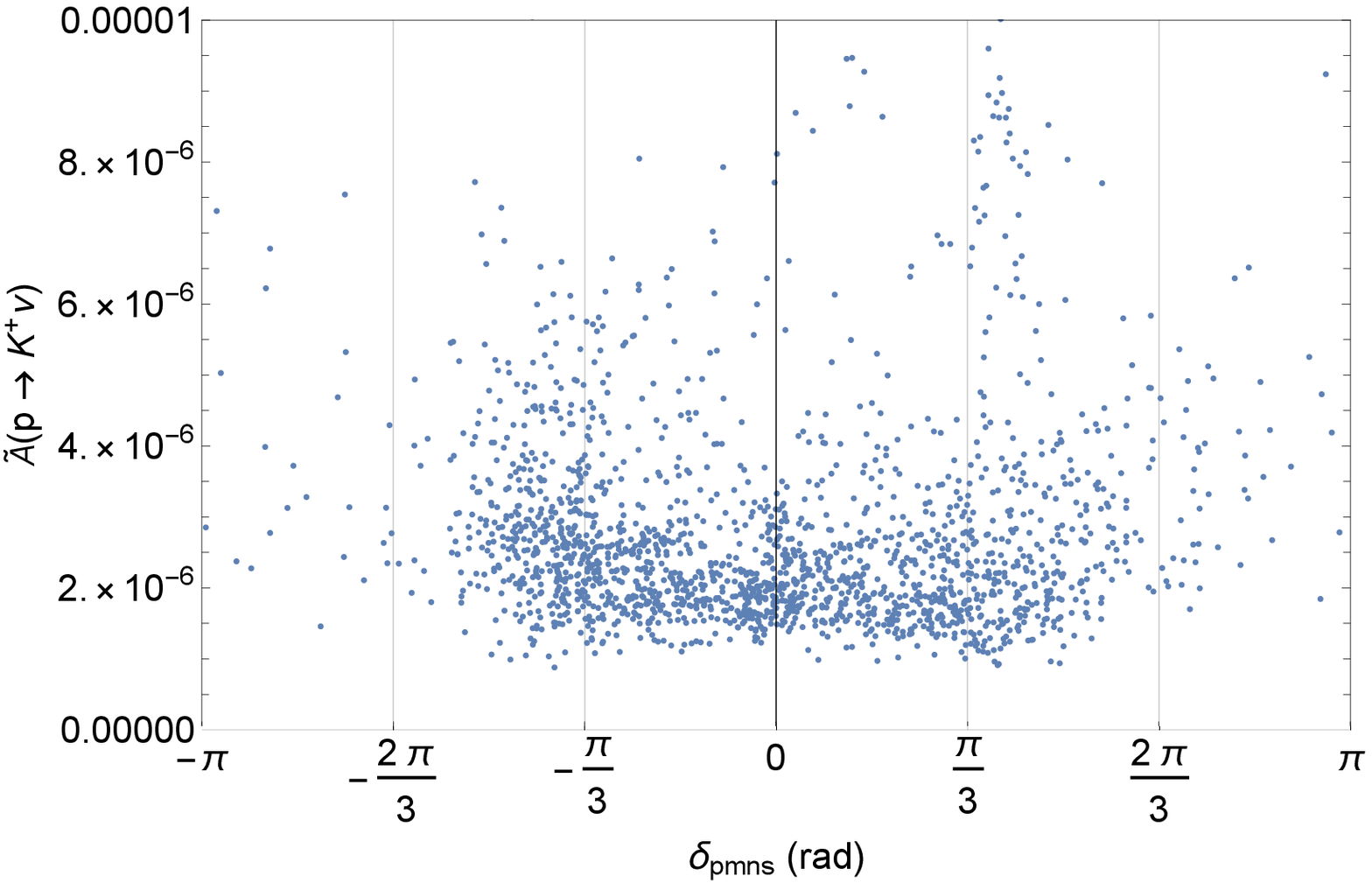}
\includegraphics[width=80mm]{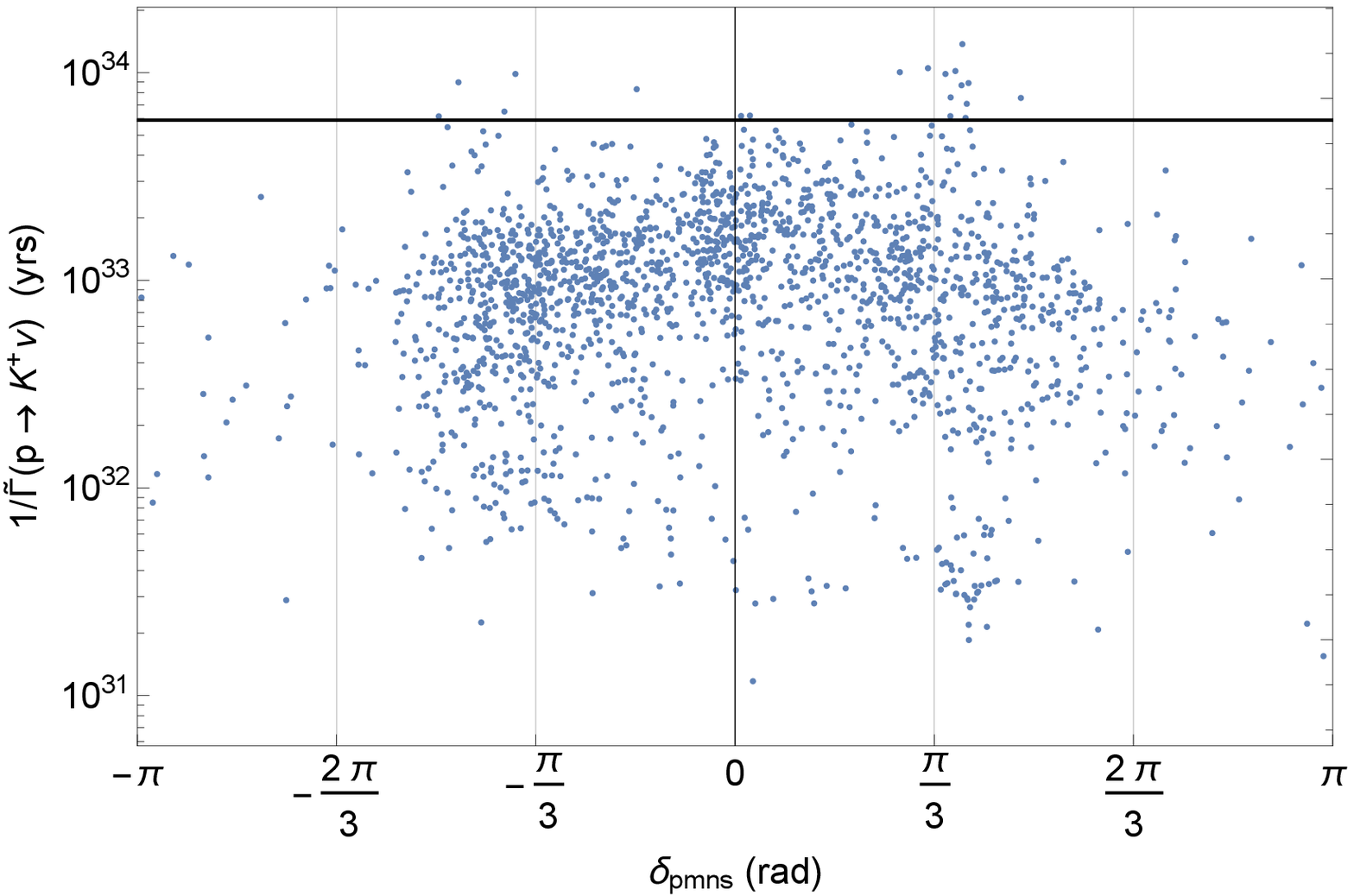}
\\
\includegraphics[width=80mm]{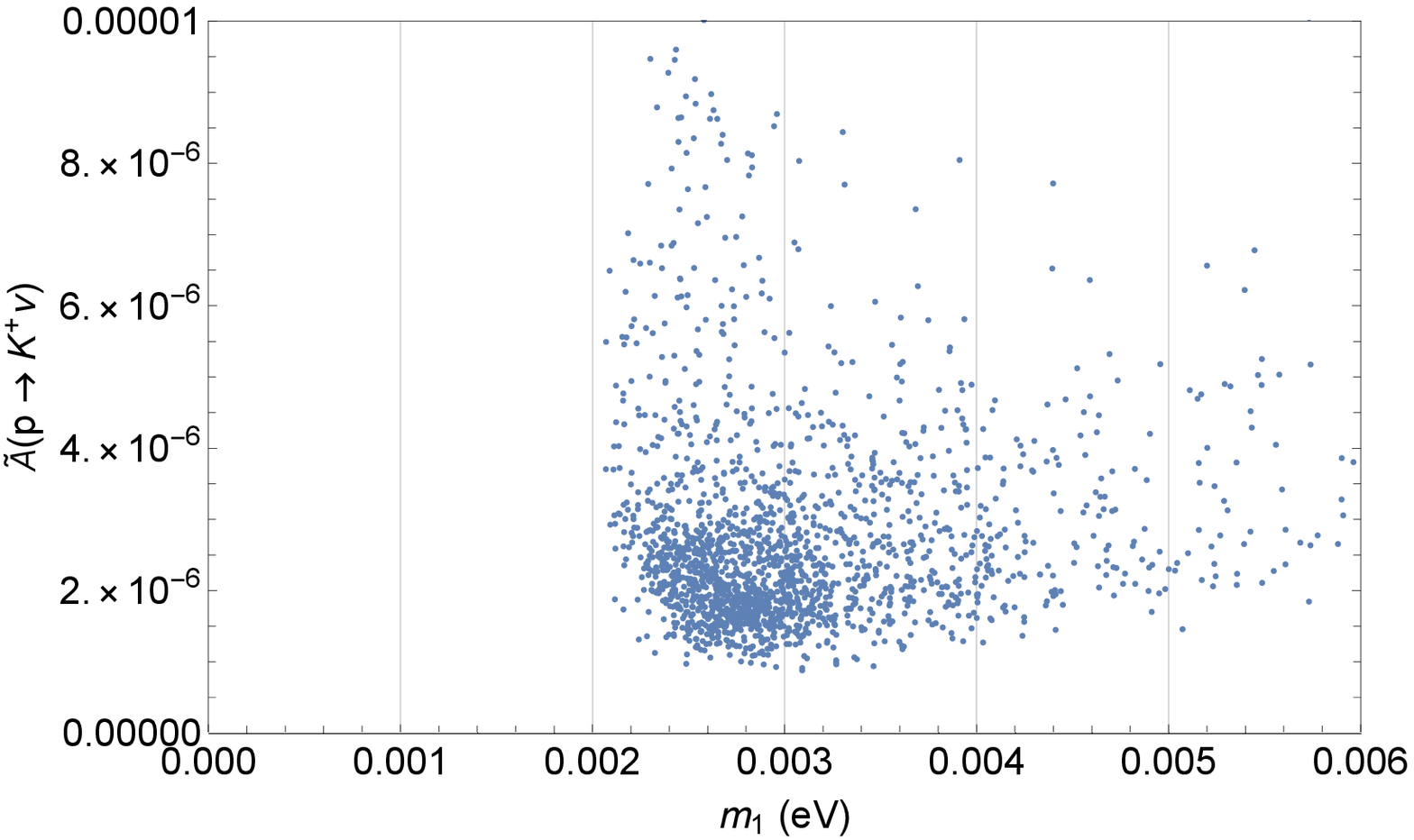}
\includegraphics[width=80mm]{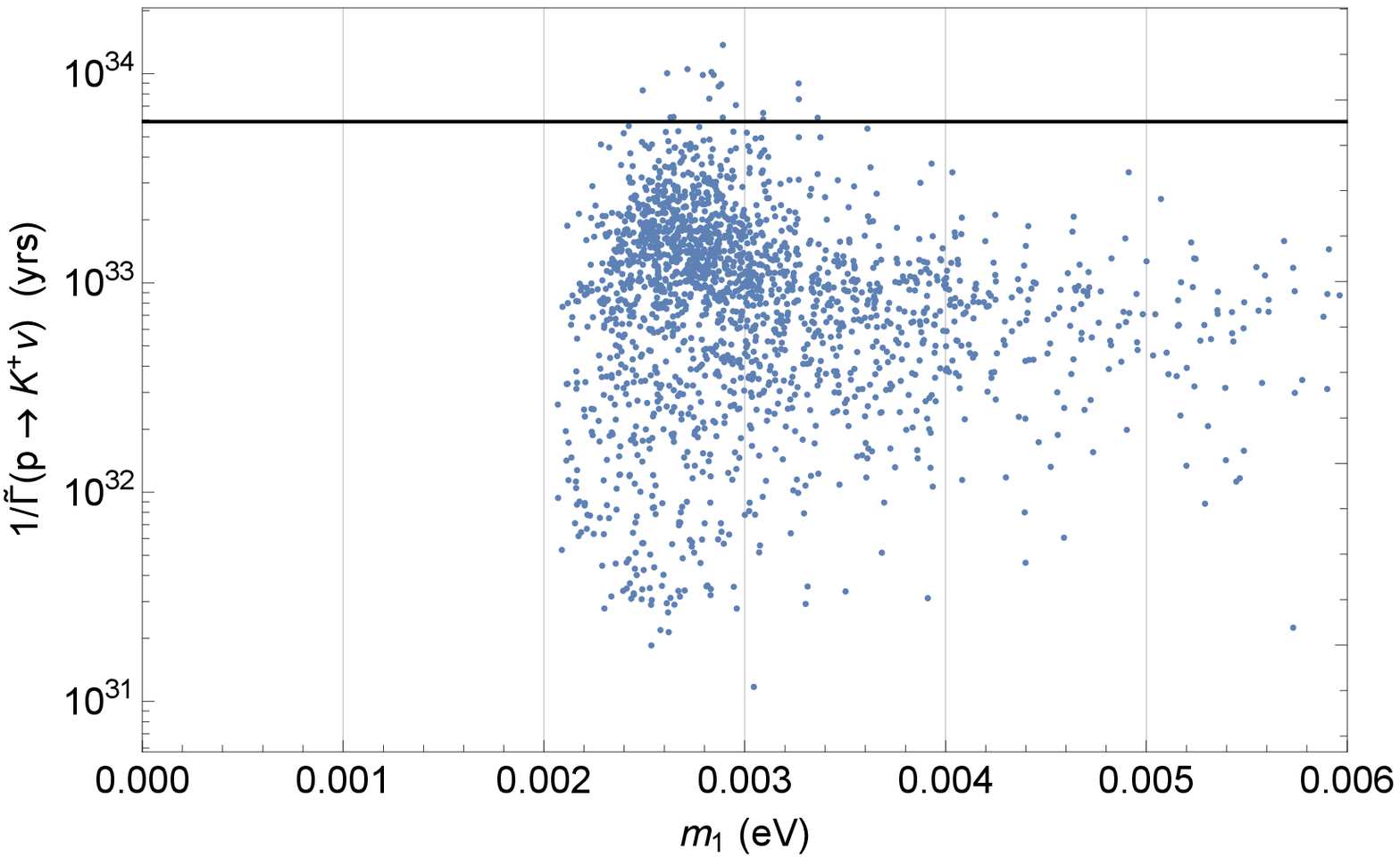}
\\
\includegraphics[width=80mm]{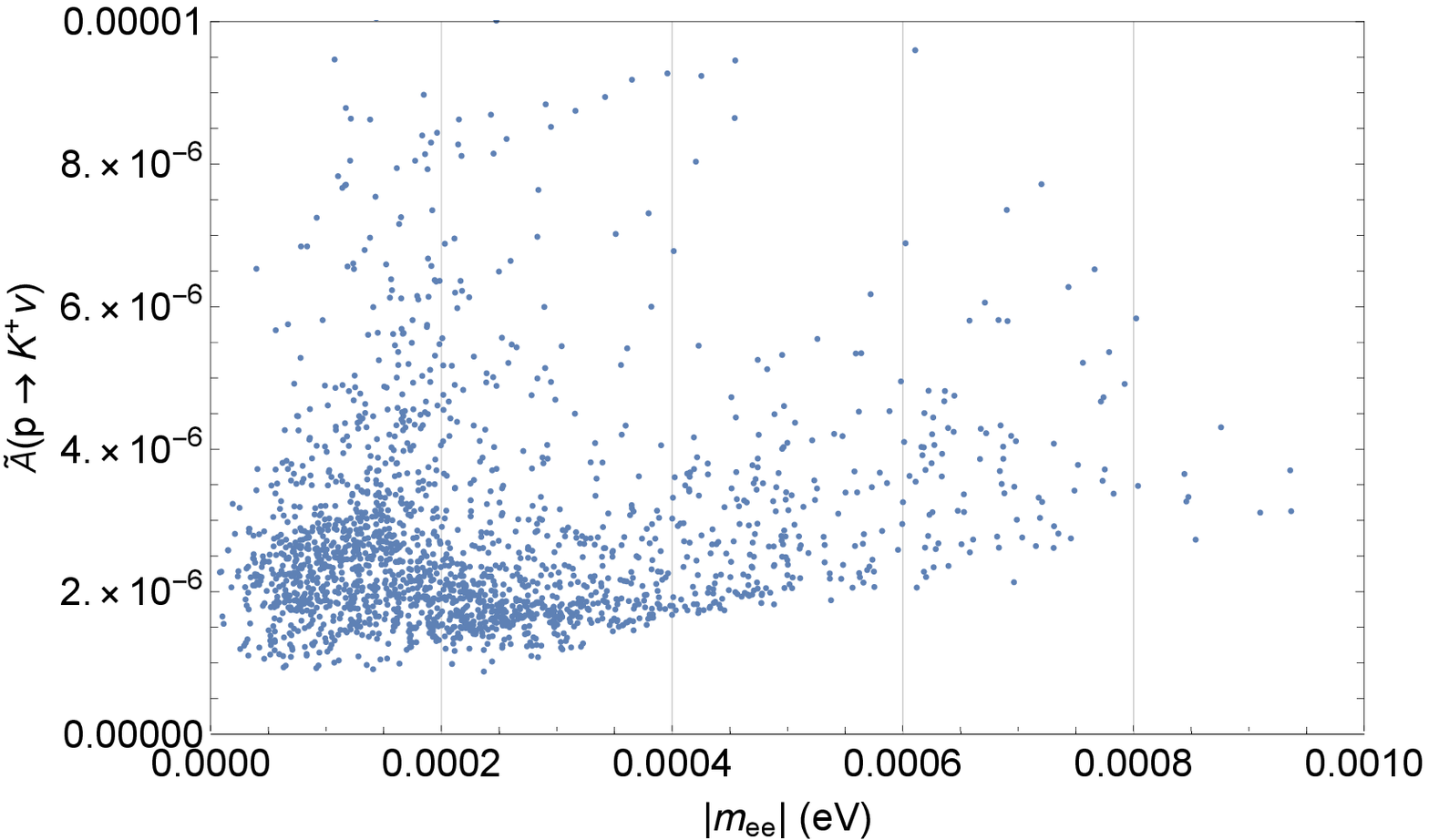}
\includegraphics[width=80mm]{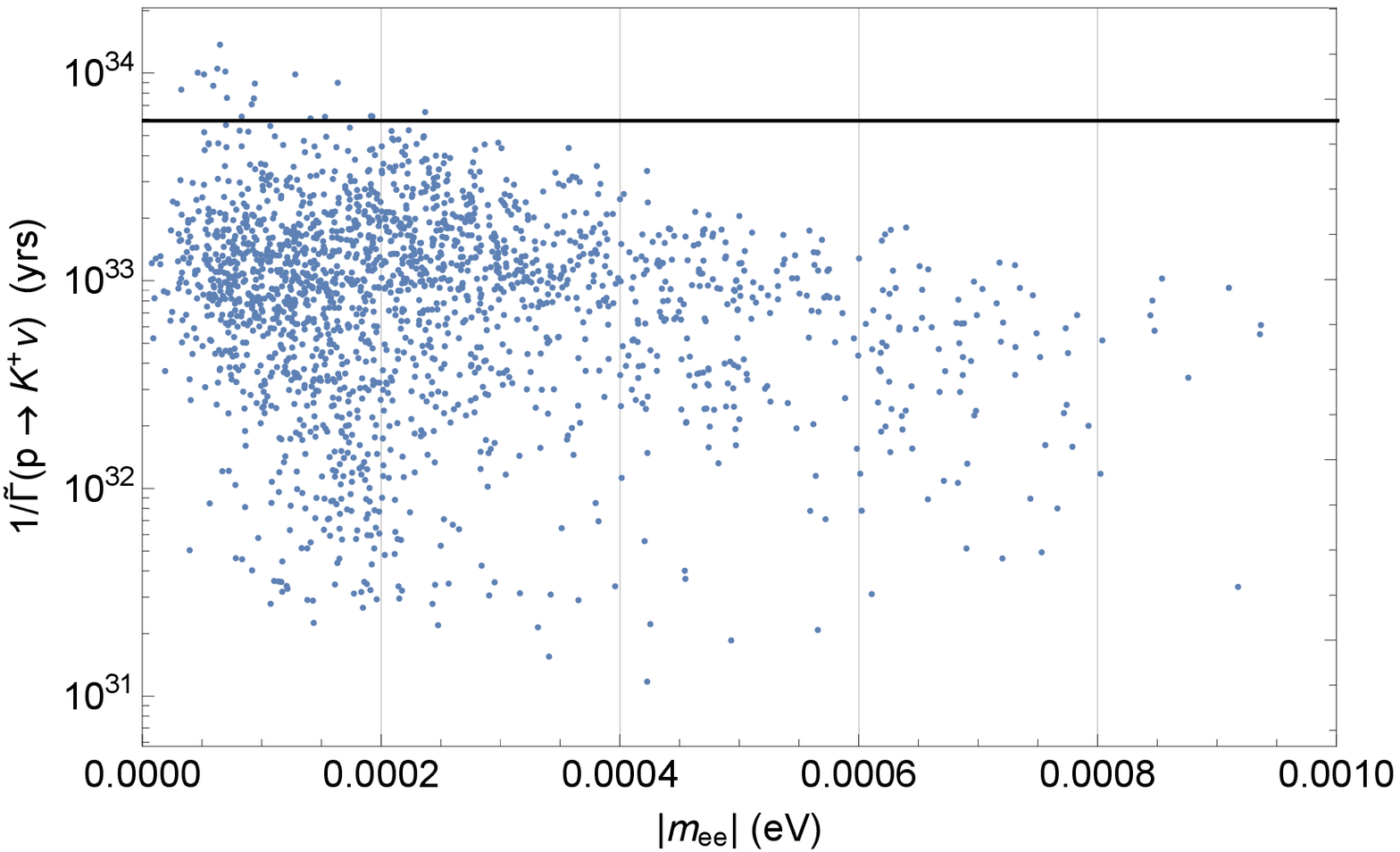}
\caption{
Results of the fitting and minimization analysis in Section~\ref{procedures},
 where the quantity~Eq.~(\ref{tobeminimized}) is minimized within the constraints of Table~\ref{fitting}.
Here we choose the higher-octant benchmark where $\sin^2\theta_{23}^{\rm pmns}=0.55\pm0.01$ in Table~\ref{fitting}.
Each dot corresponds to the result of one analysis starting from
 a different set of random values of $y_d/r_1,\,y_s/r_1,\,y_b/r_1$, $d_2,d_3$, $V_{dR}$, $r_2,r_e$.
The vertical line of the left three panels indicates the ``maximal proton decay amplitude" $\tilde{A}(p\to K^+ \nu)$ defined in Eq.~(\ref{maximal}),
 and that of the right three panels indicates the ``minimal proton partial lifetime" $1/\tilde{\Gamma}(p\to K^+ \nu)$ defined in Eq.~(\ref{samplepartiallifetime}).
From the upper to the lower panels, the horizontal line indicates the neutrino Dirac CP phase $\delta_{\rm pmns}$, the lightest neutrino mass $m_1$, and the absolute value of the (1,1)-component of the neutrino mass matrix in the charged-lepton-diagonal basis $|m_{ee}|$.
The horizontal line in each of the right panels corresponds to the current experimental 90\% CL bound on the $p\to K^+ \nu$ partial lifetime,
 $1/\Gamma(p\to K^+ \nu)=5.9\times10^{33}$~yrs.
}
\label{055}
\end{center}
\end{figure}

\begin{figure}[H]
\begin{center}
\includegraphics[width=80mm]{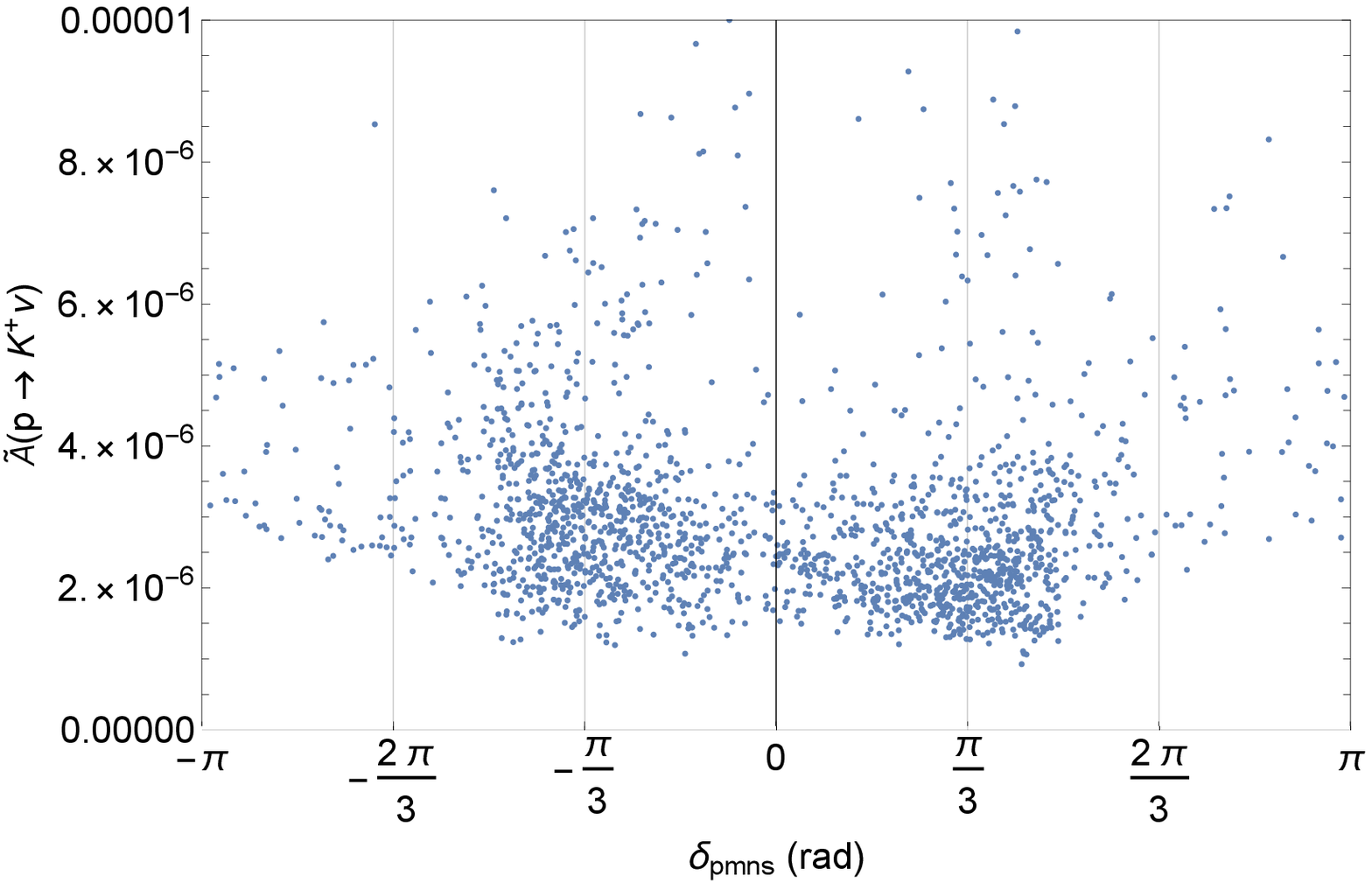}
\includegraphics[width=80mm]{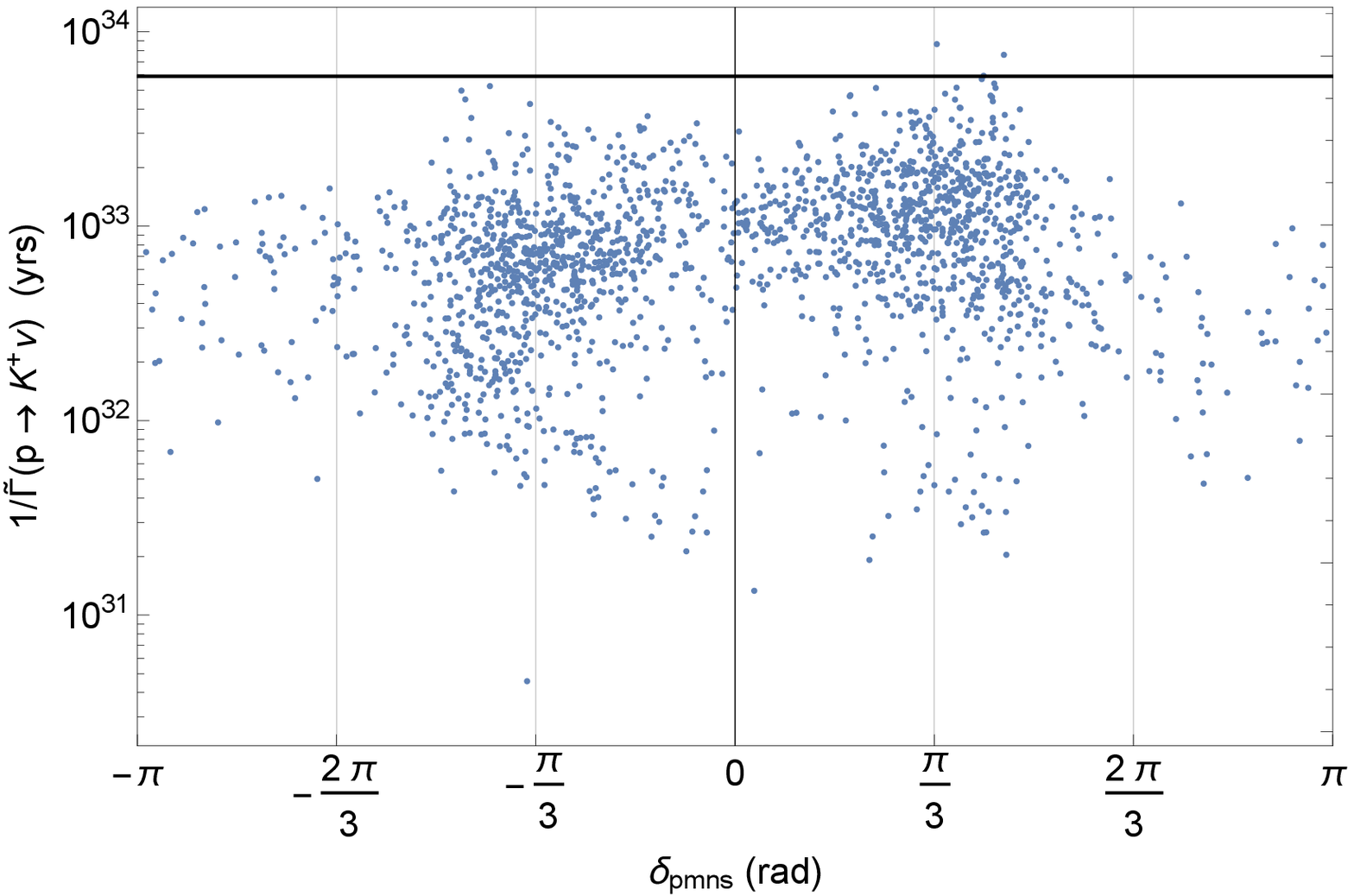}
\\
\includegraphics[width=80mm]{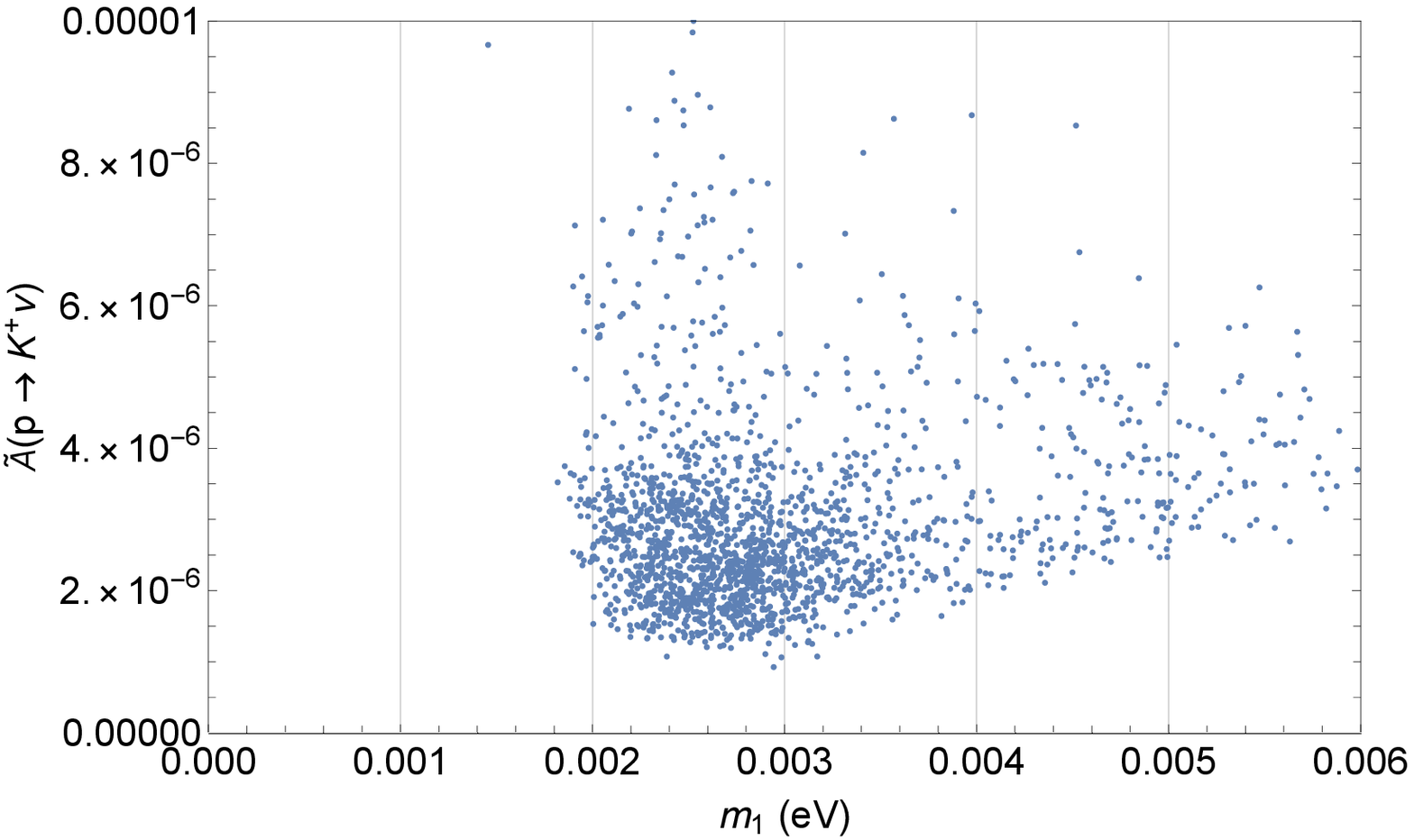}
\includegraphics[width=80mm]{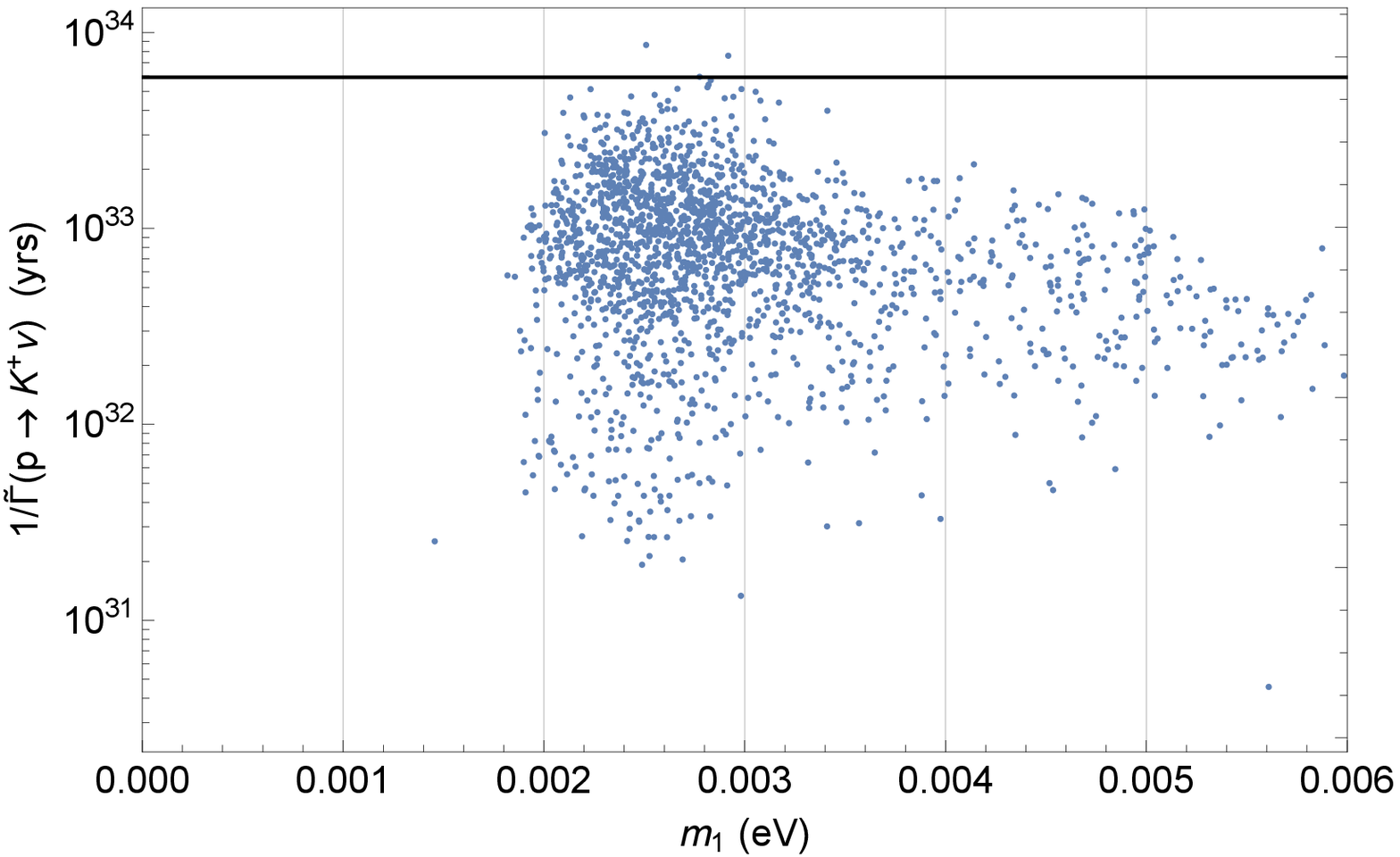}
\\
\includegraphics[width=80mm]{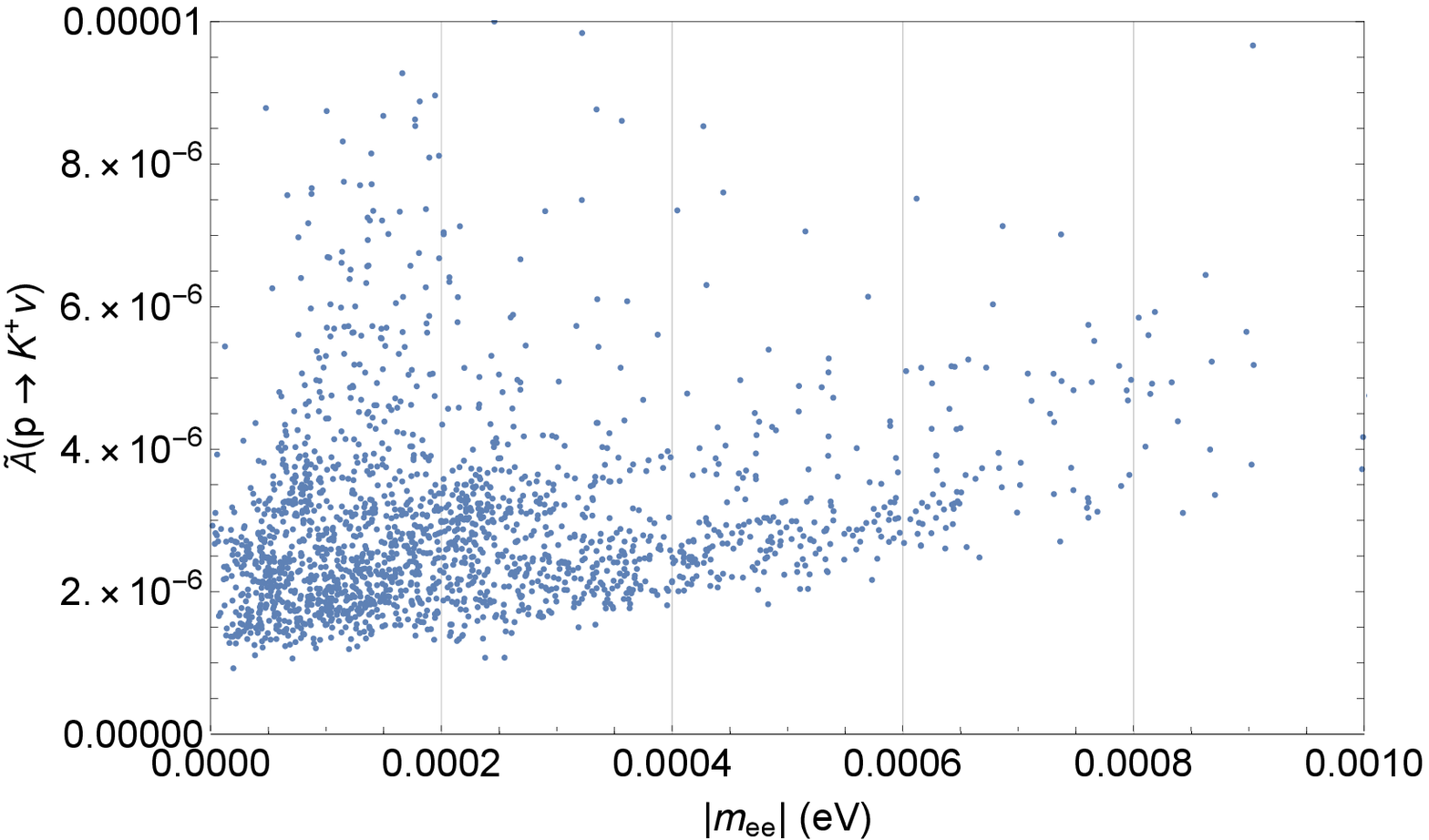}
\includegraphics[width=80mm]{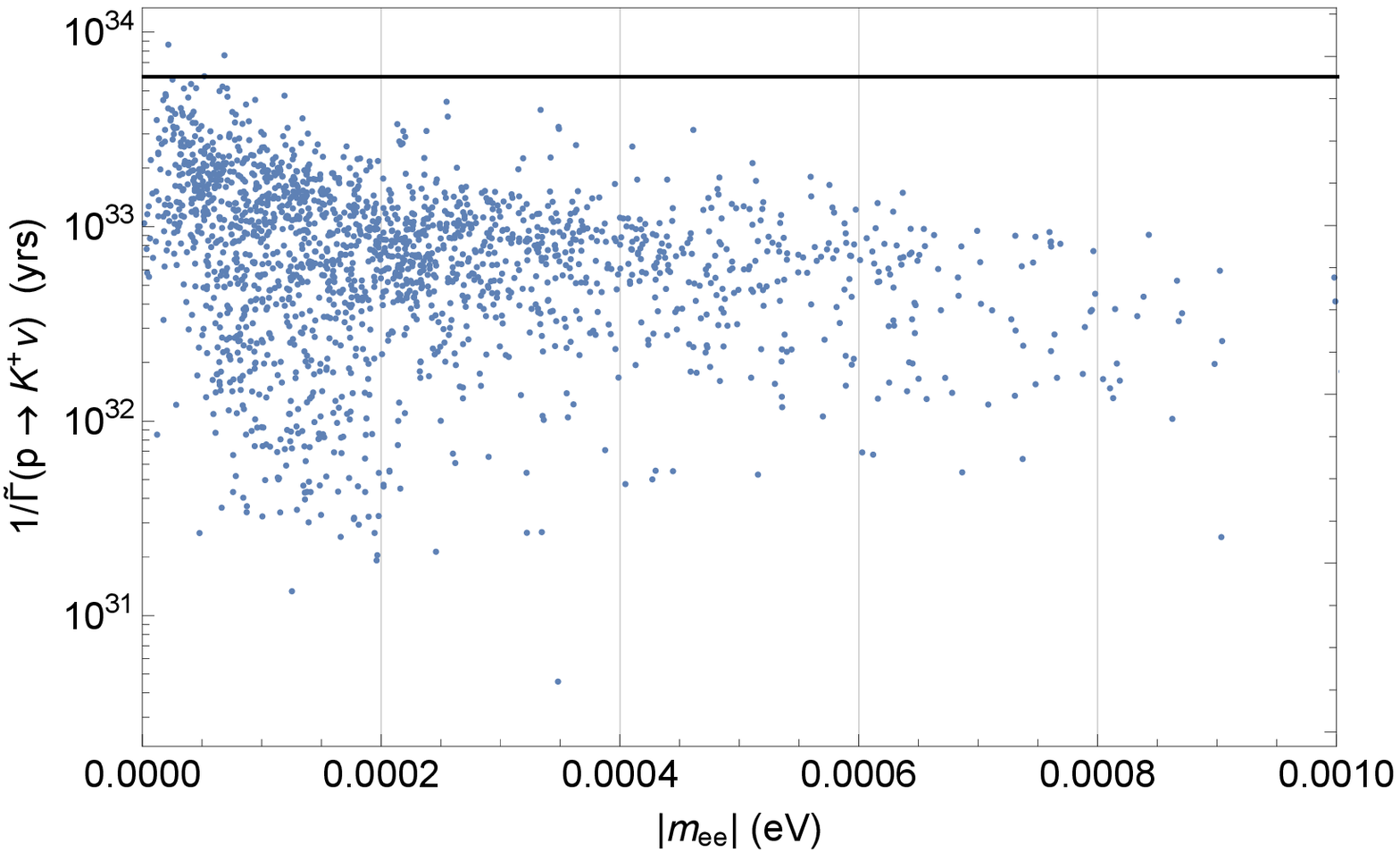}
\caption{
The same as Fig.~\ref{055} except that we choose the lower-octant benchmark where $\sin^2\theta_{23}^{\rm pmns}=0.45\pm0.01$ in Table~\ref{fitting}.
}
\label{045}
\end{center}
\end{figure}

Fig.~\ref{055} displays the results with the higher-octant benchmark where $\sin^2\theta_{23}^{\rm pmns}=0.55\pm0.01$,
 and Fig.~\ref{045} displays those with the lower-octant benchmark where $\sin^2\theta_{23}^{\rm pmns}=0.45\pm0.01$.
The left panels show $\delta_{\rm pmns},m_1,|m_{ee}|$ versus $\tilde{A}(p\to K^+ \nu)$, and the right ones show
 $\delta_{\rm pmns},m_1,|m_{ee}|$ versus $1/\tilde{\Gamma}(p\to K^+ \nu)$.
In the plots, each dot corresponds to the result of one fitting and minimization analysis starting from a different random set of values of
 $y_d/r_1,\,y_s/r_1,\,y_b/r_1$, $d_2,d_3$, $V_{dR}$, $r_2,r_e$.
The horizontal line in each of the right panels corresponds to the current experimental 90\% CL bound on the $p\to K^+ \nu$ partial lifetime,
 $1/\Gamma(p\to K^+ \nu)=5.9\times10^{33}$~yrs~\cite{Abe:2014mwa}.

From the upper panels of Figs.~\ref{055},\ref{045}, we observe that the dimension-5 proton decays are most suppressed
 for the neutrino Dirac CP phase satisfying $\pi/2\gtrsim \delta_{\rm pmns} \gtrsim -\pi/2$.
From the middle panels, we find that the dimension-5 proton decays are most suppressed for the lightest neutrino mass around $m_1\simeq0.003$~eV.
From the lower panels, we see that the dimension-5 proton decays are most suppressed when the 
 (1,1)-component of the neutrino mass matrix in the charged lepton basis satisfies $|m_{ee}|\lesssim0.0002$~eV.
The distributions of the fitting and minimization results are qualitatively the same for
 the higher-octant benchmark with $\sin^2\theta_{23}^{\rm pmns}=0.55\pm0.01$
 and the lower-octant benchmark with $\sin^2\theta_{23}^{\rm pmns}=0.45\pm0.01$,
 which suggests that the results do not depend on the precise value of $\sin^2\theta_{23}^{\rm pmns}$.
The predicted range of the Dirac CP phase $\pi/2\gtrsim \delta_{\rm pmns} \gtrsim -\pi/2$
 will be confirmed or falsified in long baseline neutrino oscillation experiments.
At present, NuFit~5.0~\cite{Esteban:2020cvm} reports a slight contradiction between the results of the T2K and the NOvA long baseline experiments
 on the Dirac CP phase in the normal mass hierarchy case. 
Therefore, we cannot currently state that the above predicted range is experimentally favored or disfavored.
The predicted values of $m_1$ and $|m_{ee}|$, with the normal neutrino mass hierarchy,
 are beyond the reach of on-going and future cosmological and low-energy experiments.

In the right panels, the dots above the black horizontal line are those fitting and minimization results which always satisfy the current experimental bounds
 on proton decays for the sample SUSY particle mass spectrum of Eq.~(\ref{massspectrum-new})
 and
 when the colored Higgs mass matrix satisfies $c=f=0$,
 $a\, (Y_{10})_{d_L s_L}+d\, (Y_{126})_{d_L s_L}=0$ and $b\, (Y_{10})_{d_L s_L}+e\, (Y_{126})_{d_L s_L}=0$,
 and when coefficients $a,b,d,e,g,h,j$ have $O(1)$ absolute value with $M_{H_C}=2\cdot10^{16}$~GeV.
That is to say,
 the dots above the black horizontal line are {\bf predictions} of 
 the $SO(10)$ GUT with the SUSY particle spectrum of Eq.~(\ref{massspectrum-new})
 and the texture of the colored Higgs mass matrix with $c=f=0$,
 $a\, (Y_{10})_{d_L s_L}+d\, (Y_{126})_{d_L s_L}=0$ and $b\, (Y_{10})_{d_L s_L}+e\, (Y_{126})_{d_L s_L}=0$.
The dots slightly below the black horizontal line by $O(1)$ factor are also viable,
 because coefficients $a,b,d,e,g,h,j$ can vary by $O(1)$.
As the dots above or slightly below the black horizontal line satisfy 
 $\pi/2\gtrsim \delta_{\rm pmns} \gtrsim -\pi/2$, 0.004~eV$\gtrsim m_1\gtrsim$0.002~eV and
 $|m_{ee}|<0.0006$~eV,
 these values of $\delta_{\rm pmns}$, $m_1$ and $|m_{ee}|$
 are {\bf predictions} of the current model.
However, it should be reminded that the SUSY particle spectrum of Eq.~(\ref{massspectrum-new})
 is for a reference purpose only,
 and the most important message of the present paper is not that
 the $SO(10)$ GUT with Eq.~(\ref{massspectrum-new}) predicts the above values of $\delta_{\rm pmns}$, $m_1$ and $|m_{ee}|$,
 but that dimension-5 proton decays are most suppressed for these values of $\delta_{\rm pmns}$, $m_1$ and $|m_{ee}|$ independently of details of the SUSY particle spectrum.
\\

\begin{figure}[H]
\begin{center}
\includegraphics[width=80mm]{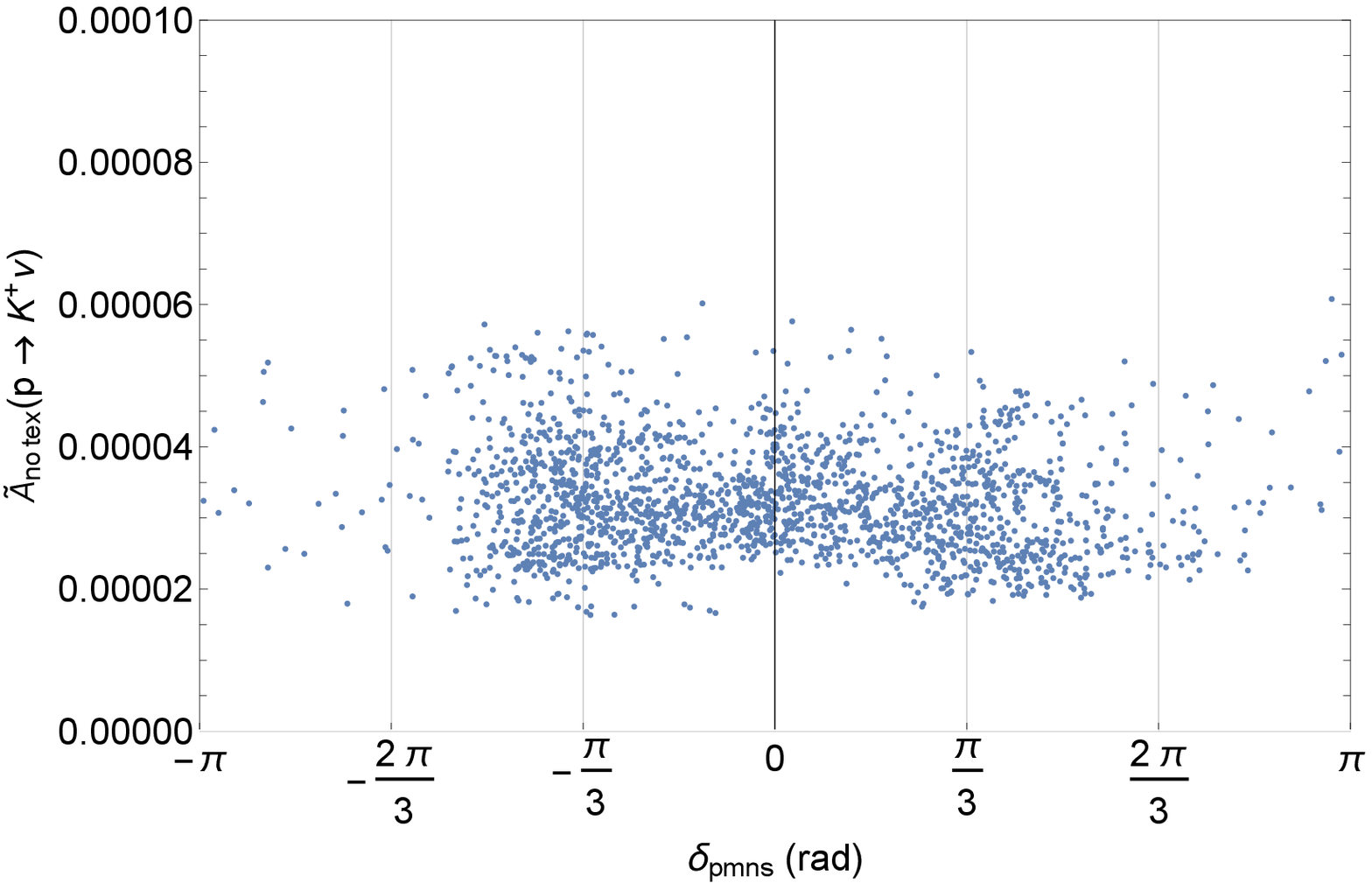}
\includegraphics[width=80mm]{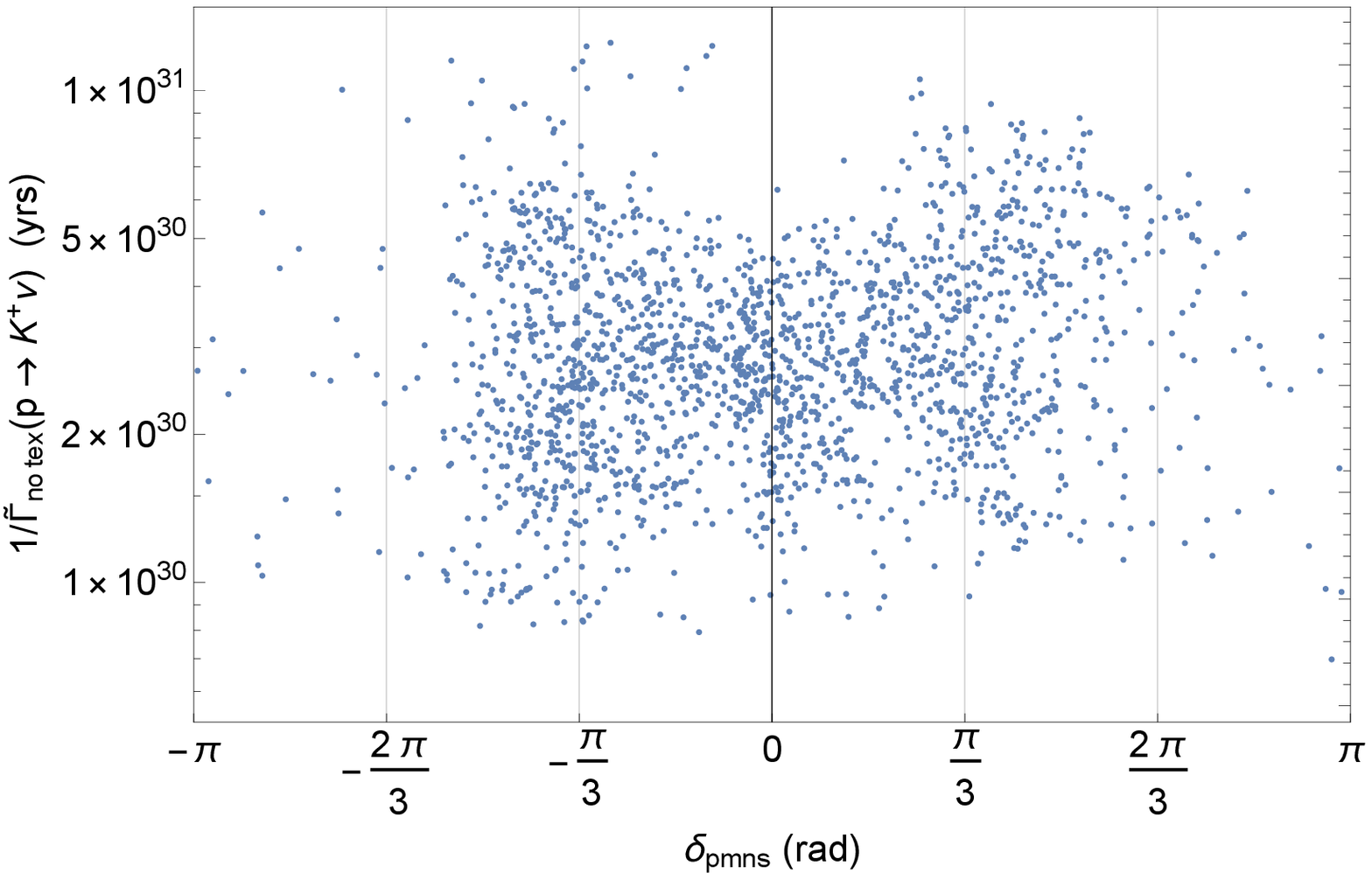}
\\
\includegraphics[width=80mm]{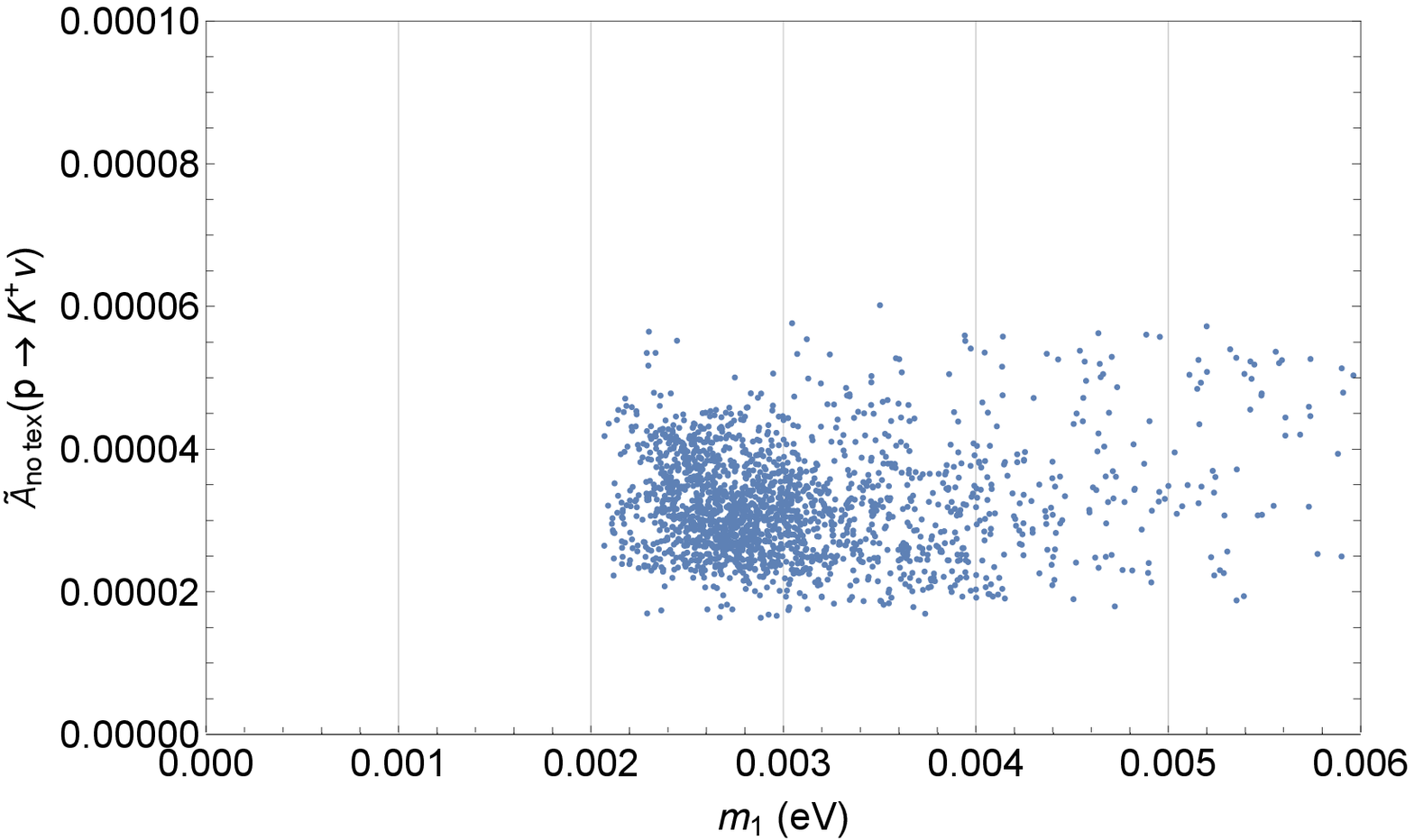}
\includegraphics[width=80mm]{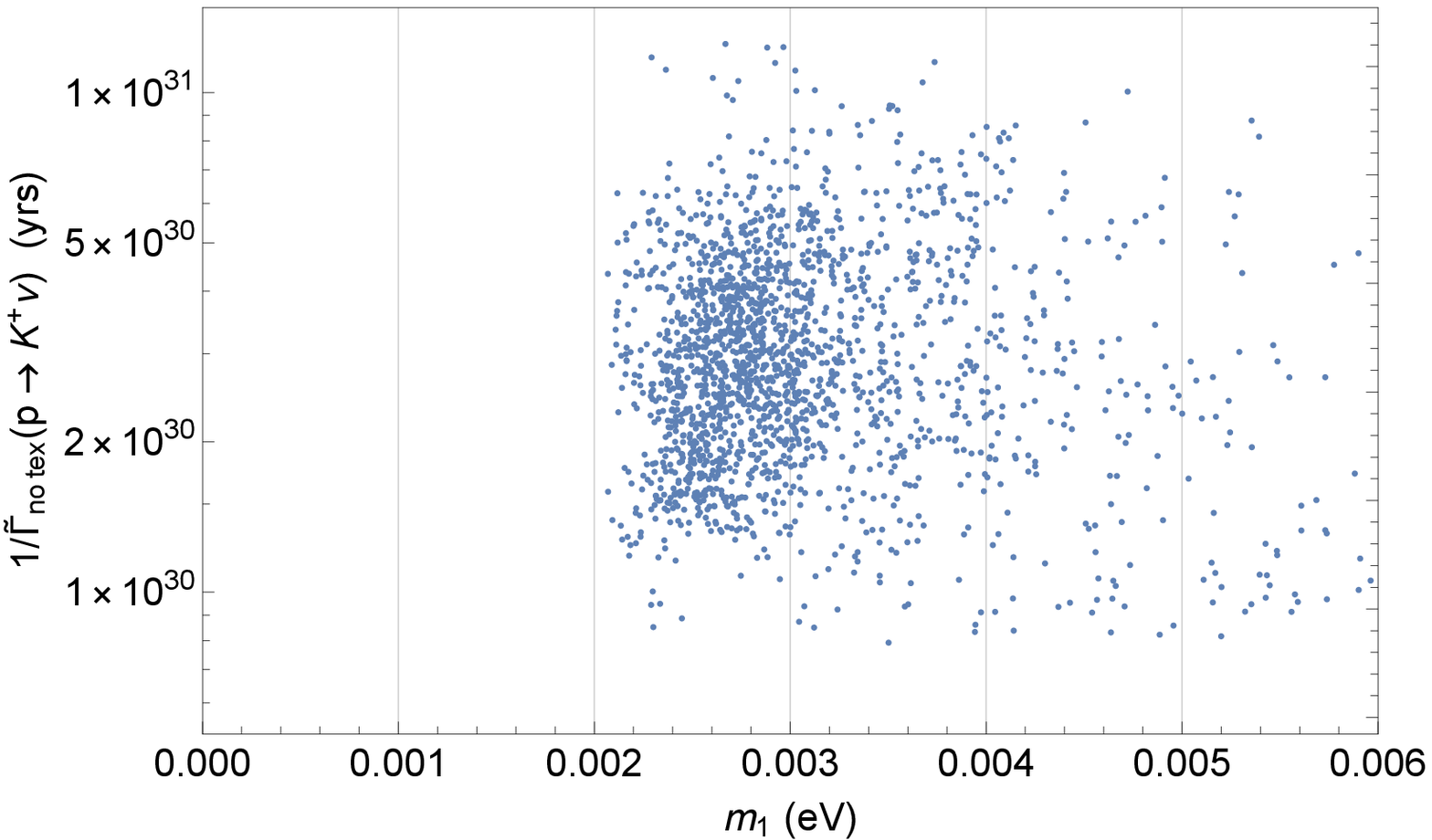}
\\
\includegraphics[width=80mm]{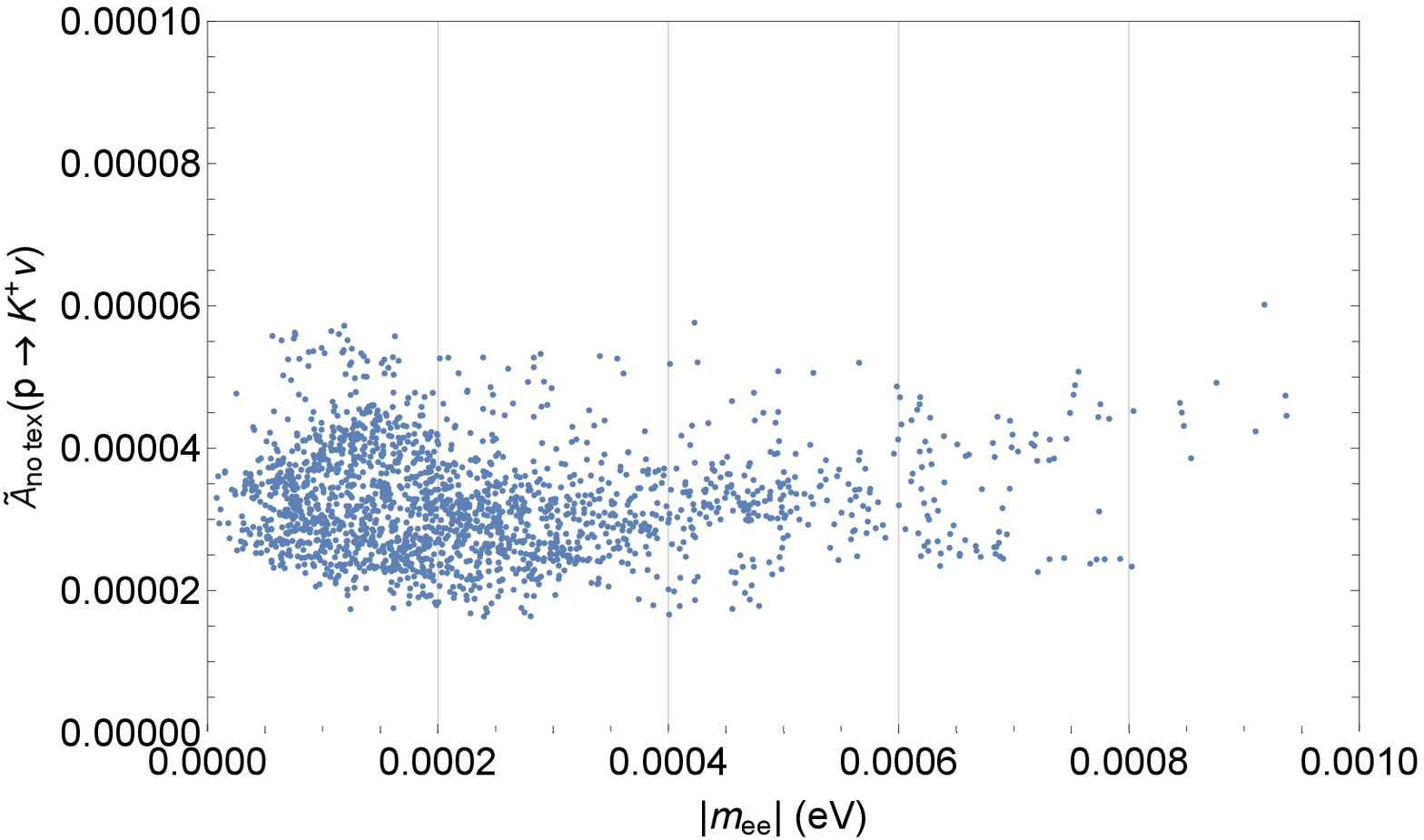}
\includegraphics[width=80mm]{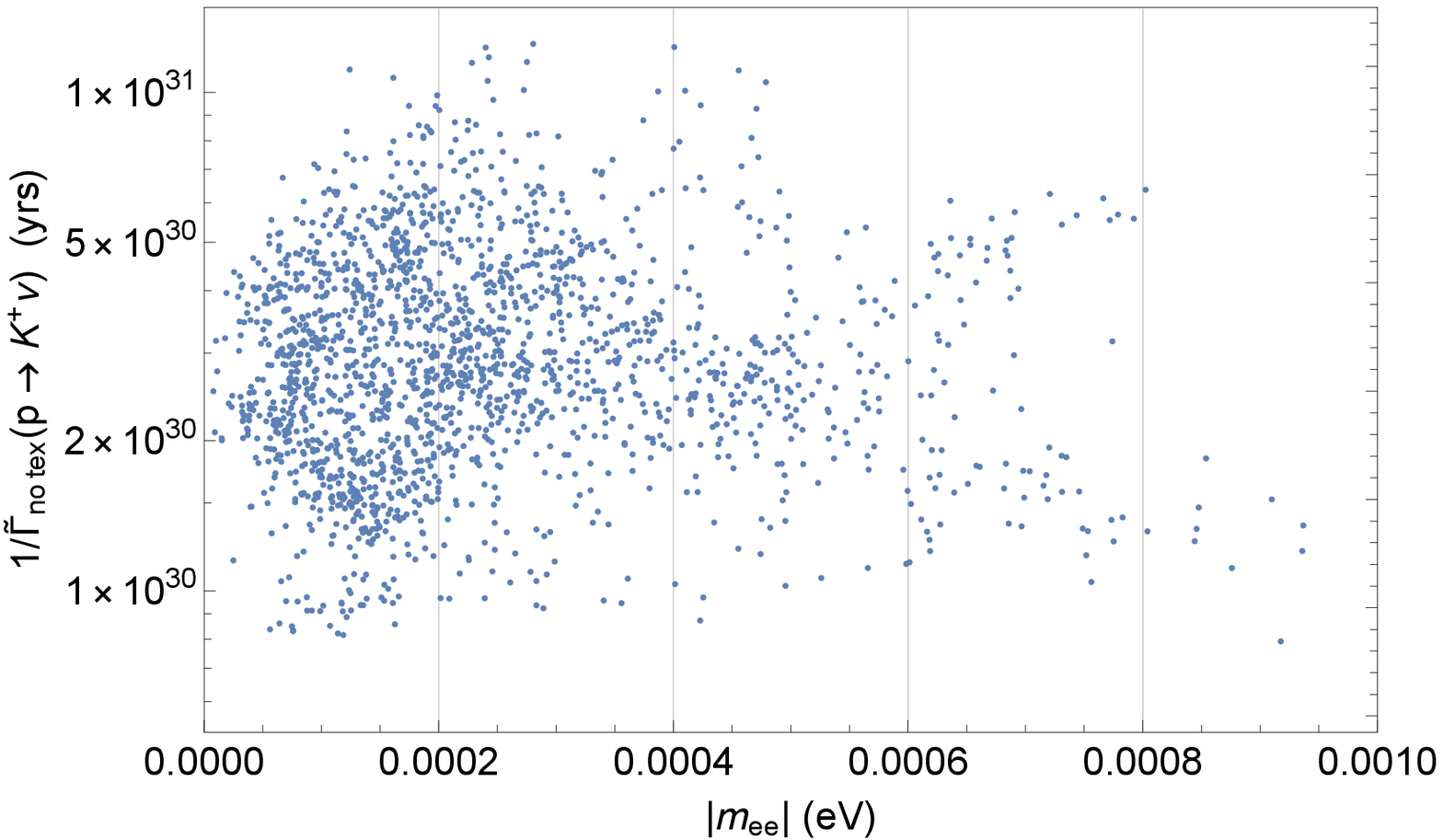}
\caption{
The same as Fig.~\ref{055} except that we assume no texture for the colored Higgs mass matrix, and 
 accordingly, we assume $|c|=|f|=1$ and that all the terms, including those with coefficient $c,f$, interfere maximally constructively.
}
\label{055-no}
\end{center}
\end{figure}

\begin{figure}[H]
\begin{center}
\includegraphics[width=80mm]{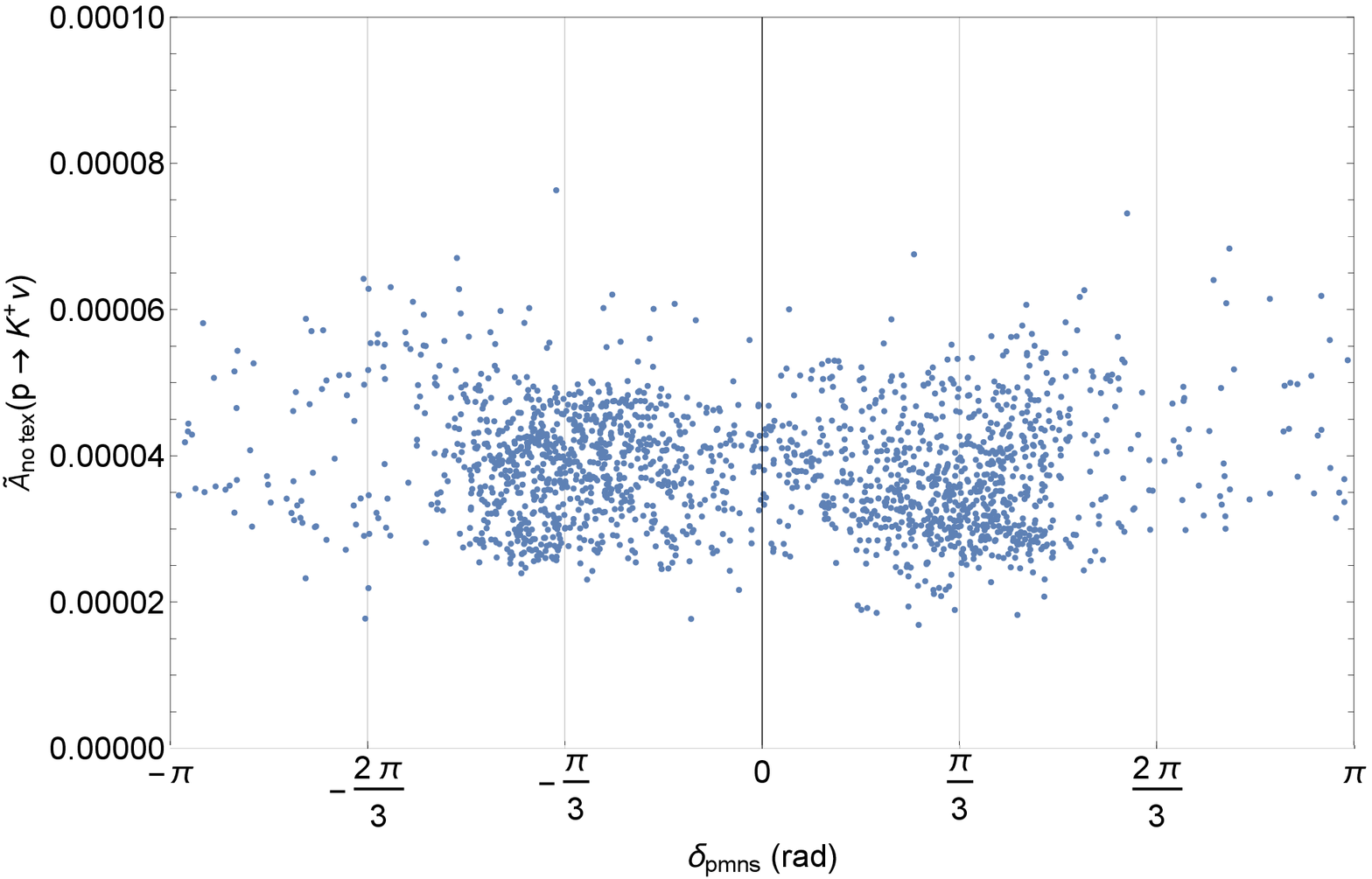}
\includegraphics[width=80mm]{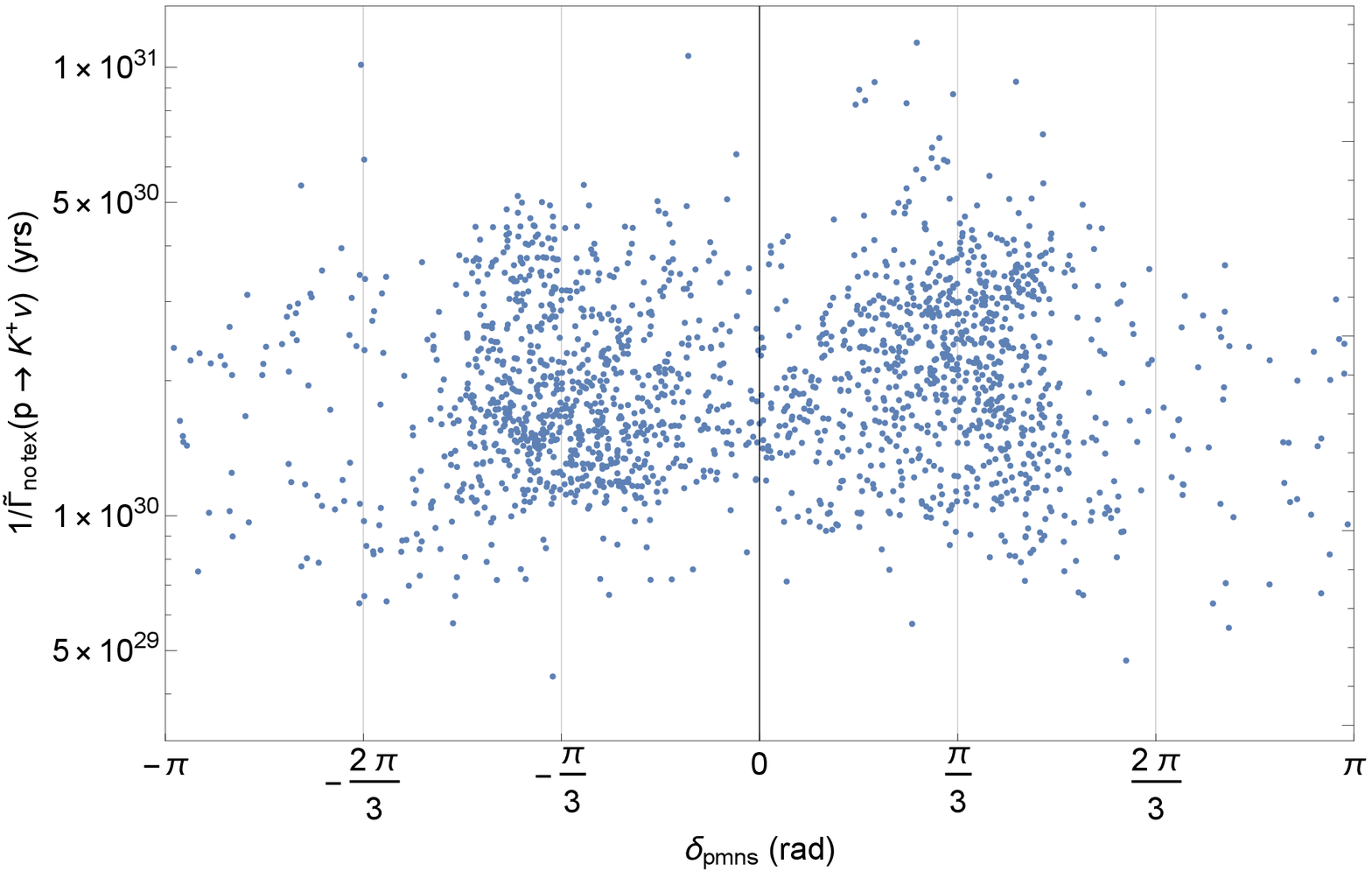}
\\
\includegraphics[width=80mm]{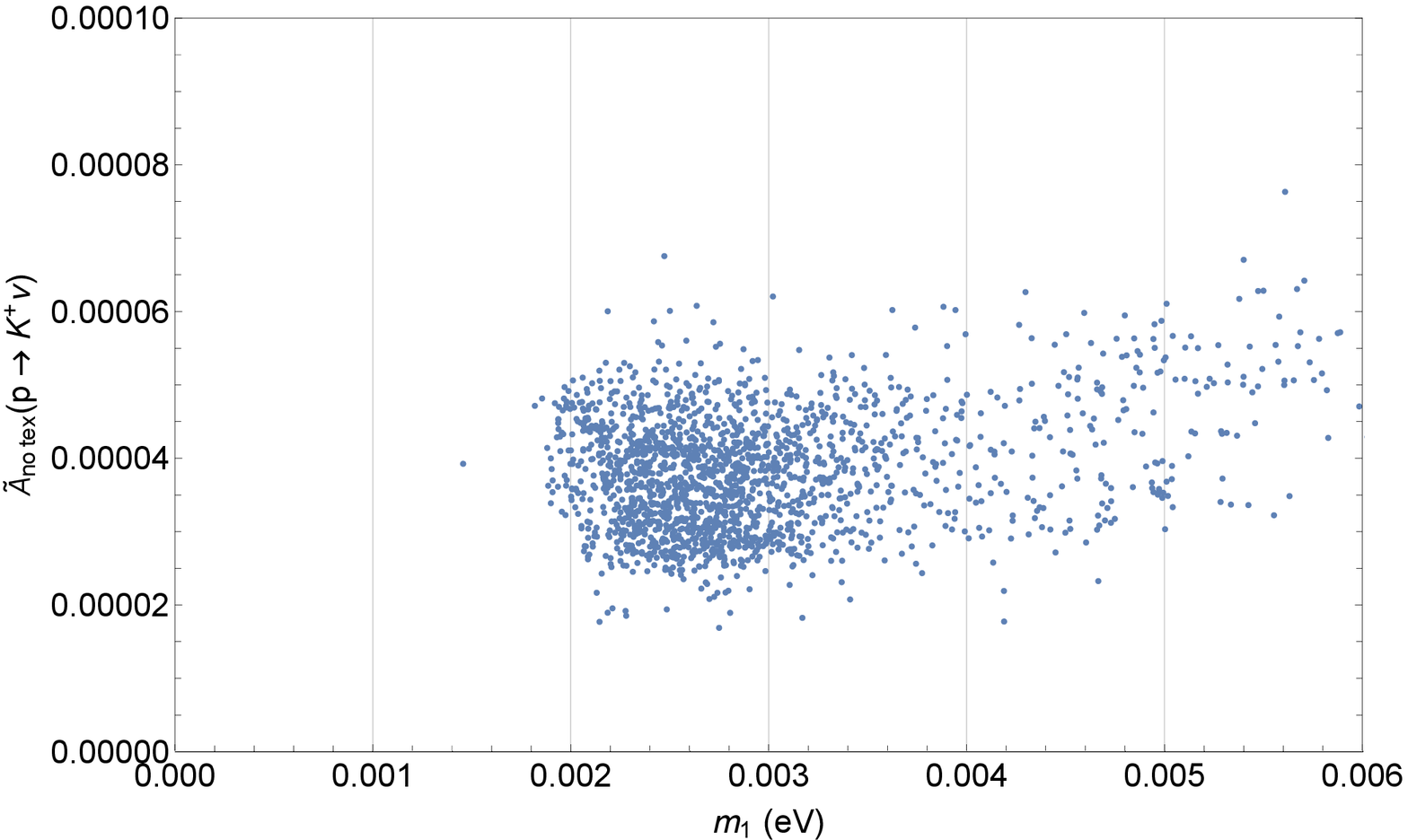}
\includegraphics[width=80mm]{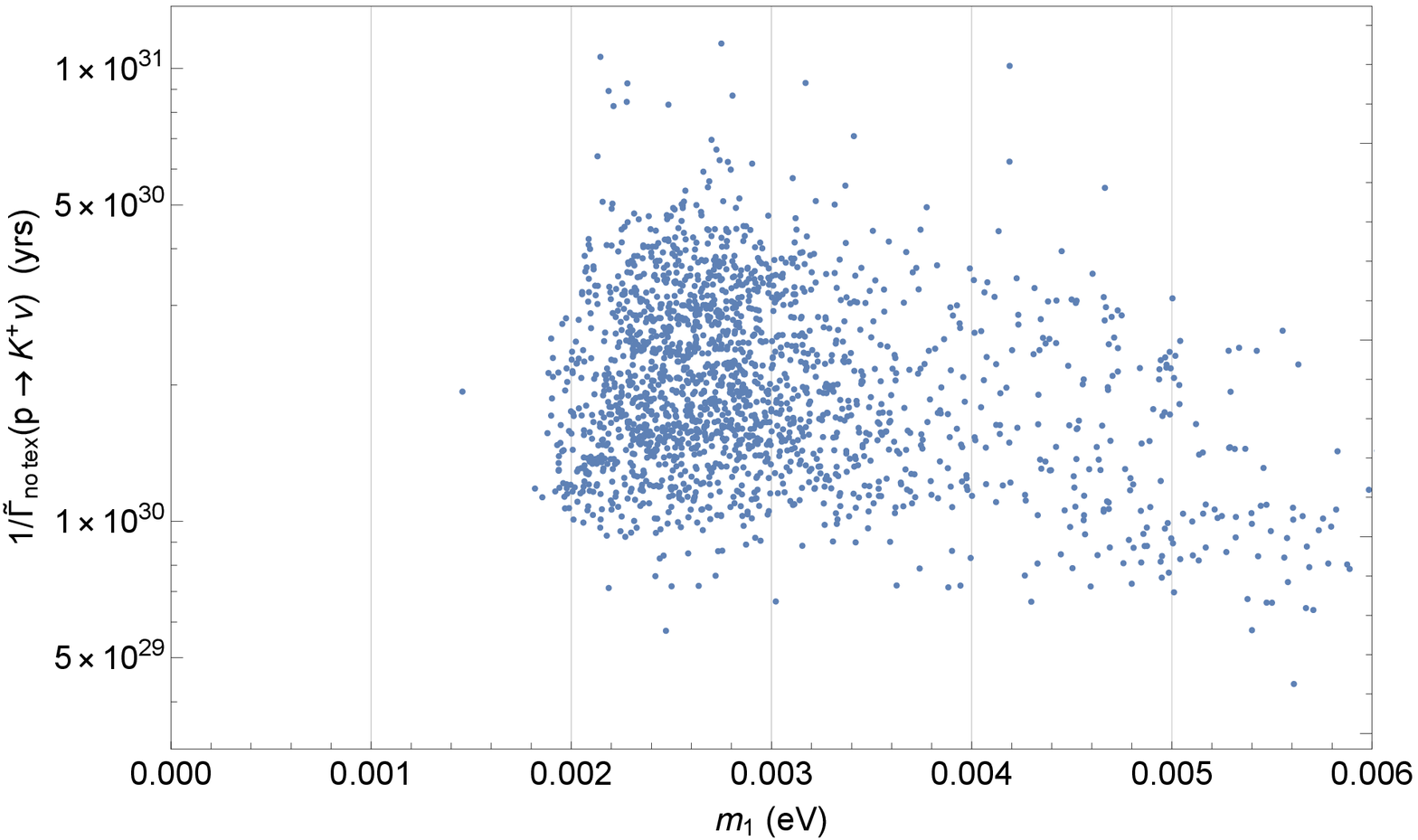}
\\
\includegraphics[width=80mm]{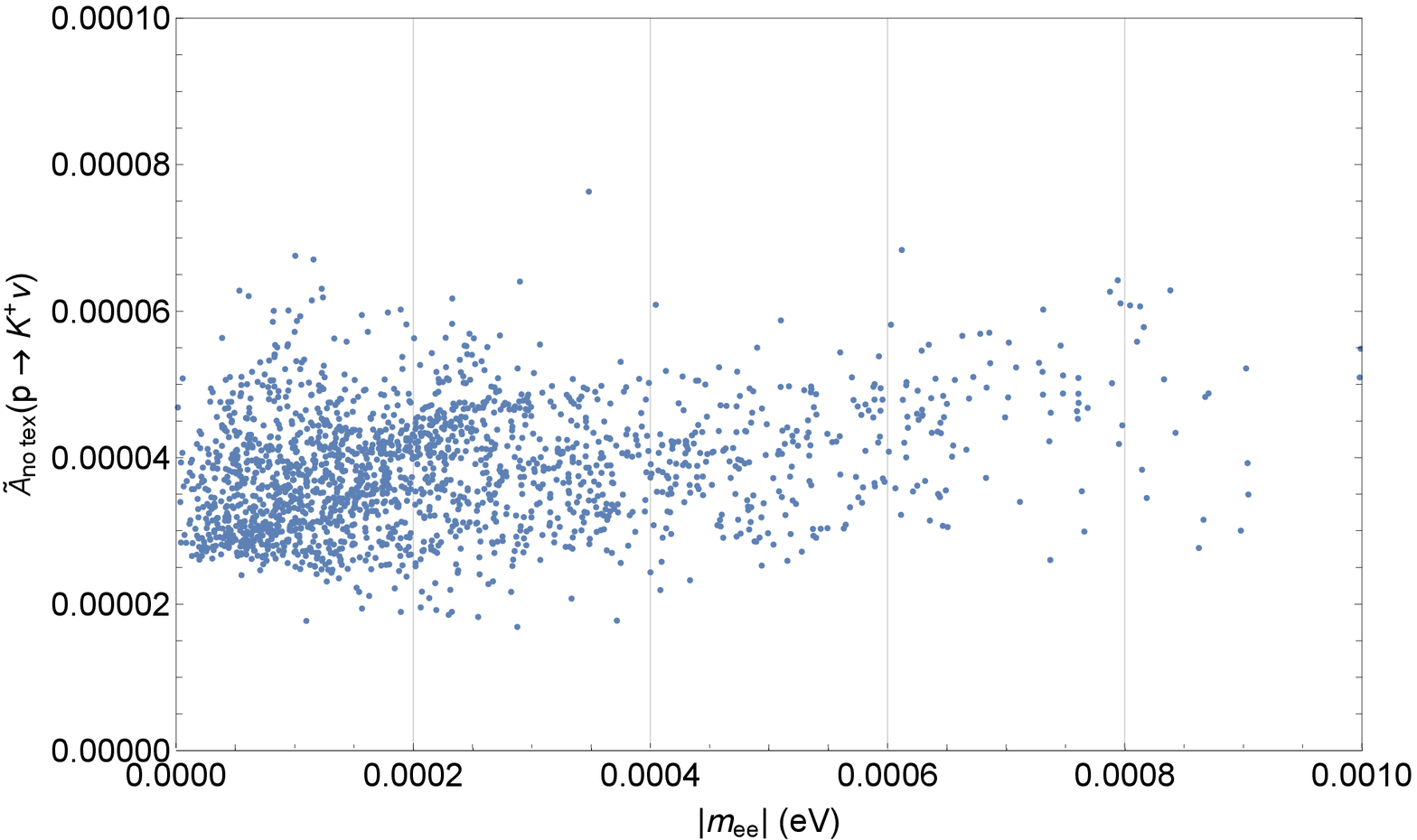}
\includegraphics[width=80mm]{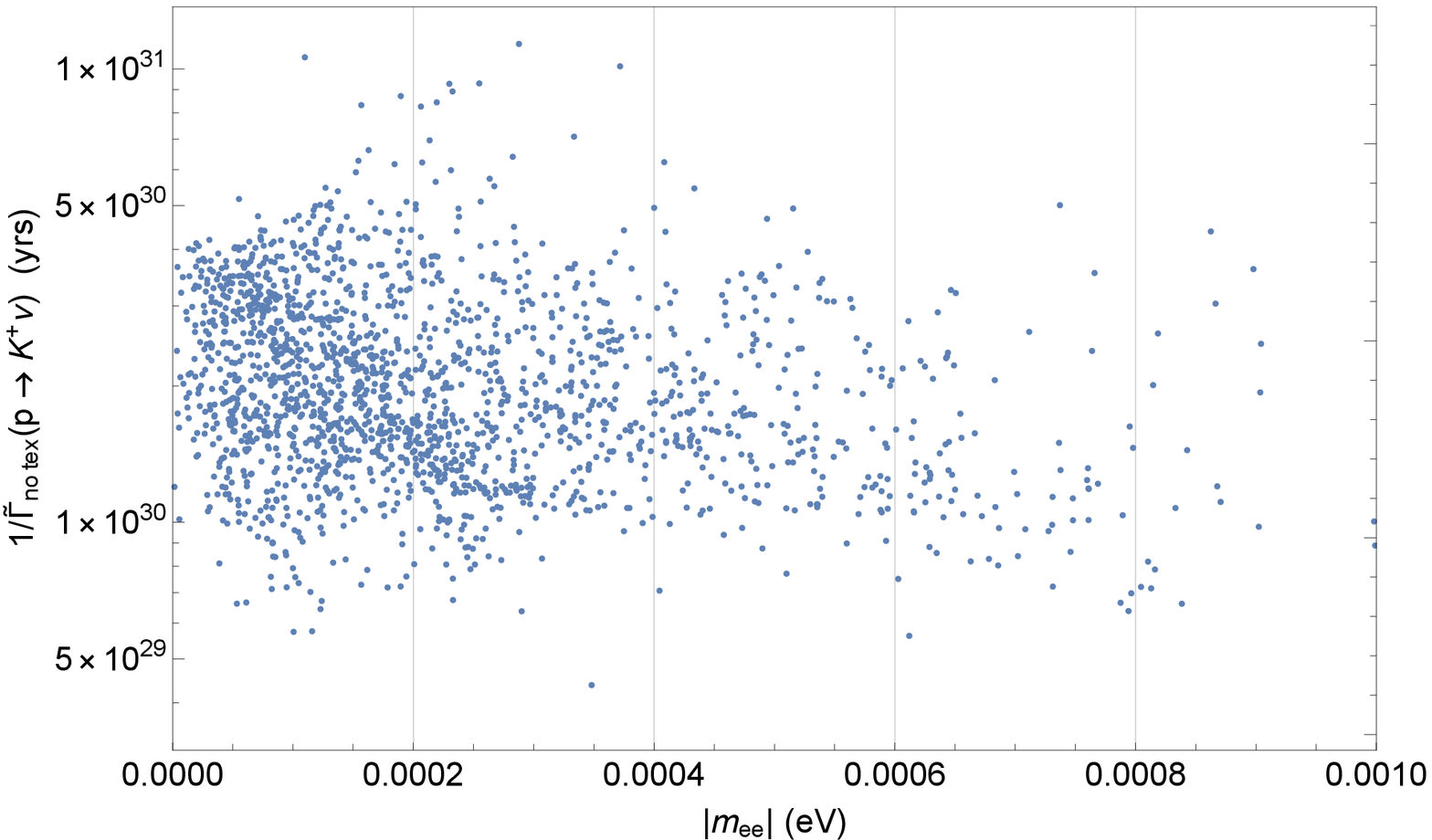}
\caption{
The same as Fig.~\ref{045} except that we assume no texture for the colored Higgs mass matrix, and 
 accordingly, we assume $|c|=|f|=1$ and that all the terms, including those with coefficient $c,f$, interfere maximally constructively.
}
\label{045-no}
\end{center}
\end{figure}

Finally, we discuss the effectiveness of the texture of the colored Higgs mass matrix
 (texture that gives $c=f=0$,
 $a\, (Y_{10})_{d_L s_L}+d\, (Y_{126})_{d_L s_L}=0$ and $b\, (Y_{10})_{d_L s_L}+e\, (Y_{126})_{d_L s_L}=0$)
 on the suppression of dimension-5 proton decays.
To this end, we compare the ``maximal proton decay amplitudes" with and without the colored Higgs mass texture,
 and the ``minimal proton partial lifetimes" with and without that texture,
 and study how the fitting and minimization results yield different values.
We define the ``maximal proton decay amplitude without colored Higgs mass texture",
 $\tilde{A}_{\rm no\,tex}(p\to K^+ \nu)$,
 and the corresponding ``minimal proton partial lifetime without colored Higgs mass texture",
 $1/\tilde{\Gamma}_{\rm no \, tex}(p\to K^+ \nu)$, as follows:
\begin{align}
\tilde{A}_{\rm no \, tex}(p\to K^+ \nu)^2 \ = \ 
&\left\{\tilde{A}_{\rm no \, tex}(p\to K^+ \bar{\nu}_\tau)|_{{\rm from} \ C_{5R}}+
\tilde{A}_{\rm no \, tex}(p\to K^+ \bar{\nu}_\tau)|_{{\rm from} \ C_{5L}}\right\}^2
\nn\\
& \ \ \ \ \ \ \ \ \ \ \ \ +\tilde{A}_{\rm no \, tex}(p\to K^+ \bar{\nu}_\mu)^2|_{{\rm from} \ C_{5L}}+
\tilde{A}_{\rm no \, tex}(p\to K^+ \bar{\nu}_e)^2|_{{\rm from} \ C_{5L}},
\nn\\
\label{maximal-no}
\end{align}
 where
\begin{align}
\tilde{A}_{\rm no \, tex}(p\to K^+ \bar{\nu}_\tau)|_{{\rm from} \ C_{5R}} \ = \ &y_ty_\tau\sum_{A,B}
\left|(1+\frac{D}{3}+F)\,V_{ts}^{\rm ckm}\,\{(Y_A)_{\tau_R t_R}(Y_B)_{u_Rd_R}-(Y_A)_{\tau_R u_R}(Y_B)_{t_Rd_R}\}\right.
\nn\\
&\left. \ \ \ \ \ \ \ \ \ \ \ \ \ \ \ \ \ \ +\frac{2D}{3}\,V_{td}^{\rm ckm}\,\{(Y_A)_{\tau_R t_R}(Y_B)_{u_Rs_R}-(Y_A)_{\tau_R u_R}(Y_B)_{t_Rs_R}\}\right|
\nn\\
&{\rm ( \ sum \ over} \ A=10,126,120 \ {\rm and} \ B=10,26,120),
\nn\\\label{maximal-no-2}
\\
\tilde{A}_{\rm no \, tex}(p\to K^+ \bar{\nu}_\alpha)|_{{\rm from} \ C_{5L}} \ = \ &g_2^2\sum_{A,B} \frac{3}{2}
\left|(1+\frac{D}{3}+F)\{
(Y_A)_{u_L d_L}(Y_B)_{s_L \alpha_L}-(Y_A)_{d_L s_L}(Y_B)_{u_L \alpha_L}\}\right.
\nn\\
&\left. \ \ \ \ \ \ \ \ \ \ \ \ \ \ \ \ \ \ +\frac{2D}{3}\{
(Y_A)_{u_L s_L}(Y_B)_{d_L \alpha_L}-(Y_A)_{d_L s_L}(Y_B)_{u_L \alpha_L}\}
\right|
\nn\\
&{\rm ( \ sum \ over} \ A=10,126 \ {\rm and} \ B=10,26,120)
\nn\\\label{maximal-no-3}
\end{align}
 with $\alpha=e,\mu,\tau$, and
\begin{align}
&\tilde{\Gamma}_{\rm no \, tex}(p\to K^+ \nu) = \frac{m_N}{64\pi}\left(1-\frac{m_K^2}{m_N^2}\right)^2 \times
\nn\\
& \ \ \ \ \ \ \ \ \left[ \ \left\{
\left|
\frac{\alpha_H(\mu_{\rm had})}{f_\pi}
A_{RL}(\mu_{\rm had},\mu_{\rm SUSY}) \frac{\mu_H}{m_{\tilde{t}_R}^2} {\cal F}' \, A_R^{\tau t}(\mu_{\rm SUSY},\mu_{H_C}) 
\frac{1}{M_{H_C}}\tilde{A}_{\rm no \, tex}(p\to K^+ \bar{\nu}_\tau)|_{{\rm from} \ C_{5R}}\right|
\right.\right.
\nn\\
&\left. \ \ \  \ \ \ \ \ \ \ \ \ \ +\left|
\frac{\beta_H(\mu_{\rm had})}{f_\pi}
A_{LL}(\mu_{\rm had},\mu_{\rm SUSY}) \frac{M_{\widetilde{W}}}{m_{\tilde{q}}^2} {\cal F} \, A_L^\tau(\mu_{\rm SUSY},\mu_{H_C}) 
\frac{1}{M_{H_C}}\tilde{A}_{\rm no \, tex}(p\to K^+ \bar{\nu}_\tau)|_{{\rm from} \ C_{5L}}\right|
\right\}^2
\nn\\
& \ \ \  \ \ \ \ \ +\left|\frac{\beta_H(\mu_{\rm had})}{f_\pi} A_{LL}(\mu_{\rm had},\mu_{\rm SUSY}) \frac{M_{\widetilde{W}}}{m_{\tilde{q}}^2} {\cal F}\,
A_L(\mu_{\rm SUSY},\mu_{H_C}) \frac{1}{M_{H_C}}\tilde{A}_{\rm no \, tex}(p\to K^+ \bar{\nu}_\mu)|_{{\rm from} \ C_{5L}}\right|^2
\nn\\
&\left. \ \ \  \ \ \ \ \ +\left|\frac{\beta_H(\mu_{\rm had})}{f_\pi} A_{LL}(\mu_{\rm had},\mu_{\rm SUSY}) \frac{M_{\widetilde{W}}}{m_{\tilde{q}}^2} {\cal F}\,
A_L(\mu_{\rm SUSY},\mu_{H_C}) \frac{1}{M_{H_C}}\tilde{A}_{\rm no \, tex}(p\to K^+ \bar{\nu}_e)|_{{\rm from} \ C_{5L}}\right|^2 \ \right]
\label{samplepartiallifetime-no}
\end{align}
 with the SUSY particle spectrum of Eq.~(\ref{massspectrum-new}).
It is important to note that the terms with coefficient $c,f$ are revived in Eqs.~(\ref{maximal-no-2}),(\ref{maximal-no-3}) and the $(Y_A)_{d_Ls_L}(Y_B)_{u_L\alpha_L}$ terms are revived in Eq.~(\ref{maximal-no-3}),
 corresponding to the situation where no texture is assumed for the colored Higgs mass matrix,
 and $|c|=|f|=O(1)$ and all the terms, including those with coefficient $c,f$, interfere maximally constructively.
We plot the fitting and minimization results on the planes of $\delta_{\rm pmns}$, $m_1$, $|m_{ee}|$ versus
 ``maximal proton decay amplitude without colored Higgs mass texture"
 $\tilde{A}_{\rm no\,tex}(p\to K^+ \nu)$ or ``minimal proton partial lifetime without colored Higgs mass texture"
 $1/\tilde{\Gamma}_{\rm no \, tex}(p\to K^+ \nu)$
  in Figs.~\ref{055-no},\ref{045-no}.
Each dot corresponds to one fitting and minimization result that has appeared in Figs.~\ref{055},\ref{045}.
Figs.~\ref{055-no},\ref{045-no} tell us that without the texture of the colored Higgs mass matrix,
 the fitting and minimization results always give $\tilde{A}(p\to K^+\nu)>10^{-5}$ and 
 $1/\tilde{\Gamma}(p\to K^+\nu)<2\times10^{31}$~yrs 
 (for the SUSY particle spectrum of Eq.~(\ref{massspectrum-new})),
  which means that the $p\to K^+\nu$ decay is highly enhanced compared to the case with that texture.
We thus conclude that the texture of the colored Higgs mass matrix with $c=f=0$,
 $a\, (Y_{10})_{d_L s_L}+d\, (Y_{126})_{d_L s_L}=0$ and $b\, (Y_{10})_{d_L s_L}+e\, (Y_{126})_{d_L s_L}=0$
 plays an important role in the suppression of dimension-5 proton decays.
\\

\section{Summary}
\label{summary}

In the renormalizable SUSY $SO(10)$ GUT model which includes single ${\bf 10}$, single ${\bf \overline{126}}$ and single ${\bf 120}$ fields
 and where the renormalizeble terms $\tilde{Y}_{10}\,{\bf 16}\, {\bf 10}\, {\bf 16}+\tilde{Y}_{126}\,{\bf 16}\, {\bf \ov{126}}\, {\bf 16}+\tilde{Y}_{120}\,{\bf 16}\, {\bf 120}\, {\bf 16}$ account for the quark and lepton Yukawa couplings and neutrino mass matrix,
 we have pursued the possibility that a texture of the fundamental Yukawa couplings $\tilde{Y}_{10},\tilde{Y}_{126},\tilde{Y}_{120}$
 suppresses dimension-5 proton decays while reproducing the correct fermion data.
Here we have assumed that the active neutrino mass comes mostly from the Type-2 seesaw mechanism.
First, we have spotted those components of the Yukawa matrices $Y_{10}(\propto\tilde{Y}_{10})$, $Y_{126}(\propto\tilde{Y}_{126})$, $Y_{120}(\propto\tilde{Y}_{120})$ which can be reduced to suppress dimension-5 proton decays without conflicting the requirement that
 they reproduce the correct quark and lepton Yukawa couplings and neutrino mass matrix.
Next, we have performed a numerical search for the texture of $Y_{10},Y_{126},Y_{120}$
 by fitting the data on the quark and lepton masses, CKM and PMNS matrices
 and neutrino mass differences and at the same time minimizing the above-spotted components of the Yukawa matrices.
We have investigated implications of the texture on unknown neutrino parameters and found that
 the ``maximal proton decay amplitude" $\tilde{A}(p\to K^+ \nu)$, which quantifies how much dimension-5 proton decays are suppressed by the Yukawa couplings,
 is minimized in the region where the neutrino Dirac CP phase satisfies $\pi/2\gtrsim \delta_{\rm pmns} \gtrsim -\pi/2$,
 the lightest neutrino mass is around $m_1\simeq0.003$~eV, and the 
 (1,1)-component of the neutrino mass matrix in the charged lepton basis satisfies $|m_{ee}|\lesssim0.0002$~eV.
The above results do not depend on the precise value of $\theta_{23}$ neutrino mixing angle.
Additionally, we present ``minimal proton partial lifetime" $1/\tilde{\Gamma}(p\to K^+ \nu)$, which corresponds to the ``maximal proton decay amplitude" and is computed for a sample SUSY particle mass spectrum.
\\

\section*{Acknowledgement}
This work is partially supported by Scientific Grants by the Ministry of Education, Culture, Sports, Science and Technology of Japan,
Nos.~17K05415, 18H04590 and 19H051061 (NH), and No.~19K147101 (TY).
\\

\section*{Appendix~A}

We review our definition of the coupling constants and masses for
 $H$, $\Delta$, $\overline{\Delta}$, $\Sigma$, $\Phi$, $A$ fields in 
 ${\bf 10}$, ${\bf 126}$, ${\bf \overline{126}}$, ${\bf 120}$, ${\bf 210}$, ${\bf 45}$ representations,
 which follows Eq.~(2) of Ref.~\cite{Fukuyama:2004ps}.
The couplings are defined in the same way as Eq.~(3) of Ref.~\cite{Fukuyama:2004ps}.
Note that ${\bf 120}$ representation field is written as $D$ in Ref.~\cite{Fukuyama:2004ps}, while we write it as $\Sigma$.
The coupling constants are defined as
\bea
W &=& \frac{1}{2}m_1\Phi^2 + m_2 \overline{\Delta}\Delta + \frac{1}{2}m_3H^2
\nn\\ &+& \frac{1}{2}m_4A^2 + \frac{1}{2}m_6\Sigma^2
\nn\\ &+& \lambda_1\Phi^3 + \lambda_2\Phi\overline{\Delta}\Delta + (\lambda_3\Delta+\lambda_4\overline{\Delta})H\Phi
\nn\\ &+& \lambda_5 A^2\Phi - i\lambda_6A\overline{\Delta}\Delta + \frac{\lambda_7}{120}\varepsilon A\Phi^2
\nn\\ &+& \lambda_{15}\Sigma^2\Phi
\nn\\ &+& \Sigma\{\lambda_{16}HA + \lambda_{17}H\Phi + (\lambda_{18}\Delta+\lambda_{19}\overline{\Delta})A + (\lambda_{20}\Delta+\lambda_{21}\overline{\Delta})\Phi\}
\label{gutsuperpotential}
\eea
 where $\varepsilon$ denotes the antisymmetric tensor in $SO(10)$ space.
\\

\section*{Appendix~B}

We present an example of VEV configurations that realize the dominance of the Type-2 seesaw contribution to the active neutrino mass
 without affecting the gauge coupling unification.
Specifically, we elaborate a VEV configuration that renders one $({\bf 1},{\bf 3},1)$ particle, one $({\bf 6},{\bf 1},-\frac{2}{3})$ particle and one $({\bf 3},{\bf 2},\frac{1}{6})$
 particle much lighter than the GUT scale (their masses are shown in Eqs.~(\ref{massof131})-(\ref{massof3216})).
The $({\bf 1},{\bf 3},1)$ particle comes from $\Delta,\ov{\Delta}$ fields, and it realizes the Type-2 seesaw mechanism through its coupling with $H_uH_u$ 
  generated by the $\lambda_3 \Delta H\Phi$ term and coupling with $L_iL_j$ generated by the $(\tilde{Y}_{126})_{ij}\Psi_i\ov{\Delta}\Psi_j$ term
  (as shown in Eq.~(\ref{type2formula})).
Since $({\bf 1},{\bf 3},1)$, $({\bf 6},{\bf 1},-\frac{2}{3})$ and $({\bf 3},{\bf 2},\frac{1}{6})$ representations complete {\bf 15} representation of $SU(5)$ subgroup,
 that they are lighter than the GUT scale does not affect the gauge coupling unification.
Therefore, by making the mass of these particles sufficiently small and by increasing the VEV of $\Delta,\ov{\Delta}$,
 we can achieve the dominance of the Type-2 seesaw contribution without spoiling the gauge coupling unification.

Our example VEV configuration is given by
\bea
&&\left|\lambda_2\Phi_1 - 30\sqrt{6}i\,\frac{\lambda_2\lambda_{19}}{\lambda_6\lambda_{21}}m_2 + 20\sqrt{6}m_2\right| \ \sim \ M_{\rm int}, 
\nn\\
&&|\lambda_2\Phi_2 - 30\sqrt{2}i\,\frac{\lambda_2\lambda_{19}}{\lambda_6\lambda_{21}}m_2| \ \sim \ M_{\rm int},
\nn\\
&&\left|\lambda_2\Phi_3 + 60i\,\frac{\lambda_2\lambda_{19}}{\lambda_6\lambda_{21}}m_2\right| \ \sim \ M_{\rm int},
\nn\\
&&M_{\rm int} \ \ll \ 2\cdot10^{16}~{\rm GeV} \ \sim \ |\lambda_2\Phi_1| \ \sim \ |\lambda_2\Phi_2| \ \sim \ |\lambda_2\Phi_3|,
\nn\\
&&|v_R| \ > \ 10^{17}~{\rm GeV}.
\label{texture2}
\eea
The first four lines of Eq.~(\ref{texture2}) mean that $\lambda_2\Phi_1 - 30\sqrt{6}i\frac{\lambda_2\lambda_{19}}{\lambda_6\lambda_{21}}m_2 + 20\sqrt{6}m_2$,
 $\lambda_2\Phi_2 - 30\sqrt{2}i\frac{\lambda_2\lambda_{19}}{\lambda_6\lambda_{21}}m_2$, and $\lambda_2\Phi_3 + 60i\frac{\lambda_2\lambda_{19}}{\lambda_6\lambda_{21}}m_2$ are fine-tuned to nearly 0 as compared to the GUT scale,
 while $\lambda_2\Phi_1$, $\lambda_2\Phi_3$ are about the GUT scale.
These fine-tunings are not based on any symmetry, but are natural at the quantum level due to the non-renormalization theorem.
Eq.~(\ref{texture2}) and Eq.~(\ref{texture})
 can simultaneously be consistent with the the $F$-flatness conditions (displayed in Eq.~(28) of Ref.~\cite{Fukuyama:2004ps})
 if one tunes $m_1,m_4,\lambda_1,\lambda_2,\lambda_5$ appropriately.
The relation $|v_R| > 10^{17}~{\rm GeV}$ does not raise the mass of GUT-scale particles much above $2\cdot10^{16}~{\rm GeV}$
 as long as we take $\lambda_2,\lambda_3,\lambda_4,\lambda_6,\lambda_{18},\lambda_{19},\lambda_{20},\lambda_{21}$ sufficiently smaller than 1.
We have numerically confirmed that even with the highly restricted VEV configuration and coupling constants of Eqs.~(\ref{texture}),(\ref{texture2}),
 the ratio of the masses of various fields (other than the pair of (${\bf 1}$, ${\bf2}$, $\pm\frac{1}{2}$) fields with zero mass eigenvalue)
 and the values of
 $a/d,r_2,r_3,r_e$ vary in a wide range and there is no strong correlation among them.

From Eq.~(\ref{texture2}) and Eq.~(\ref{texture}), we obtain
\begin{align}
&\left|m_2-\frac{1}{10\sqrt{6}}\lambda_2\Phi_1+\frac{1}{10\sqrt{2}}\lambda_2\Phi_2+\frac{1}{5}\sqrt{\frac{3}{2}}\lambda_6A_2\right| \ \sim \ M_{\rm int},
\label{massof131}\\
&\left|m_2+\frac{1}{10\sqrt{6}}\lambda_2\Phi_1-\frac{1}{30\sqrt{2}}\lambda_2\Phi_2+\frac{1}{30}\lambda_2\Phi_3+\frac{1}{5}\lambda_6A_1+\frac{1}{5\sqrt{6}}\lambda_6A_2\right| \ \sim \ M_{\rm int},
\label{massof61m23}\\
&\left|m_2+\frac{1}{30\sqrt{2}}\lambda_2\Phi_2+\frac{1}{60}\lambda_2\Phi_3+\frac{1}{10}\lambda_6A_1+\frac{1}{5}\sqrt{\frac{2}{3}}\lambda_6A_2\right| \ \sim \ M_{\rm int}
\label{massof3216}
\end{align}
 with $M_{\rm int}\ll2\cdot10^{16}$.
The left-hand side of Eq.~(\ref{massof131}) is the mass of the $({\bf 1},{\bf 3},1)$ particle, which comes from $\Delta,\ov{\Delta}$.
That of Eq.~(\ref{massof61m23}) is the mass of the $({\bf 6},{\bf 1},-\frac{2}{3})$ particle, which also comes from $\Delta,\ov{\Delta}$.
Eq.~(\ref{massof3216}) and the relation $\lambda_{18}=\lambda_{20}=0$ guarantee that one eigenvalue of the mass matrix of the $({\bf 3},{\bf 2},\frac{1}{6})$ fields
 is about $M_{\rm int}$ (refer to Eq.~(65) of Ref.~\cite{Fukuyama:2004ps}).
Since $({\bf 1},{\bf 3},1)+({\bf 6},{\bf 1},-\frac{2}{3})+({\bf 3},{\bf 2},\frac{1}{6})$ completes {\bf 15} representation of $SU(5)$,
 the presence of one $({\bf 1},{\bf 3},1)$, one $({\bf 6},{\bf 1},-\frac{2}{3})$ and one $({\bf 3},{\bf 2},\frac{1}{6})$ particles with mass of order $M_{\rm int}$
 does not affect the gauge coupling unification irrespectively of the value of $M_{\rm int}$.
Also, the unified gauge coupling is perturbative around the GUT scale for any value of $M_{\rm int}$.
Therefore, we can take $M_{\rm int}$ arbitrarily small.

\begin{figure}[H]
\begin{center}
\includegraphics[width=80mm]{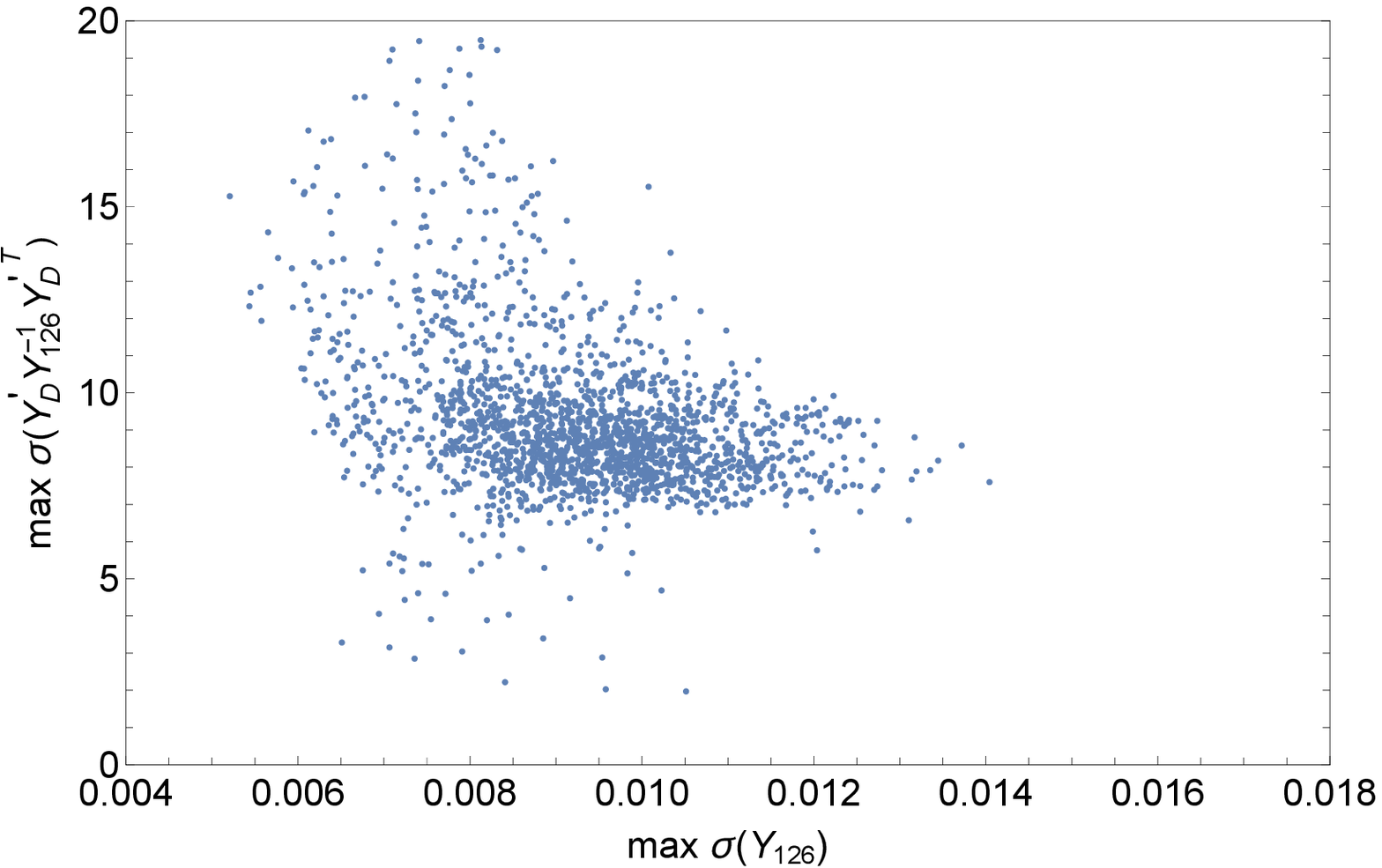}
\includegraphics[width=80mm]{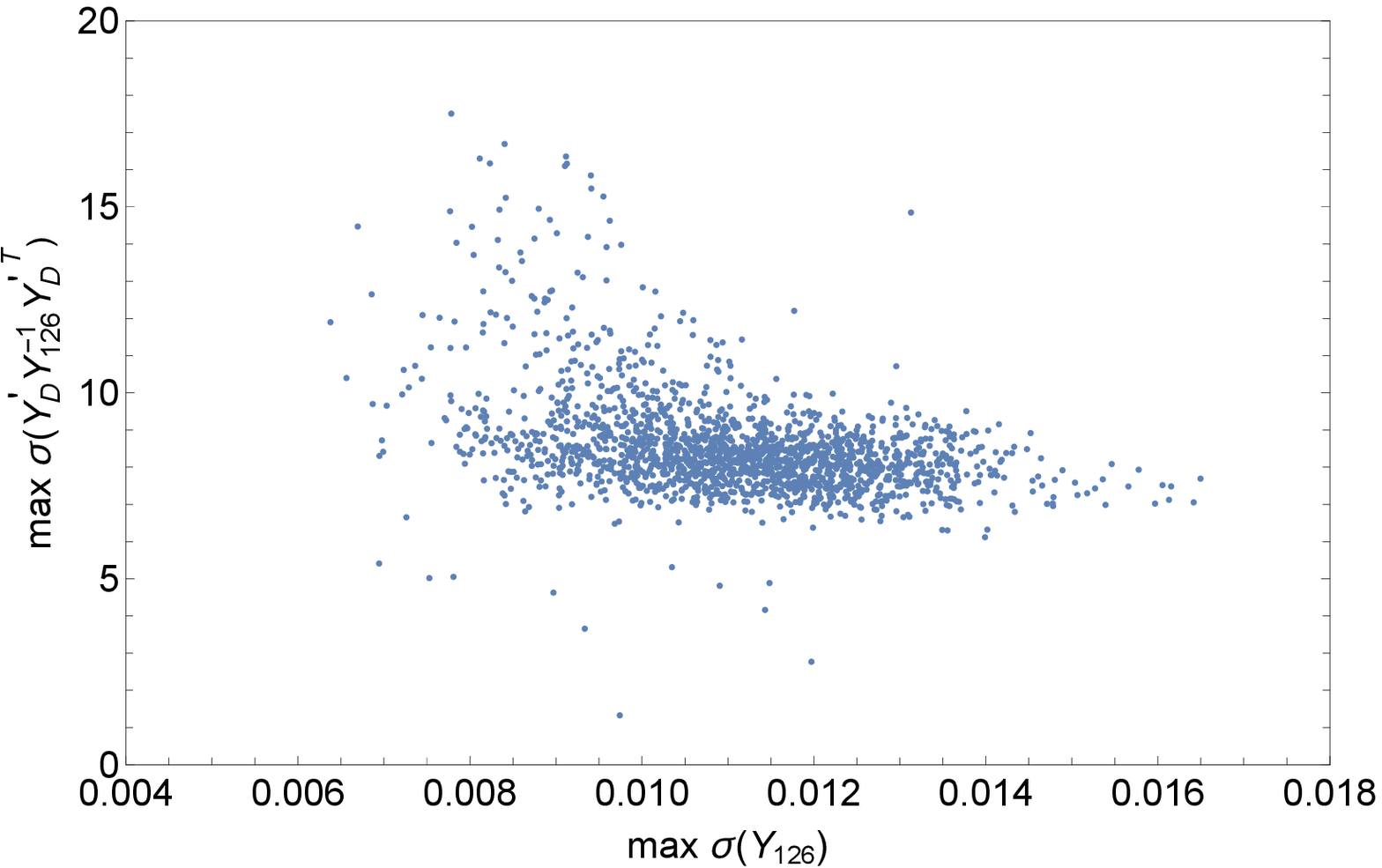}
\caption{
The largest singular value of $Y_{126}$, max\,$\sigma(Y_{126})$, and that of $Y_D'Y_{126}^{-1}(Y_D')^T$ with $Y_D'=Y_{10}-3r_2Y_{126}$, 
 max\,$\sigma(Y_D'Y_{126}^{-1}(Y_D')^T)$.
Each dot corresponds to the result of one fitting and minimization analysis, and hence to one dot in Figs.~\ref{055},\ref{045}.
The left panel shows the results for the higher-octant benchmark where $\sin^2\theta_{23}^{\rm pmns}=0.55\pm0.01$,
 and the right panel shows those for the lower-octant benchmark where $\sin^2\theta_{23}^{\rm pmns}=0.45\pm0.01$.
}
\label{type21}
\end{center}
\end{figure}

The $({\bf 1},{\bf 3},1)$ particle, coming from $\Delta,\ov{\Delta}$, generates the Type-2 seesaw contribution to the active neutrino mass through the couplings of 
 $\lambda_3\,\Delta H\Phi$ and $(\tilde{Y}_{126})_{ij}\Psi_i\ov{\Delta}\Psi_j$.
The Type-2 seesaw contribution is estimated to be
\bea
m_\nu^{\rm Type\mathchar`-2} \ \sim \ Y_{126} \, \lambda_3\, U_{H}U_{\Phi} \, \frac{v^2}{M_{\rm int}}
\label{type2formula}
\eea
 where $v=246$~GeV, and $U_{H},U_{\Phi}$ denote the ratio of MSSM Higgs $H_u$ in $H$ and $\Phi$ fields, respectively.
We have found numerically that $|U_{H}|$ is around 1 and that $|U_{\Phi}|$ varies in a wide range.
The Type-1 seesaw contribution is estimated to be
\bea
m_\nu^{\rm Type\mathchar`-1} \ \sim \ Y_DY_{126}^{-1}Y_D^T \, \frac{v^2}{|v_R|}
\eea
 where $Y_D$ has been given in Eq.~(\ref{ydirac}).
To compare the Type-2 and Type-1 seesaw contributions,
 we evaluate the largest singular value of $Y_{126}$ and that of $Y_DY_{126}^{-1}Y_D^T$ for each fitting result 
 (corresponding to each dot in Figs.~\ref{055},\ref{045}).
In fact, since we have not performed a fitting of coefficient $r_D$ in $Y_D$,
 we substitute $Y_D'Y_{126}^{-1}(Y_D')^T$ with $Y_D'=Y_{10}-3r_2Y_{126}$ for $Y_DY_{126}^{-1}Y_D^T$.
We think that the singular values of $Y_D'Y_{126}^{-1}(Y_D')^T$ well approximate those of $Y_DY_{126}^{-1}Y_D^T$.
The plots for the largest singular values of $Y_{126}$ and $Y_D'Y_{126}^{-1}(Y_D')^T$ are in Fig.~\ref{type21}.
From Fig.~\ref{type21}, we see that the largest singular value of $Y_{126}$ ranges from 0.004 to 0.018, while that of $Y_D'Y_{126}^{-1}(Y_D')^T$ is below 20,
 for both benchmarks.
Since we have assumed $|v_R|>10^{17}$~GeV, the Type-1 seesaw contribution is negligible for the active neutrino mass.
On the other hand, if we take
\bea
M_{\rm int} \ \sim \ |\lambda_3U_{H}U_{\Phi}| \cdot 10^{13}~{\rm GeV},
\eea
 the Type-2 seesaw contribution accounts for the active neutrino mass.
\\

We can achieve the dominance of the Type-2 seesaw contribution also by adding a {\bf 54} representation field.
Its coupling with $\Delta^2$ gives rise to the mixing of $({\bf1},{\bf3},1)$ components of the {\bf 54} field and $\Delta$ when $\Delta$ develops a VEV.
Also, the coupling of the {\bf 54} field with $H^2$ and other fields generates the coupling of the $({\bf1},{\bf3},1)$ particles with $H_u^2$.
One can decrease the mass of the $({\bf1},{\bf3},1)$ particles without affecting the gauge coupling unification
 by fine-tuning the VEVs and coupling constants as studied in Ref.~\cite{Goh:2004fy}.
In this way, the Type-2 seesaw contribution can be made dominant.
\\


\end{document}